\documentclass[fleqn,usenatbib,useAMS,twocolumn]{mnras}

\usepackage{graphicx}	% Including figure files
\usepackage{amsmath}	% Advanced maths commands
\usepackage{amssymb}	% Extra maths symbols
\usepackage{multicol}        % Multi-column entries in tables
\usepackage{bm}		% Bold maths symbols, including upright Greek
\usepackage{pdflscape}	% Landscape pages
\usepackage{subcaption}
\usepackage[T1]{fontenc}
\usepackage{ae,aecompl}

\usepackage{newtxtext,newtxmath}
\usepackage{etoolbox}
\makeatletter
\patchcmd\@combinedblfloats{\box\@outputbox}{\unvbox\@outputbox}{}{%
  \errmessage{\noexpand\@combinedblfloats could not be patched}%
}%
\makeatother
\newcommand{\py}{\textsc{Python}}
\newcommand{\cld}{\textsc{Cloudy}}

\newcommand{\phin}{$\phi_H$-$n_H$}
\newcommand{\civline}{C\textsc{iv}~1550\AA}
\newcommand{\ciiiline}{C\textsc{iii}]~1909\AA}
\newcommand{\mgline}{Mg\textsc{ii}~2798\AA}

\newcommand{\heiiuv}{He\textsc{ii}~1640\AA}
\newcommand{\heiiopt}{He\textsc{ii}~4686\AA}
\title[Stratified disc wind models for the AGN BLR]
{
Stratified disc wind models for the AGN broad-line region: ultraviolet, optical and X-ray properties
}

\author[J.~H. Matthews et al.]
{James~H. Matthews,$^{1,2}$\thanks{E-mail: matthews@ast.cam.ac.uk}
Christian Knigge,$^{3}$
Nick Higginbottom,$^{3}$
Knox S. Long,$^{4,5}$\newauthor
Stuart A. Sim,$^{6}$
Samuel W. Mangham,$^{3}$
Edward J. Parkinson$^{3}$
and Henrietta A. Hewitt$^{6}$
\\
$^1$Institute of Astronomy, University of Cambridge, Madingley Road, Cambridge CB3 0HA, UK\\
$^2$Astrophysics, Department of Physics, University of Oxford, Keble Road, Oxford OX1 3RH, UK\\
$^{3}$School of Physics and Astronomy, University of Southampton, Highfield, Southampton, SO17 1BJ, UK\\
$^{4}$Space Telescope Science Institute, 3700 San Martin Drive, Baltimore, MD, 21218, USA\\
$^{5}$Eureka Scientific Inc., 2542 Delmar Avenue, Suite 100, Oakland, CA, 94602-3017, USA\\
$^{6}$School of Mathematics and Physics, Queen's University Belfast, University Road, Belfast, BT7 1NN, UK
}

\date{Accepted 2020 January 9. Received 2020 January 9; in original form 2019 December 12.}

\pubyear{2019}

\begin{document}
%\captionsetup{aboveskip=0pt}

\label{firstpage}
\pagerange{\pageref{firstpage}--\pageref{lastpage}}
\maketitle

\begin{abstract}
The origin, geometry and kinematics of the broad line region (BLR) gas in quasars and active galactic nuclei (AGN) are uncertain. We demonstrate that clumpy biconical disc winds illuminated by an AGN continuum can produce BLR-like spectra. We first use a simple toy model to illustrate that disc winds make quite good BLR candidates, because they are self-shielded flows and can cover a large portion of the ionizing flux-density ($\phi_H$-$n_H$) plane. We then conduct Monte Carlo radiative transfer and photoionization calculations, which fully account for self-shielding and multiple scattering in a non-spherical geometry. The emergent model spectra show broad emission lines with equivalent widths and line ratios comparable to those observed in AGN, provided that the wind has a volume filling factor of $f_V\lesssim0.1$. Similar emission line spectra are produced for a variety of wind geometries (polar or equatorial) and for launch radii that differ by an order of magnitude. The line emission arises almost exclusively from plasma travelling below the escape velocity, implying that `failed winds' are important BLR candidates. The behaviour of a line-emitting wind (and possibly any `smooth flow' BLR model) is similar to that of the locally optimally-emitting cloud (LOC) model originally proposed by Baldwin et al (1995), except that the gradients in ionization state and temperature are large-scale and continuous, rather than within or between distinct clouds. Our models also produce UV absorption lines and X-ray absorption features, and the stratified ionization structure can partially explain the different classes of broad absorption line quasars.
\end{abstract}


%%%%%%%%%%%%%%%%%%%%%%%%%%%%%%%%%%%%%%%%%%%%%%%%%%

%%%%%%%%%%%%%%%%% BODY OF PAPER %%%%%%%%%%%%%%%%%%

\section{Introduction}
The spectra of unobscured, radiatively efficient (type 1) quasars and active galactic nuclei (AGN) consist of a blue continuum and a series of broad and narrow emission lines. The broad emission lines  originate in the broad-line region (BLR), thought to be composed of approximately virialised gas in the vicinity of the black hole. The BLR is important. It allows us to measure black hole masses via virial estimates \citep[e.g.][]{krolik_ultraviolet_1991,peterson_keplerian_1999,mclure_measuring_2002,greene_estimating_2005,peterson_measuring_2014} and to carry out reverberation mapping \citep[RM;][]{blandford_reverberation_1982,peterson_reverberation_1993}. Broad emission lines also appear to be intrinsically linked to the accretion process \citep[c.f. `changing-look' quasars;][]{lamassa_discovery_2015,macleod_systematic_2016} and form a crucial part of unified models for AGN and quasars \citep[e.g.][]{antonucci_unified_1993,urry_unified_1995}. The classic AGN unification picture consists of BLR `clouds' that are obscured by some kind of dusty torus at high inclinations. The physical origin, geometry and dynamics of these putative BLR clouds are still uncertain, however. 

So, what do we know about the broad emission lines? {We know that the UV/optical spectrum is fairly universal -- the usual emission lines include \civline, Ly~$\alpha$, N\textsc{v}~1240\AA, \mgline\ and the Balmer series. The presence of the intercombination line \ciiiline\ with critical density of $10^{9.5}$cm$^{-3}$ implies there must be a significant volume of line-emitting gas with electron density below this value. We also know that the lines {\em reverberate} and respond to continuum changes \citep[e.g.][]{blandford_reverberation_1982,gaskell_line_1986,peterson_reverberation_1993,goad_response_1993} except in rare circumstances \citep{goad_space_2016}. Tracking exactly how they respond allows us to estimate the characteristic size of the BLR, $R_{\mathrm{BLR}}$, from reverberation delays. This exercise reveals a close relationship between the BLR size and the central luminosity consistent with $R_{\mathrm{BLR}} \propto L^{1/2}$ \citep{koratkar_radius-luminosity_1991,bentz_radius-luminosity_2006,bentz_radius-luminosity_2009,bentz_low-luminosity_2013,du_radius-luminosity_2019} -- roughly the form one would expect if the BLR is photoionized by the AGN continuum \cite[although see also][]{galianni_test_2013}. When observed in a single object, this phenomenon is known as the `breathing' of the BLR \citep{netzer_agn_1990,korista_what_2004}. Velocity-resolved RM allows transfer and response functions to be inferred for individual lines, which in principle allows the kinematics and geometry of the BLR to be recovered. Diverse signatures have been reported, including all of inflow, outflow and rotation \citep{gaskell_direct_1988,koratkar_emission-line_1989,ulrich_month_1996,denney_diverse_2009,gaskell_line_2013,du_supermassive_2016}. Caution must be exercised when interpreting these signatures  \citep{waters_reverberation_2016,mangham_reverberation_2017,yong_kinematics_2017}, but velocity-resolved RM is nevertheless one of the most powerful tools available for learning about the geometry and motion of BLR gas. Recently, the GRAVITY collaboration reported the first spatially resolved measurement of rotation in a quasar, showing that the Paschen-$\alpha$ BLR gas in 3C~273 is rotating in a manner consistent with Keplerian motion of a thick line-emitting disc around a BH \citep{gravity_collaboration_detection_2018}. Reviews of broad emission line phenomenology are provided by \cite{netzer_agn_1990}, \cite{sulentic_phenomenology_2000} and \cite{gaskell_what_2009}, among others.

A key advance in the understanding of the BLR was the discovery that the emission line spectrum arises naturally if one invokes a series of moderately optically thick clouds with a wide range of densities placed at a wide range of radii \citep[][hereafter BFKD95]{baldwin_locally_1995}. This model tends to average to the correct output spectrum due to strong selection effects: as long as a sufficiently wide range of cloud densities is available at all radii, there will almost always be clouds that are optimally efficient at reprocessing any given continuum into any given line. Under these conditions, each line is formed primarily in the clouds that optimally produce it. The resulting spectrum does a good job of replicating the fairly uniform BLR spectrum observed from object to object. We refer to this model and its derivatives as locally optimally-emitting cloud (LOC) models. LOC models have been applied relatively successfully to observations of AGN, both in general and for individual objects such as NGC 5548 \citep{korista_locally_2000,korista_what_2004,hamann2002,nagao2006,baldwin_origin_2004,lawther_quantifying_2018}.

It is also possible that the broad emission lines are produced by outflowing material driven from the accretion disc -- a disc wind. Smoking-gun signatures of winds are seen in large subsets of AGN and quasar populations. The clearest examples occur in the ultraviolet band in a class of objects known as broad absorption line (BAL) quasars, which make up at least 20\% of all luminous quasars \citep{weymann_comparisons_1991,knigge_intrinsic_2008,dai_2mass_2008,allen_strong_2011}. At X-ray wavelengths, ultra-fast outflows (UFOs) have been detected in a number of AGN \citep{pounds_high-velocity_2003,reeves_massive_2003,chartas_xmm-newton_2003,gofford_suzaku_2013}, and the so-called warm absorbers \citep[WAs;][]{fabian_asca_1994,otani_variable_1996,cappi_warm_1996,orr_soft_1997} may also have a disc wind or outflow origin \citep{reynolds_warm_1995,krolik_observable_1995,krolik_warm_2001,bottorff_dynamics_2000,blustin_nature_2005,mizumoto_thermally_2019}. Further evidence for outflow comes from blueshifts (or blue asymmetries) seen in the C\textsc{iv}~1550~\AA\ emission line in many quasar spectra \citep{gaskell1982,richards_broad_2002,baskin_what_2005,sulentic_c_2007,richards_unification_2011,coatman_c_2016}. All in all, winds appear to be ubiquitous in luminous AGN and quasars. However, despite a number of attempts to `join the dots' between the various wind phenomena \citep[e.g.][]{elvis_structure_2000,gallagher_stratified_2007,tombesi_unification_2013,elvis_quasar_2017,hamann_does_2018,giustini_global_2019} the extent to which they are connected -- to each other, or to the BLR -- is not well known. 

In BAL quasars, the blue-shifted absorption is observed in many of the same lines that are also observed in emission. It is therefore tempting to associate the BAL and BEL phenomena with the same gas \citep[although see e.g.][]{borguet_major_2012,chamberlain_strong_2015,arav_evidence_2018}. As a result, disc wind models for the BLR are fairly popular \citep[e.g.][]{shlosman_active_1985,emmering_magnetic_1992,murray_accretion_1995,bottorff_dynamics_1997,young_rotating_2007,richards_unification_2011,kollatschny_accretion_2013,chajet_magnetohydrodynamic_2013,chajet_magnetohydrodynamic_2017,waters_reverberation_2016,yong_using_2018,lu_active_2019} and may unify much of AGN phenomenology  \citep[e.g.][]{elvis_structure_2000}. Disc wind models often involve some kind of cloud/clump structure within the flow \citep{emmering_magnetic_1992,de_kool_radiation_1995,elvis_quasar_2017}, meaning the distinction between wind and cloud models for the BLR is somewhat blurred. A recent model that has gained some attention is the failed dust-driven wind model proposed by \cite{czerny_origin_2011}, and discussed further by various authors \citep{czerny_dust_2014,czerny_dust_2015,czerny_failed_2017, galianni_test_2013,baskin_dust_2018}. This model is quite attractive compared to line-driven wind models, as there is observational evidence for a link between the BLR size and the dust sublimation radius \citep{galianni_test_2013,gandhi_dust_2015}.

\subsection{Aims of this study}
The optical and UV emission line spectra in type 1 quasars and AGN are remarkably universal; our main aims are to (i) explore this specific property in detail with particular focus on disc winds, outflows and failed winds as BLR candidates and (ii) investigate the more general X-ray, UV and optical absorption and emission characteristics of disc winds compared to observations. The key insight of BFKD95 was that the line ratios we observe in typical quasar spectra are similar to the ratios expected from assuming that each line is only produced by clouds with conditions that produce that line optimally (i.e. create the largest equivalent width in that line). Cloud conditions are broadly described by ionizing flux and density. Thus, to avoid a fine-tuning problem, BLR material must somehow cover a substantial portion of the ionizing flux/density plane, as well as physically covering a large fraction of the continuum source. We will show that disc wind models for the BLR naturally populate this ionizing flux/density plane, and this results in a BLR-like spectra. Of course, this does not mean that the BLR {\em is} a disc wind. However, it does mean that disc winds should be considered as a viable alternative to LOC/cloud models. Furthermore, although our starting point is the type of winds that produce BALs in ultraviolet spectra, our results are quite generally applicable to flow and outflow models for the BLR. We begin by exploring some of these general properties, using simple toy models, before conducting full Monte Carlo radiative transfer and photoionization calculations in the latter part of the paper. We discuss our results in section~\ref{sec:discussion} and offer some concluding thoughts in section~\ref{sec:conclusion}.

\section{Analytic Models and background theory}
The observed intensity of an emission line from the BLR can be written, following, e.g., \cite{bottorff_he_2002}, as
\begin{equation}
I = \int_{\mathrm{BLR}} ({\cal H}~j) dV ,
\end{equation}
where $j$ describes the line emissivity of the plasma, and ${\cal H}$ is a complicated function describing what fraction of the line radiation reaches the observer, in cm$^{-2}$~sr$^{-1}$. The quantity ${\cal H}$ depends on the details of the radiative transfer through the surrounding medium and on the viewing angle of the observer. Obtaining the mapping from $h$ to the parameters describing the BLR geometry and the illuminating SED is non-trivial, because it involves solving a coupled radiative transfer-photoionization problem in a non-spherical geometry.

In the Sobolev escape probability formalism (see e.g. Castor 1970), $j$ for a given transition $u\rightarrow l$ is given by 
\begin{equation}
j = \beta_{ul} A_{ul} n_u h \nu_{ul},
\end{equation}
where $n_u$ is the upper level number density, $A_{ul}$ is the usual Einstein coefficient and $\nu_{ul}$ is the frequency of the transition. Here, $\beta_{ul}$ is the Sobolev escape probability, a quantity which governs how much of the radiation escapes from the (local) Sobolev region for a given line interaction. Most of the complexity regarding gradients in ionization state, excitation state, temperature and density is encoded in $n_u$, since
\begin{equation}
n_u = f_{\mathrm{lev}} (J_\nu, n_H, T_e)~
f_{\mathrm{ion}} (J_\nu, n_H, T_e)~
\frac{n_{\mathrm{elem}}}{n_H}~
n_H,
\end{equation}
where $f_{\mathrm{lev}}=n_u/n_i$ is the fraction of ions in the upper state of the transition, $f_{\mathrm{ion}}=n_i/n_{\mathrm{elem}}$ is the relevant ionization fraction, $n_{\mathrm{elem}}$ is the density of the element in question, and $n_{\mathrm{i}}$ is the ion density. 

In general, the quantities $f_{\mathrm{ion}}$ and $f_{\mathrm{lev}}$ are complicated functions of the variables describing the physical conditions of the plasma: $J_\nu$ (the local mean intensity), $n_H$ (the Hydrogen number density), and $T_e$ (the electron temperature). These functions are different for each level in each ion and must really be obtained using a fully self-consistent radiative transfer and ionization framework, and such a calculation is presented in Sections 3 and 4. However, we can elucidate the key physics by considering just the density $n_H$ and the number of Hydrogen (Lyman) ionizing photons per unit area, $\phi_H$. The latter is related to $J_\nu$ via the equation
\begin{equation}
\phi_H = \int_{13.6\mathrm{eV}/h}^{\infty} 
\frac{4 \pi J_\nu}{h \nu} d\nu,
\label{eq:phi_h}
\end{equation}
while the optically thin equivalent for a source that produces $N_\gamma$ Hydrogen (Lyman) ionizing photons is
\begin{equation}
\phi_{H,\mathrm{thin}} (r) = \frac{N_\gamma}{4\pi r^2}.
\label{eq:phi_thin}
\end{equation}
We also introduce the two usual Hydrogen ionization parameters
\begin{equation}
U_H= \frac{\phi_H}{n_H c}
\label{eq:ip}
\end{equation}
and
\begin{equation}
\xi=\int_{13.6\mathrm{eV}/h}^{1000\mathrm{eV}/h} 
\frac{4 \pi J_\nu}{n_H} d\nu,
\label{eq:xi}
\end{equation}
as well as the emission measure, which we define as
\begin{equation}
\mathrm{EM} = \int f_V~n_e~n_H dV,
\label{eq:EM}
\end{equation}
where $n_e$ is the electron density and $f_V$ is the volume filling factor. These quantities will be referred to in the subsequent discussion.

\subsection{The ionizing flux-density plane}
The ionizing flux-density (or \phin) plane is a commonly used parameter space for photoionization modelling of AGN broad emission lines \citep[e.g. BFKD95;][]{korista_atlas_1997,hamann2002,baldwin_origin_2004,leighly2007a,ruff_new_2012,lawther_quantifying_2018}. The position of a parcel of gas in this plane is only a crude measure of its ionization state, since it encodes no information about the shape of the ionizing spectral energy distribution (SED) and says little about the absolute emissivity of the parcel. However, it is still a powerful diagnostic diagram.

For a BLR model to display optimally emitting behaviour, it must cover a significant portion of the \phin\ plane, such that shifting the region that it occupies over the contours of emissivity still results in an efficiently emitting region being covered. This `shifting' can occur due to both changes in luminosity and SED {\em within} a particular source, but also {\em between} different sources that produce similar emission line spectra. In LOC models, every cloud in the BLR is assumed to see the true (intrinsic) SED in these models, subject only to geometric dilution. The requisite \phin\ coverage is then achieved by assuming that clouds covering a wide range of densities, $n_H$, exist at any given location, $\vec{r}$. By contrast, in  ``flow'' models of the BLR, there is a one-to-one mapping between density and location, $n_H(\vec{r})$, but the SED shape varies with position due to attenuation and other radiative transfer effects.
One can envision many situations where there are aspects of both classes of models -- for example, in the fractal cloud model discussed by \cite{bottorff_fractal_2001} or the turbulent outflow model of \cite{waters_2019} -- but the above statements capture the essence of the distinction.

To illustrate how a flow model effectively fills out the \phin\ plane, we consider a parameterised model for an axisymmetric, biconical disc wind. We initially focus on the relationship between $\phi_H$ and $n_H$ in two simple, but physically relevant cases: (i) the optically thin limit; and (ii) the grey opacity limit. These cases capture the key principles, while the radiative transfer simulations presented in later sections include a more detailed treatment of the physical processes involved.

\begin{figure*}
    \centering
	\includegraphics[width=\textwidth]{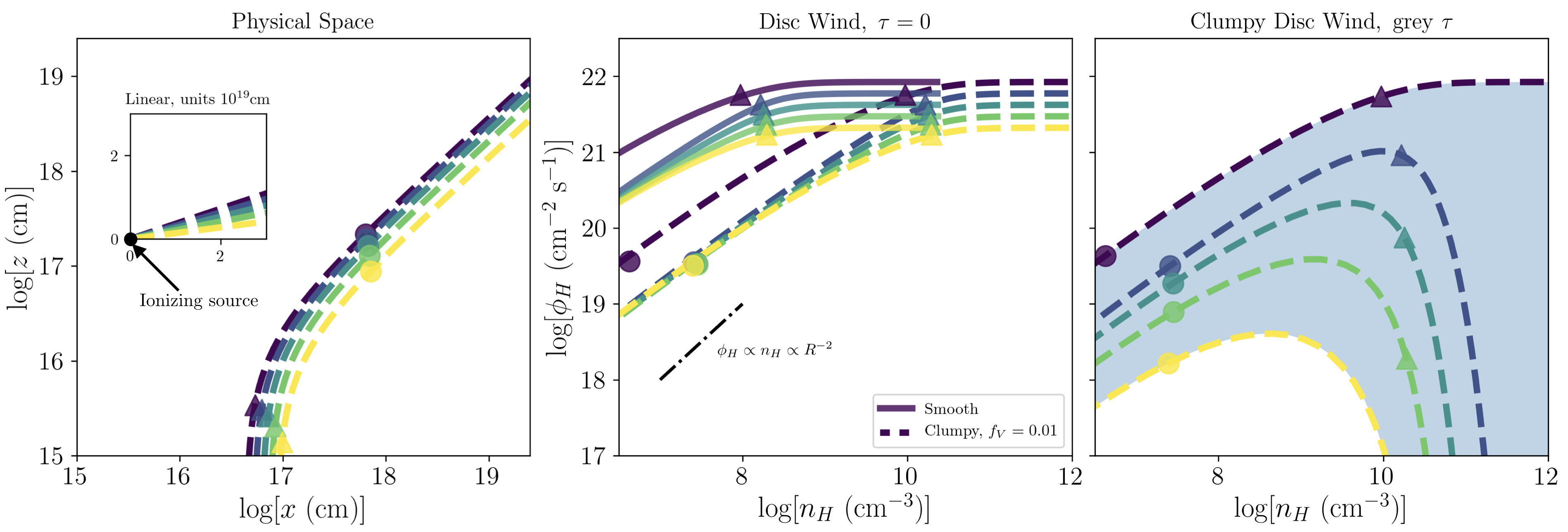}
    \caption{A demonstration of how disc winds fill out the \phin\ plane. {\sl Left:} Five streamlines in physical space for a model with the same parameters as the fiducial model from M16, where an equatorial wind rises from an accretion disc with an acceleration length of $R_v=10^{19}$cm. Ten locations in the wind are marked with triangles and circles, and their locations in \phin\ space are shown with the same symbols in the right hand panel. {\sl Middle:} The same five streamlines shown in \phin\ space for the optically thin limit with two clumping factors, $f_V=1,0.01$.
    {\sl Right:} An absorbed clumpy disc wind, with $f_V=0.01$. A grey opacity $\kappa=\sigma_T n_H$ has been applied to attenuate the ionizing flux. The effect is to smear the curves out across the \phin\ plane. 
    }
    \label{fig:phi-nh}
\end{figure*}

\subsubsection{Biconical Disc Wind}
We adopt a simple model for a biconical disc wind in which we specify a velocity law and mass-loss rate and the streamlines rise from a disc. Specifically, we use the parameterisation of a biconical disc wind described by \cite{shlosman_winds_1993}. Streamlines originate at radii $r_0$ and cover launch radii from $r_{\mathrm{min}}$ to $r_{\mathrm{max}}$. The opening angles of the flow are defined with respect to the disc normal and denoted $\theta_1$ and $\theta_2$, such that the corresponding angle for each streamline is given by 
\begin{equation}
\theta(r_0) = \theta_1 + (\theta_2 - \theta_1) \left(\frac{r_0-r_{\mathrm{min}}}{r_{\mathrm{max}}-r_{\mathrm{min}}} \right)^\gamma
\end{equation}
In this work, we set the exponent $\gamma=1$. An example of the path traversed by 5 streamlines in $x$ and $z$ is shown in the left-hand panel of Fig.~\ref{fig:phi-nh}. The equation for Hydrogen number density at a position vector $\vec{r}$ in a biconical wind is given by \citep{shlosman_winds_1993,knigge_application_1995}
\begin{equation}
n_H(\vec{r}) = \frac{\rho_0(r_0)}{\mu m_H}~\frac{r_0}{x}\frac{d r_0}{d x} \frac{v_0(r_0)}{v_l(r_0)},
\label{eq:density_law}
\end{equation}
where $m_H$ is the mass of a Hydrogen atom and $\mu$ is the mean molecular mass. We assume that the local sound speed in the disc at the streamline base, $c_s(r_0)$, sets $v_0$, and $v_\infty$ is a constant multiple of the escape velocity, $v_\mathrm{esc}$, at the streamline base. The poloidal velocity at a distance $l$ along a streamline is 
 \begin{equation}
v_l (r_0) = v_0+(v_\infty-v_0)
\left(
\frac{(l/R_v)^\alpha}{(l/R_v)^\alpha +1} 
\right),
\label{eq:sv93_velocity}
\end{equation}
where $R_v$ is the acceleration length and $\alpha$ is an exponent governing the rate of acceleration. The density along a streamline is set by this velocity law and the mass-loss rate at the streamline base, $\dot{m}(r_0)$. We can rewrite equation~\ref{eq:density_law} as a function of $l$, $\dot{m}(r_0)$ and $\theta$: 
\begin{equation}
n_H(l,r_0,\theta) = 
\left[\frac{\dot{m} (r_0)}{\mu m_H \cos \theta}\right] 
\left[\frac{r_0}{x} \frac{dr_0}{dx} \right]  
\left[ v_0+(v_\infty-v_0) \left(
\frac{(l/R_v)^\alpha}{(l/R_v)^\alpha +1} \right) \right].
\label{eq:density_law2}
\end{equation}
this equation illustrates some interesting properties of the flow. The first term is the initial density, which is set by the local mass-loss rate, angle of the streamline and local sound speed, which can all vary with $r_0$. The second term encapsulates the divergence of the flow.  The third term is the velocity law, which depends on $c_s, R_v, \alpha$ and $v_\mathrm{esc}$. For a Keplerian $\alpha$-disc, the critical velocities depend on launch radius as $c_{s} \propto r_0^{-3/8}$ and  $v_{\mathrm{esc}} \propto r_0^{-1/2}$. 

We now seek a relationship between $\phi_H$ and $n_H$ in the optically thin limit; in other words we require them as a function of position, $\vec{r}$. We assume that all ionizing photons come from a central point source (ignoring probable effects of an extended emission region or disc anisotropy). Once again, the optically thin value for $\phi_H$ is given by equation~\ref{eq:phi_thin}. The density at a given position can be found by finding the footpoint of the relevant streamline and solving equation~\ref{eq:density_law2} numerically. We conduct calculations for a smooth disc wind and a clumpy wind. In the latter case, the density is enhanced by a factor of $1/f_V$, and we follow \cite{matthews_testing_2016} in setting $f_V=0.01$. The result is plotted in the middle panel of Fig.~\ref{fig:phi-nh}, which shows that, even for the optically thin limit, streamlines intersect different ionization states. This is because the initial velocities and streamline angles depend on $r_0$, which leads to the covering of a small portion of the \phin\ parameter space. Furthermore, the density along a streamline in an accelerating biconical wind typically drops faster than $r^{-2}$. As a result, the \phin\ curves do not tend to a constant ionization state, even asymptotically.This effect can be seen in Fig.~\ref{fig:phi-nh} and means that a streamline will always intersect regions of the \phin\ plane that correspond to different ionization state. This advantageous behaviour is specific to axisymmetric geometries; spherically symmetric models such as `pressure-law' models \citep{rees_small_1989,netzer_quasar_1992,goad_response_1993,lawther_quantifying_2018} or a simple spherical wind instead produce single power-law slopes. 

The final thing to do is to consider how absorption in the flow affects $\phi_H$. The simplest case is the pure absorption limit with a grey opacity, $\kappa=\sigma_T n_H$. For this case, a simple ray-tracing technique can be used to find $\phi_H$ as a function of position. The result of such a calculation is shown in the right-hand panel in Fig.~\ref{fig:phi-nh} for the clumpy wind.\footnote{Note that, for an opacity that depends linearly on density, $\tau$ is actually the same for smooth and clumpy winds, since the enhanced density in the clumpy flow cancels with the filling factor (see sections~\ref{sec:clumping} and \ref{sec:limitations}).} The approximate effect of absorption in the flow is to smear the curves downwards in the \phin\ plane, particularly in the region close to the launch point of the streamline where the column density towards the origin is highest.\footnote{The effect of shielding and absorption by a disc wind is discussed in detail by \cite{leighly2004} and \cite{leighly2007a}, where it is referred to as ``filtering'' of the continuum.}
Thus, two effects (absorption, and density changes along the outflow) conspire to cover large regions of the \phin\ plane, with the effect maximised when the directions are closest to orthogonal, that is, for the shallowest relationship beween $\phi_H$ and $n_H$. 

\section{Numerical Model}
In the previous section, we used a grey opacity to illustrate the approximate effect of absorption in the flow. In this section, we instead calculate the radiative transfer through a disc wind self-consistently. To do so, we use a Monte Carlo radiative transfer (MCRT) and ionization code \citep{long_modeling_2002}, confusingly known as \py, to simulate emission line spectra from a kinematic prescription for a biconical disc wind\footnote{More information about and the source code for \py\ can be found at \url{https://github.com/agnwinds/python}.}. The code uses the Sobolev approximation to treat line transfer, and operates in 2.5D, in that photon trajectories and velocity shifts are calculated in 3D, but the wind grid is assumed to be azimuthally symmetric. Our methods are described in detail in a number of other studies. Here we provide a brief overview of the techniques with the relevant references and discuss some of the more unique aspects of the work. The general principles behind MCRT are described in a series of papers by \cite{abbott_multiline_1985}, \cite{lucy_multiline_1993} and \cite{lucy_improved_1999,lucy_monte_2002,lucy_monte_2003}, and are reviewed by \cite{noebauer_monte_2019}.

\subsection{Kinematics, Geometry and Input SED}
For our MCRT simulations we use the same flexible wind model of \cite[][hereafter SV93]{shlosman_winds_1993} that we described in the previous section. SV93  originally applied their model to disc winds in accreting white dwarfs, but the model has since been applied to AGN by \cite{higginbottom_simple_2013}, \citet[][hereafter M16]{matthews_testing_2016} and \cite{yong_black_2016}. The biconical geometry is similar to those commonly invoked in disc wind unification models \citep[e.g.][]{emmering_magnetic_1992,de_kool_radiation_1995,murray_accretion_1995,elvis_structure_2000}.

In the SV93 geometry, the basic wind geometry is specified, and the poloidal velocity in the flow is obtained by identifying the appropriate streamline. The density at a given point is then determined by mass conservation along that streamline according to equation~\ref{eq:density_law2}. We modify the initial velocity assumption from previous work, as we now set the initial velocity to the sound-speed at the streamline base rather than using a fixed value. Specific angular momentum is conserved along a streamline such that the rotational velocity is given by $v_\phi = v_k(r_0) r_0/r$, where $r$ is the radial coordinate, and $v_k(r_0)$ is the Keplerian velocity at the streamline base. The flexibility built in to the SV93 prescription allows us to investigate a variety of wind geometries spanning a range of densities and velocities. Our adopted biconical geometry has successfully reproduced the absorption and emission properties of high-state accreting white dwarfs \citep{shlosman_winds_1993, long_modeling_2002,noebauer_geometry_2010,matthews_impact_2015}, as well as the absorption troughs of BAL quasars \citep{higginbottom_simple_2013}. It has also been applied more generally to AGN emission lines \citep{yong_black_2016}, while a similar prescription \citep{knigge_application_1995} has also been used to model AGN at X-ray wavelengths \citep{sim_multidimensional_2008,sim_multidimensional_2010,sim_synthetic_2012,hagino_disc_2016,mizumoto_line-driven_2018}.
However, clearly the choice of parameterisation of a disc wind can affect the resulting calculations, and so we caution that some of our conclusions may be specific to this choice. 

In previous work, we demonstrated that it is quite difficult to build a unified model for BAL and non-BAL quasars using an equatorial wind rising from a limb-darkened, optically thick accretion disc (M16). This is because the emission line EWs are much smaller at low inclinations than at high inclinations, whereas the observed emission line properties of BAL and non-BAL quasars are very similar (\citealt{matthews_quasar_2017}; see also \citealt{weymann_comparisons_1991,dipompeo_rest-frame_2012,runnoe_bals_2013,tuccillo_multiwavelength_2017,yong_using_2018,rankine2019}. In this work, we instead illuminate the wind with an isotropic SED of the same spectral shape as M16, which consists of a multi-colour blackbody disc component and an X-ray power-law with spectral index $\alpha_X = -0.9$. The angular distribution of the SED affects the observed spectra in a number of ways \citep[e.g.][]{netzer_discs_1987,matthews_quasar_2017}. In addition, the geometry of the flow in line-driven and dust-driven winds is also sensitive to the SED anisotropy \citep{dyda_geometry_2018,williamson_3d_2019} and motivates further work involving multi-frequency radiation hydrodynamics. Unlike M16, we do not have an emissivity profile with radius along the disc midplane, and all photon packets are launched from the central source.  The disc midplane then acts only to specularly reflect photon packets. We briefly discuss the impact of disc anisotropy and SED shape in section~\ref{sec:params} and Appendix A. The SEDs used in this work are shown in Fig.\ref{fig:seds}.

\begin{figure}
	\includegraphics[width=\linewidth]{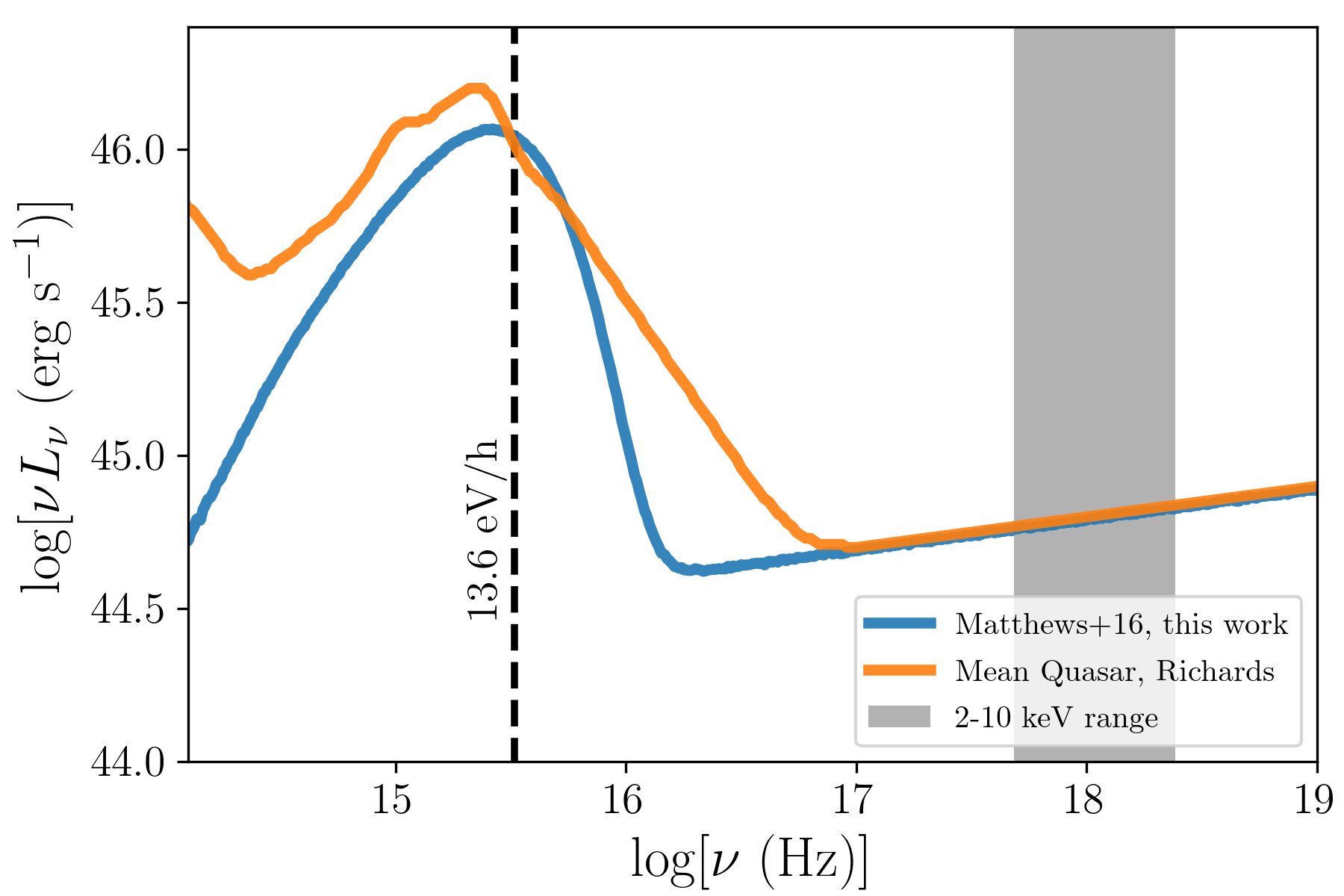}
    \caption{The spectral energy distribution used in this work. The blue curve shows the SED from M16, consisting of a multi-temperature blackbody `disc' component and an X-ray power law. We compare to the mean quasar SED from \protect\cite{richards_spectral_2006} (orange). The grey band marks the 2-10 keV frequency range and a dashed line marks the Lyman edge.}
    \label{fig:seds}
\end{figure}

\subsection{Photoionization and Radiative Transfer}
Our code works in two stages. First, the ionization state is calculated by tracking randomly generated photon packets through a predefined outflow geometry, discretized into plasma cells. As these photons pass through the simulation grid, their heating and ionizing effect on the plasma is recorded through the use of Monte Carlo estimators. This process continues until the code converges on a solution in which the heating and cooling processes are balanced and the temperature stops changing significantly (see section~\ref{sec:conv}). Once the ionization and temperature structure of the outflow has been calculated, the spectrum is synthesized by tracking photons through the plasma until sufficient signal to noise is achieved in the output spectrum for lines to be easily identified. 

Our radiative transfer scheme is based on techniques described by \cite{long_modeling_2002} and \cite{matthews_impact_2015}; in particular, we treat H \& He as `macro-atoms' \citep{lucy_monte_2002,lucy_monte_2003} as implemented in \py\ by \cite{sim_two-dimensional_2005}, but retain a fast `two-level atom' treatment for other atomic species (the `metals': C, N, O, Ne, Na, Mg, Al, Si, S, Ar, Ca and Fe). The treatment of level populations and ion populations is also calculated differently for H \& He and for metals. This hybrid method is originally described by \cite{matthews_impact_2015}. We enforce strict radiative equilibrium via an indivisible packet constraint, with the exception of adiabatic cooling, which can destroy $k$-packets (quantised thermal energy, see \citealt{lucy_monte_2002}) as described by M16. Code validation tests and more thorough descriptions of the code are available in various PhD theses \citep{nick_thesis,matthews_disc_2016}.

\subsubsection{Convergence}
\label{sec:conv}
Convergence criteria in the code are originally described by LK02. We consider a wind cell converged when (i) the electron temperature, $T_e$ has stopped changing significantly ($\Delta T_e / T_e \leq 5\%$) and (ii) heating and cooling are equal to within $5\%$. We record the overall convergence fraction as
\begin{equation}
f_{c} = \frac{1}{N_\mathrm{cells}} \sum_n^\mathrm{cells} C_n,
\end{equation}
where $N_\mathrm{cells}$ is the total number of cells. $C_n=1$ if cell $n$ meets the convergence criteria, and $C_n = 0$ otherwise. We also calculate a packet-weighted convergence condition, $f_w$, where each cell also has a weight applied to it based on the number of photon passages through the cell, such that
\begin{equation}
    f_{w} = 
    \left(\sum_n^\mathrm{cells} C_n N_{\mathrm{\gamma},n}
    \right)
    \bigg / 
    \left(\sum_n^\mathrm{cells} N_{\mathrm{\gamma},n} 
    \right )
\end{equation}
where $N_{\mathrm{\gamma},n}$ is the number of photon packets passing through cell $n$. We expect $f_w$ to act as an improved figure-of-merit for convergence; cells with few photons are not important in determining the emergent spectrum when the wind is close to radiative equilibrium, but these cells can be poorly converged due to poor MC statistics. In this paper, we found most models converged well with 25 ionization cycles with 15 million photon packets per cycle. This results in a run time of approximately $320$ core-hours per model. For the weighted convergence criterion, $f_w$, the mean convergence percentage was $91.5$\%, the median was 
$93.2$\%, and the minimum was $74.3$\%. For the unweighted criterion, $f_c$, the mean was $91.0$\%, the median was 
$92.0$\%, and the minimum was $77.5$\%. The models with the poorest convergence were those with the highest volume filling factors and smallest launch radii -- these are the hottest models that in any case would be expected to be too highly ionized to produce a BLR-like spectrum.

\subsection{Atomic Data}
We adopt the same atomic data as M16, except we now include a number of improved methods and more complete datasets. For macro-ions, we use a 10 level H model atom, a one level He~\textsc{i} model and a 10 level He~\textsc{ii} model, with bound-bound collision strengths calculated from the `g-bar' or \cite{van_regemorter_rate_1962} approximation. We used a one-level He~\textsc{i} ion for speed purposes, but we verified that a run with 53 levels produced near-identical spectra, so this choice does not affect our results. The macro-atom data sources are as described by M16. For simple ions, we now use collision strengths from \textsc{Chianti} \citep{dere_chianti_1997,landi_chiantiatomic_2013}, only defaulting to the g-bar approximation when they are not available or where our atomic model does not have sufficient level information. We also use dielectronic recombination rates, radiative recombination rates and collisional ionization rates from \textsc{Chianti}. For the semiforbidden C\textsc{iii}]~1909~\AA\ line we combine the $J=0,1,2$ upper levels into one `J-mixed' level with the three relevant \textsc{Chianti} collision strengths and one downwards radiative transition from the $J=1$ state. This allows us to model the line within a two-level atom with the multiplicities adjusted accordingly. We include a treatment of Auger ionization using inner shell cross-sections from \cite{verner_analytic_1995} and electron yield data from \cite{kaastra_x-ray_2000}. We use solar abundances from \cite{verner1994} for our calculations. The atomic database is sufficiently complete to yield good agreement with codes such as \textsc{Cloudy} for ionization state calculations \citep{long_modeling_2002,higginbottom_simple_2013,matthews_disc_2016}.

\subsection{Clumping}
\label{sec:clumping}
There are relatively strong observational and theoretical arguments for clumping in quasar outflows
\citep[e.g.][]{shlosman_active_1985,emmering_magnetic_1992,de_kool_radiation_1995,hamann_extreme-velocity_2013,mccourt_characteristic_2018,mas-ribas_radiation-pressure_2019,hamann_emergence_2019}. Multiple mechanisms for wind clumping have been proposed, including the line-driven instability \citep[LDI;][]{macgregor_radiative_1979,owocki_instabilities_1984,owocki_instabilities_1985,sundqvist_2d_2018}, magnetic confinement \citep{emmering_magnetic_1992,de_kool_radiation_1995}, and various forms of thermal instability, condensation or fragmentation \citep{shlosman_active_1985,proga_cloud_2015,waters_efficient_2016,waters_synthetic_2017,elvis_quasar_2017,mccourt_characteristic_2018,waters_non-isobaric_2019,waters_2019,dannen2020}. Clumping can also be introduced by velocity perturbations in the disc \citep{dyda_axisymmetry_2018}. Over-ionization from the X-ray source is a well-known problem for both BAL formation and line-driving in disc winds \citep[e.g.][]{murray_accretion_1995,proga_dynamics_2000,proga_dynamics_2004,hamann_extreme-velocity_2013,higginbottom_simple_2013,higginbottom_line-driven_2014}. The motivation for introducing clumping in our previous work was primarily to address this issue -- it allowed BALs to be formed at realistic X-ray luminosities as an alternative to shielding of the flow by a magnetocentrifugal or failed wind \citep{proga_dynamics_2000,proga_dynamics_2004,everett_radiative_2005} or `hitch-hiking gas' \citep{murray_accretion_1995}.

In this work, we follow M16 in adopting the `microclumping' approximation commonly used in stellar-wind modelling \citep[e.g.][]{hillier_effects_1991,hillier_constraints_1999,hamann_spectrum_1998,hamann_spectrum_2008}. Within this approximation, we assume individual clumps are optically thin and the intra-clump medium is a vacuum. This means the effect of clumping can be parameterised with a single variable, the volume filling factor $f_V$. Densities are enhanced by a factor of $1/f_V$ compared to the value calculated from the SV93 model, but the volume filling factor means that the total mass in a cell is unchanged. As a result, linear opacities such as photoelectric absorption and electron scattering remain unchanged for a given ionization state, whereas processes that scale with the square of density (e.g. collisional excitation, free-free, recombination) are enhanced by a factor $1/f_V$ \citep[see][for more details on our specific implementation]{matthews_disc_2016,matthews_testing_2016}. For fixed mass-loss rate and ionizing luminosity, clumping acts to make the wind cells less ionized and increase their emission measure.

The microclumping assumption is simple and does not adequately capture the physics of both clump formation within the wind, and radiative transfer through the clumps. However, a physical, or even practical, way to parameterise the clumps in more detail is not clear at this stage, although a number of methods incorporating statistical approaches, macroclumping and/or porosity have been suggested in the context of stellar winds \citep{feldmeier_x-ray_2003,owocki_effect_2006,hamann_spectrum_2008,surlan_three-dimensional_2012} or dusty AGN tori \citep[e.g.][]{stalevski_3d_2012}. Microclumping represents a reasonable first step and avoids the introduction of additional, poorly constrained, free parameters. Optically thin clumps may actually be a reasonable approximation, if the fragmentation mechanism proposed by \cite{mccourt_characteristic_2018} operates in a disc wind. However, there are situations where we expect the approximation to break down - for example, when the clump size implied by the optically thin assumption becomes unrealistically small. Although the numerics of our calculations are clearly affected by the choice of clumping parameterisation, our general arguments are not reliant on the specific choice, as they apply to any smooth or `thin spray' type flows where the ionization stratification is primarily macroscopic and substructures are mostly optically thin. The limitations of our clumping treatment and possible avenues for future work in this area are discussed further in sections~\ref{sec:limitations} and \ref{sec:conclusion}, respectively.

\section{Results}
We conducted a number of radiative transfer simulations sampling the parameter grid defined in Table~\ref{grid_table}. This particular grid was chosen to sample a range of launch radii and wind launching angles for fixed BH mass ($10^9~M_\odot$) and Eddington fraction ($\approx0.2$), using our previous simulations (M16) as a starting point. The range of launch radii was chosen with the value from \cite{higginbottom_simple_2013} as the smallest radii (450~$r_g$), whereas the larger radii ($2250 r_g$ and $4500 r_g$) correspond more closely to the empirical BLR size (see also section~\ref{sec:limitations}). For reference, $4500~r_g \approx 0.2$pc for our adopted black hole mass. The illuminating SED is shown in Fig.~\ref{fig:seds}. We have picked two illustrative models, Models A and B, which match the observations relatively well and produce strong emission lines with large EWs. However, we also examine the rest of the spectra and line luminosities. The full set of wind spectra and physical conditions for all 54 models is made publicly available at \url{https://github.com/jhmatthews/windy-blr-2020} and summary plots are included in the supplementary material. Model A has an equatorial wind, with  $\theta_1=70^\circ,\theta_2=82^\circ,R_{\mathrm{launch}}=450~r_g$ and $f_V=0.01$. Model B has a more collimated, polar geometry and a larger launch radius, with $\theta_1=20^\circ,\theta_2=35^\circ,R_{\mathrm{launch}}=4500~r_g$ and $f_V=0.01$.  

To compare our models to observations, we make use of the \cite{selsing_x-shooter_2016} X-Shooter quasar composite. This quasar composite is chosen instead of higher signal-to-noise versions such as the \cite{vanden_berk_composite_2001} composite spectrum from SDSS because it suffers less from host galaxy contamination beyond $\sim5000$\AA, allowing us to more accurately compare the continuum shape and line strengths in that region. The overall spectrum is nevertheless extremely similar to other quasar composites. We note that our spectra are presented for a single quasar mass and Eddington fraction, whereas the composite spectrum is built from many quasars with a range of fundamental parameters -- this is worth bearing in mind when comparing the spectra. We also make use of the SDSS DR7 quasar catalog compiled by \cite{shen_catalog_2011}, which allows us to compare the emission line properties of our synthetic spectra to those from a large sample of quasars.  

\begin{table}
\centering
\begin{tabular}{p{2cm}p{2cm}p{1cm}p{1cm}p{1cm}}
\hline 
Parameter & Grid Points & Model A & Model B  \\
\hline \hline 
$R_{\mathrm{launch}}~(r_g)$ 	&	(450,2250,4500) & 450 & 4500 \\ 
$\theta_{1}$ 	& ($20^{\circ},45^{\circ},70^{\circ}$) & $70^{\circ}$ & $20^{\circ}$
&  \\ 
$\theta_{2}$ (if $>\theta_1$) 	& ($35^{\circ},60^{\circ},85^{\circ}$) & $85^{\circ}$ & $35^{\circ}$ \\
$f_V$  	        &	 ($1,0.1,0.01$) & $0.01$ & $0.01$ \\ \\
\hline
\multicolumn{5}{|l|}{Important Fixed Parameters} \\
\hline \hline
$M_{\mathrm{BH}}$ 	&	  \multicolumn{4}{|l|}{$10^9~M_\odot$} \\ 
$\dot{M}_\mathrm{acc}$ 	&	 \multicolumn{4}{|l|}{$5 M_{\odot}~\mathrm{yr}^{-1}\approx 0.2~\dot{M}_{\mathrm{Edd}}$}  \\
$\dot{M}_\mathrm{wind}$ 	&	 \multicolumn{4}{|l|}{$\dot{M}_\mathrm{acc}$}  \\
 Grid size &	 \multicolumn{4}{|l|}{$100\times100$}  \\
$L_{\mathrm{bol}}$ &	\multicolumn{4}{|l|}{$3\times10^{46}$~erg~s$^{-1}$}  \\
$L_X$ (2-10 keV) &	\multicolumn{4}{|l|}{$10^{45}$~erg~s$^{-1}$}  \\
$\alpha_{\mathrm{X}}$ &	 \multicolumn{4}{|l|}{$-0.9$}  \\
$R_v$ 	        &	  \multicolumn{4}{|l|}{$10^{19}$cm} \\
$r_{\mathrm{min}}$ 	& \multicolumn{4}{|l|}{$2/3~R_{\mathrm{launch}}$} \\ 
$r_{\mathrm{max}}$ 	& \multicolumn{4}{|l|}{$4/3~R_{\mathrm{launch}}$} \\ 
$v_0 (r_0)$ &	 \multicolumn{4}{|l|}{$c_s(r_0)$}  \\
$v_\infty (r_0)$ &	 \multicolumn{4}{|l|}{$v_\mathrm{esc} (r_0)$}  \\
\hline 
\end{tabular}
\caption{{\sl Top:} The grid points used in the parameter search, resulting in 54 models. The values of the grid parameters for the two illustrative models discussed are also shown. 
{\sl Bottom:} Various fixed parameters used in the simulations, with definitions in the text.  
}
\label{grid_table}
\end{table}

\subsection{Synthetic spectra and equivalent widths}
\label{sec:spectra}
We present synthetic spectra from Models A and B in a number of different forms. The left hand panel of Figs.~\ref{fig:modela} and \ref{fig:modelb} show a continuum-normalised spectrum at a viewing angle of $20^\circ$, compared to the XShooter quasar composite spectrum. The spectra exhibit most of the main emission lines seen in the UV in the quasar composite with comparable widths and EWs. The inclusion of the two-level treatment of \ciiiline\ results in a strong broad emission line in the spectrum. For model A, the EWs and relative intensities for the strongest lines are given in Table~\ref{ew_table}, compared to values from \cite{shen_catalog_2011} and BFKD95. To calculate the Ly-$\alpha$ EW and intensity we calculated the relevant quantity for the blue wing of the line and multiplied it by two, to avoid contamination from N\textsc{v}~1240\AA. We also show wider wavelength range plots in the right hand panel of Figs.~\ref{fig:modela} and \ref{fig:modelb}. These plots are not continuum normalised, so that they show differences in the overall spectrum. We discuss the differences between our models and the data later in this subsection.  

\begin{figure*}
	\includegraphics[width=\linewidth]{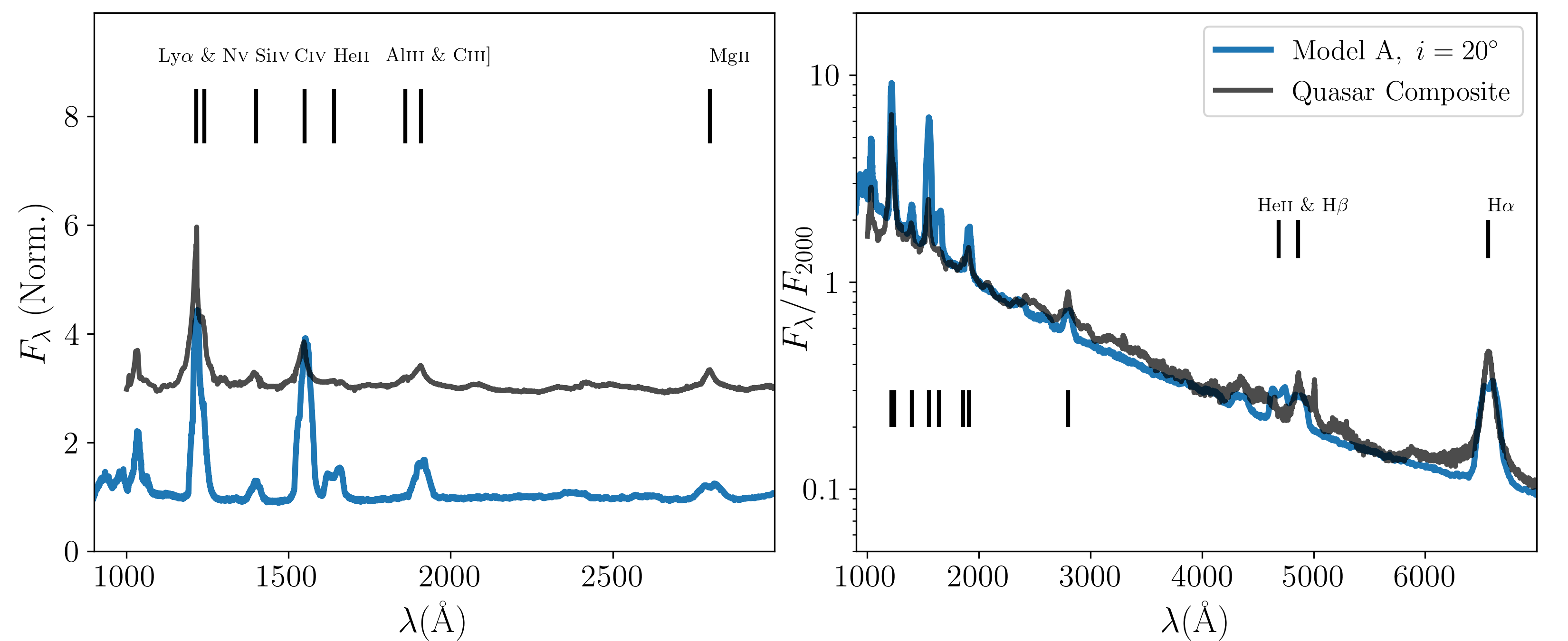}
    \caption{A comparison between spectra from model A (blue) and the XShooter quasar composite spectrum (black) from \protect\cite{selsing_x-shooter_2016}. {\sl Left:} Ultraviolet comparison. The spectra are normalised to the continuum, which is obtained from a fit to the spectrum between 800~\AA\ and 3000~\AA\ with the prominent emission lines masked. An offset of $+2$ is applied to the composite spectrum for clarity. Strong permitted UV transitions are labeled. {\sl Right:} Comparison across a wider wavelength range and without continuum normalisation -- instead the flux is scaled to that at 2000\AA\ and shown on a logarithmic scale. Strong permitted UV transitions are marked with lines and the H$\beta$, H$\alpha$ and He~\textsc{ii}~4686~\AA\ optical emission lines are labeled.}
    \label{fig:modela}
\end{figure*}

\begin{figure*}
	\includegraphics[width=\linewidth]{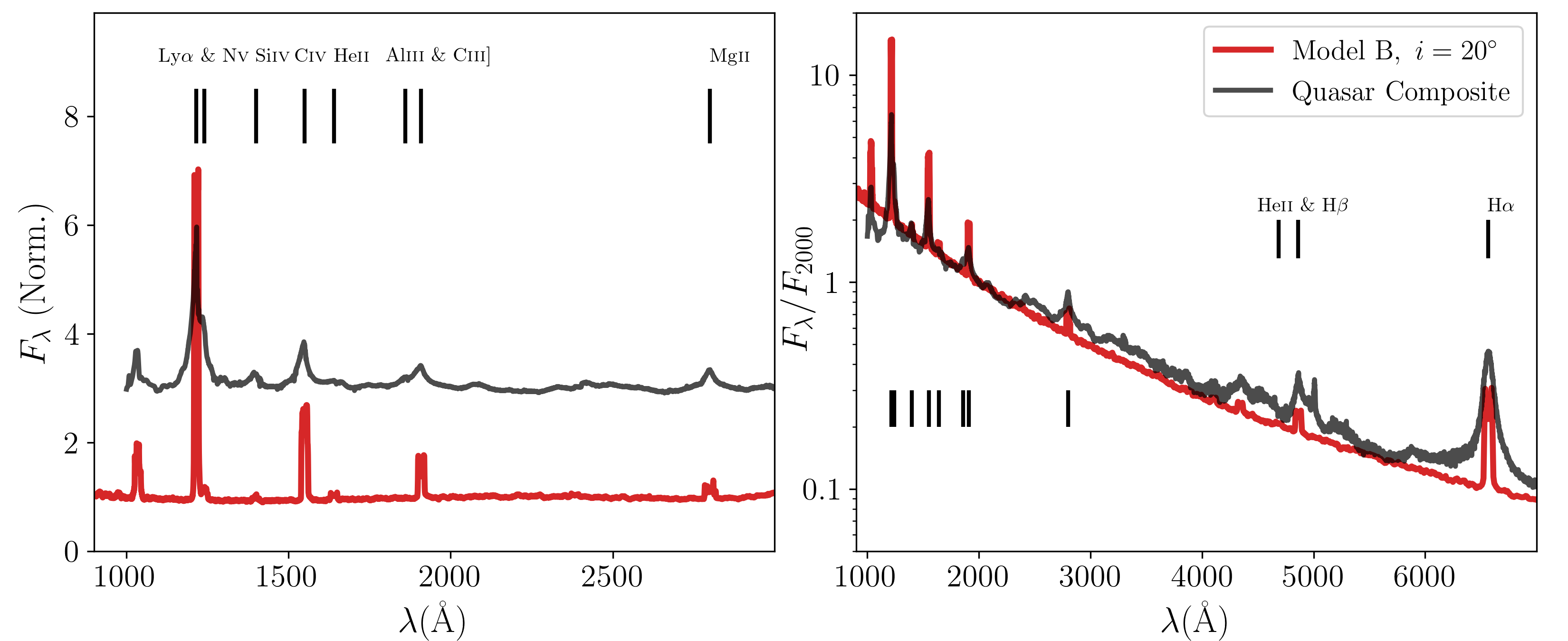}
    \caption{As Fig.~\ref{fig:modela}, but for Model B, which is a more polar wind model ($\theta_1=20^\circ,\theta_2=35^\circ$), launched from further out ($R_\mathrm{launch}=4500~r_g$).} 
    \label{fig:modelb}
\end{figure*}

To illustrate the behaviour across the simulation grid, we show the range in each wavelength bin for all spectra from our runs in Fig.~\ref{fig:mean-spec}, overlaid with the geometric mean spectrum, also calculated using all spectra from our simulation grid. This spectrum is compared once more to the XShooter quasar composite. We show this figure to give a feel for the typical spectra in the simulation grid and their diversity. Models A and B are somewhat special cases, and the mean EWs in the sample are lower, but this plot illustrates that the emission lines commonly appear across a range of parameter space. In fact, all models with $f_V \leq 0.1$ produce a detectable Ly~$\alpha$ and \civline\ line. The geometric mean spectrum from the simulation grid can be considered a reasonable approximation to a BLR spectrum. We also compare the full simulation grid to data in Fig.~\ref{fig:lums_hexbin}, where we show a hexagonally binned histogram of $\log L_{C\textsc{iv}}$ against $\log \lambda L_{1350}$ for all SDSS DR7 quasars from \cite{shen_catalog_2011}. Overplotted are points from each model in the simulation grid, for an inclination of $20^\circ$. Broadly speaking, the models span the range of line luminosities expected for a quasar of comparable luminosity, although the models with $f_V = 0.01$ systematically exceed the mean, whereas models with $f_V = 0.1$ lie systematically below it. 

Although the overall agreement between the quasar composite spectrum and our model spectra is encouraging, there are a number of more detailed aspects of the spectra that deviate from observed properties. The ratio of \civline\ to Ly~$\alpha$ EW is too high in model A, although the mean line ratio calculated from all the models is more comparable to observations. The models tend to under-produce \mgline, H$\beta$, and other low ionization lines. The weakness in these lines may indicate that there is not enough material at low ionization states, or that this material is not emitting efficiently. The spectra also generally have an excess of \heiiuv\ and \heiiopt\ emission compared to the composite spectrum, which could also indicate that the wind is slightly too ionized compared to a typical BLR. The `small blue bump', caused by Balmer continuum emission, is not prominent in our models and the model spectra lie below the composite spectrum in the $\approx2200-4000$\AA\ region. Balmer continuum emission is produced in our model, but it is not strong enough to match the observations, which again may suggest that higher densities are needed. The Fe pseudocontinuum is discussed further in section~\ref{sec:pseudo}.

Another difference between our model spectra is that double-peaked lines, produced near the base of the wind, are relatively common in the spectra, particularly in Model B and at higher inclinations (see section~\ref{sec:bals} and Fig.~\ref{fig:inclination}). Double-peaked lines are seen in a small fraction of AGN, but they are relatively rare \citep{eracleous_double-peaked_1994,eracleous_completion_2003,strateva_double-peaked_2003,storchi-bergmann_double-peaked_1993,storchi-bergmann_double-peaked_2017}. Velocity shear from a disc wind can transform double-peaked profiles to single-peaked \citep{murray_wind-dominated_1996,murray_disk_1997,flohic_effects_2012}, but the shear is not sufficient to achieve this in our models, partly due to the large acceleration length \citep[see also][]{matthews_impact_2015}.

\begin{table}
\centering
\begin{tabular}{lccccc}
\hline 
& & \multicolumn{4}{c}{Equivalent Width (\AA)} \\
Line & $\lambda$ (\AA) & A & B & Mean, all models & Mean, S11 \\
\hline \hline 
Ly~$\alpha$     & 1215 	& 113.6 & 72.8 & 39.7  &  --     \\ 
C~\textsc{iv}   & 1550 	& 118.9 & 29.2 & 29.0  &  44.4    \\ 
He~\textsc{ii}  & 1640 	& 26.1  & 2.9  & 8.7   &  --      \\ 
C~\textsc{iii}] & 1909 	& 30.4  & 10.9 & 8.0   &   --      \\ 
Mg~\textsc{ii}  & 2798 	& 14.9	& 1.7  & 3.5   &   39.8    \\ 
H$\beta$        & 4864 	& 38.9	& 13.0 & 9.8   &  73.7    \\ 
\hline 
& & \multicolumn{4}{c}{Relative Intensity to Ly~$\alpha$} \\
Line & $\lambda$ (\AA) & A  & B & Mean, all models & BFKD95 \\
\hline \hline 
C~\textsc{iv}   & 1550  & 0.81  &  0.28 & 0.54 & 0.4-0.6   \\ 
He~\textsc{ii}  & 1640 	& 0.16  &  0.03 & 0.14 & 0.09-0.2  \\ 
C~\textsc{iii}] & 1909 	& 0.14  &  0.08 & 0.08 & 0.15-0.3  \\ 
Mg~\textsc{ii}  & 2798 	& 0.04  &  0.01 & 0.02 & 0.15-0.3  \\ 
H$\beta$        & 4864 	& 0.04	&  0.02 & 0.02 & 0.07-0.2  \\ 
\hline 
\end{tabular}
\caption{Equivalent widths and relative line intensities of selected lines in models compared to observational data from BFKD95 and S11.}
\label{ew_table}
\end{table}

\begin{figure*}
	\includegraphics[width=\linewidth]{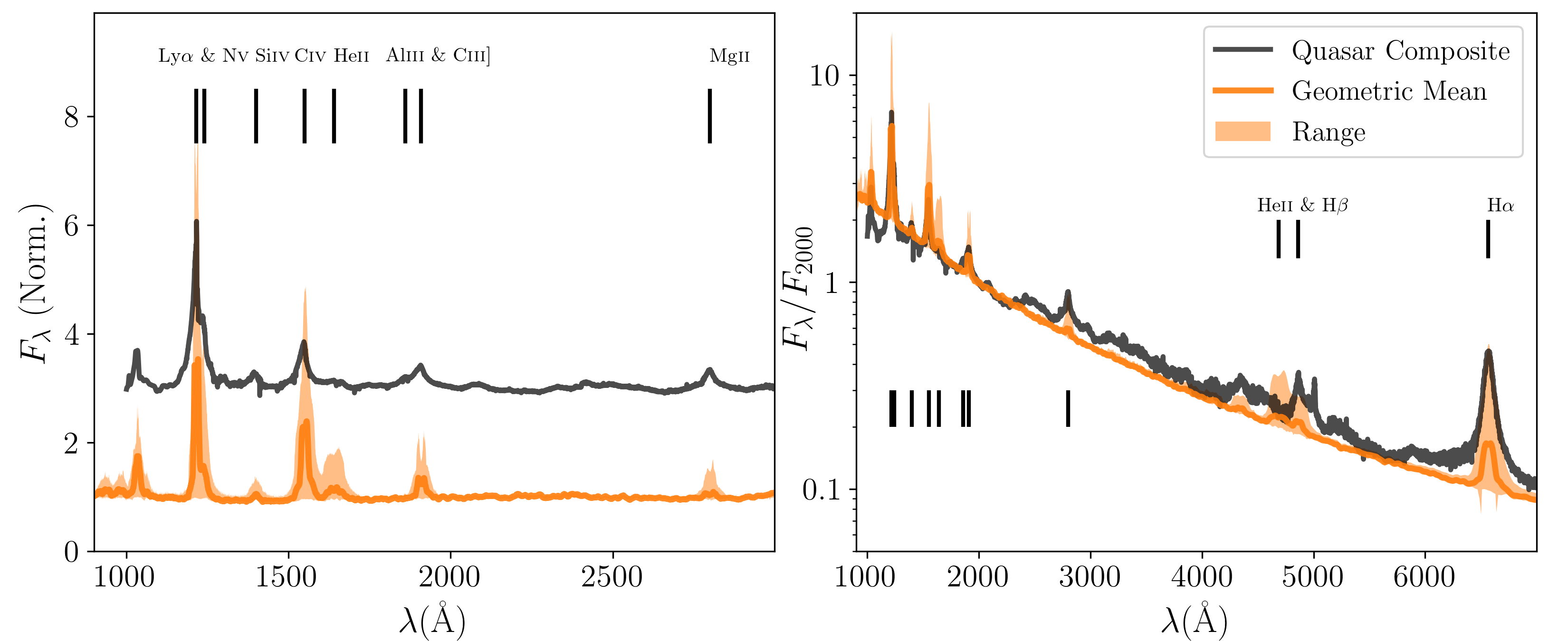}
    \caption{The geometric mean of the spectra from simulation grid (orange) compared to the S16 quasar composite (black), with the range in each flux bin shaded in translucent orange. The plot illustrated that emission lines are common, but also that the mean EW of the emission lines in the models is a little lower than for real quasars but can be considered a reasonable approximation to a BLR spectrum. We show the complete set of spectra in the supplementary material.}
    \label{fig:mean-spec}
\end{figure*}

\begin{figure*}
\centering
\begin{subfigure}{.5\textwidth}
  \centering
  \includegraphics[width=\linewidth]{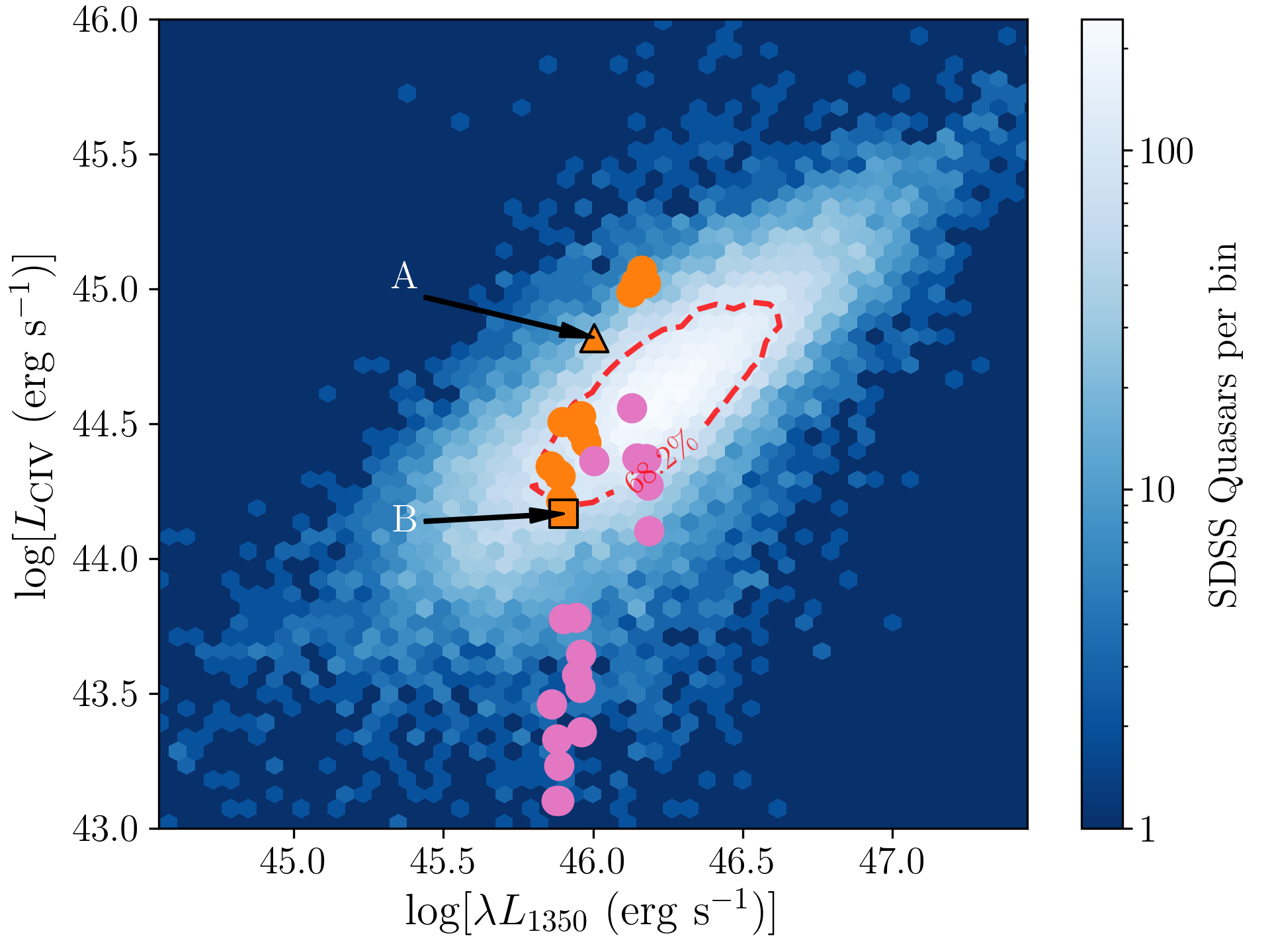}
\end{subfigure}%
\begin{subfigure}{.5\textwidth}
  \centering
  \includegraphics[width=\linewidth]{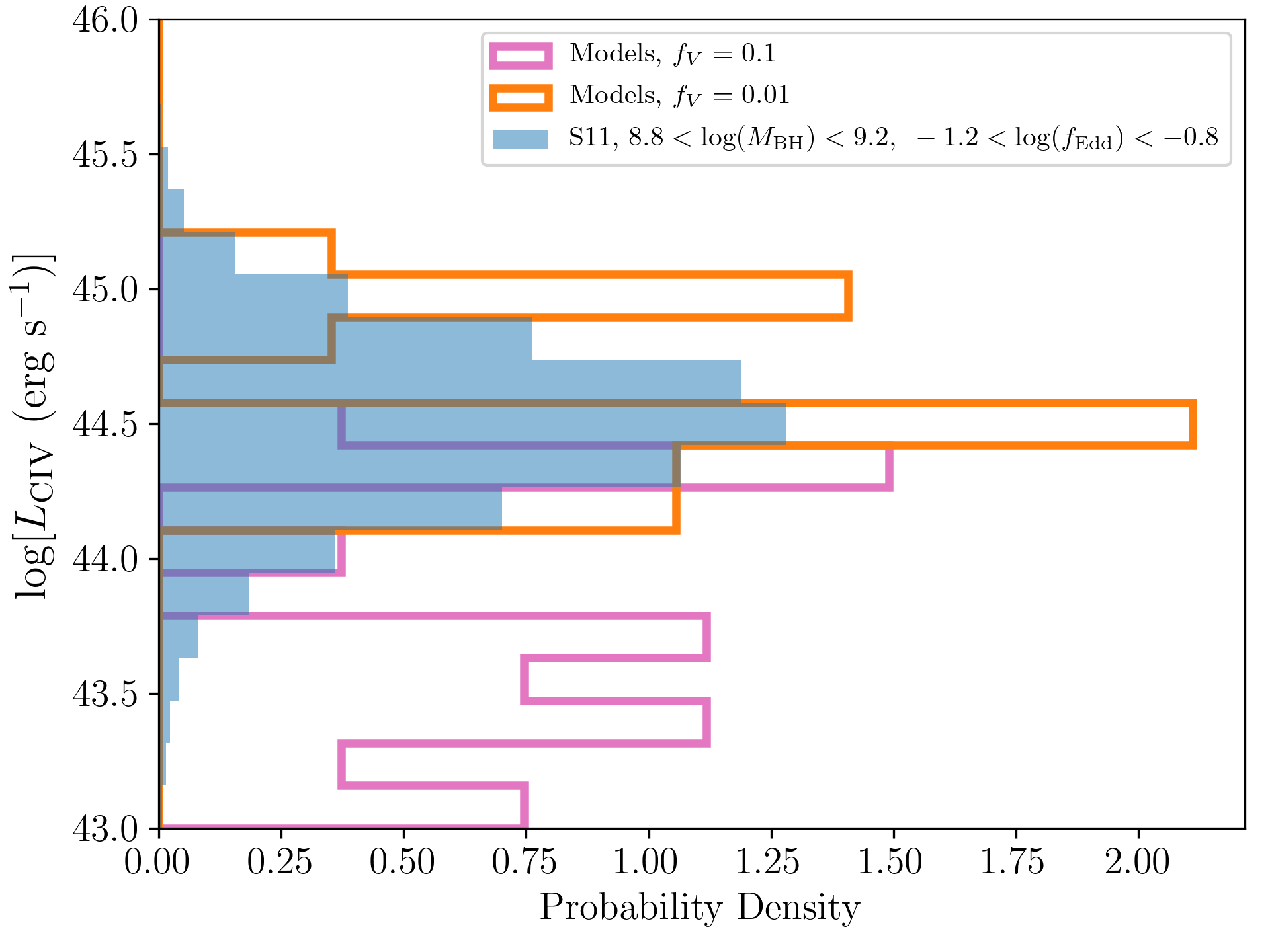}
\end{subfigure}
\caption{{\sl Left:} C\textsc{iv}~1550\AA\ line luminosity plotted against $\lambda L_{1350}$ for SDSS DR7 quasars and our simulation grid. The background histogram shows quasars from the \protect\cite{shen_catalog_2011} catalog of quasar properties from SDSS DR7 for $1.93<z<4.75$, the approximate redshift window where rest-frame wavelengths of 1350\AA\ and 1550\AA\ are both observable in SDSS DR7. The red dotted line shows the contour enclosing $68.2\%$ of objects. The circles show values from the simulation grid for models with $f_V=0.01$ (orange) and $f_V=0.1$ (pink). Models A and B are labelled. {\sl Right:} a 1D histogram of C\textsc{iv}~1550\AA\ line luminosity from the same catalog, but with the parameters limited to be comparable to our chosen simulation values: BH mass is between 8.8 and 9.2 (in logarithmic solar mass units) and the logarithm of the Eddington fraction limited between -1.2 and -0.8. Again, the simulation grid is shown for models with $f_V=0.01$ (orange) and $f_V=0.1$ (pink).
}
\label{fig:lums_hexbin}
\end{figure*}

\begin{figure*}
	\includegraphics[width=0.9\textwidth]{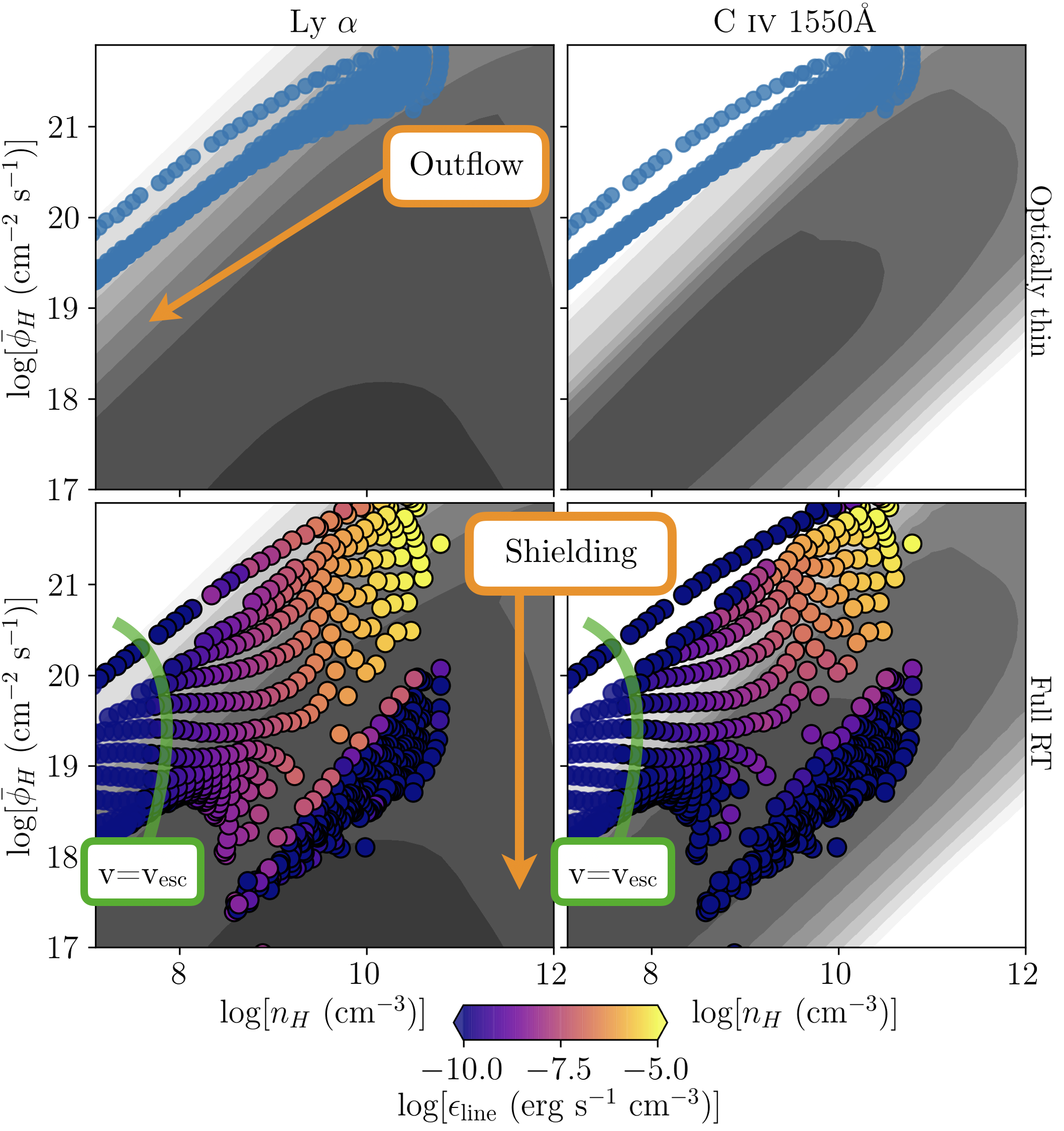}
    \caption{The \phin\ plane from a full MCRT photoionization simulation of a disc wind. In each case, the background contours show results from the \cld\ simulations described in the text, for Ly-$\alpha$ (Left) and \civline\ (right). The points show the location in the \phin\ plane in the disc wind model (for model A), both in the optically thin case (Top), where $\phi_H$ is set by equation~\ref{eq:phi_thin}, and in the full MCRT simulation where the estimator $\bar{\phi}_H$ comes from equation~\ref{eq:phibar}. In the bottom two panels, the points are colour-coded according to the logarithm of the emissivity of the respective lines. Orange arrows mark the approximate `smearing' effects of outflow and shielding as described in section~2, and the point at which $v=v_\mathrm{esc}$ is marked in green.
    }
    \label{fig:phi-nh-model}
\end{figure*}

\begin{figure*}
	\includegraphics[width=0.85\textwidth]{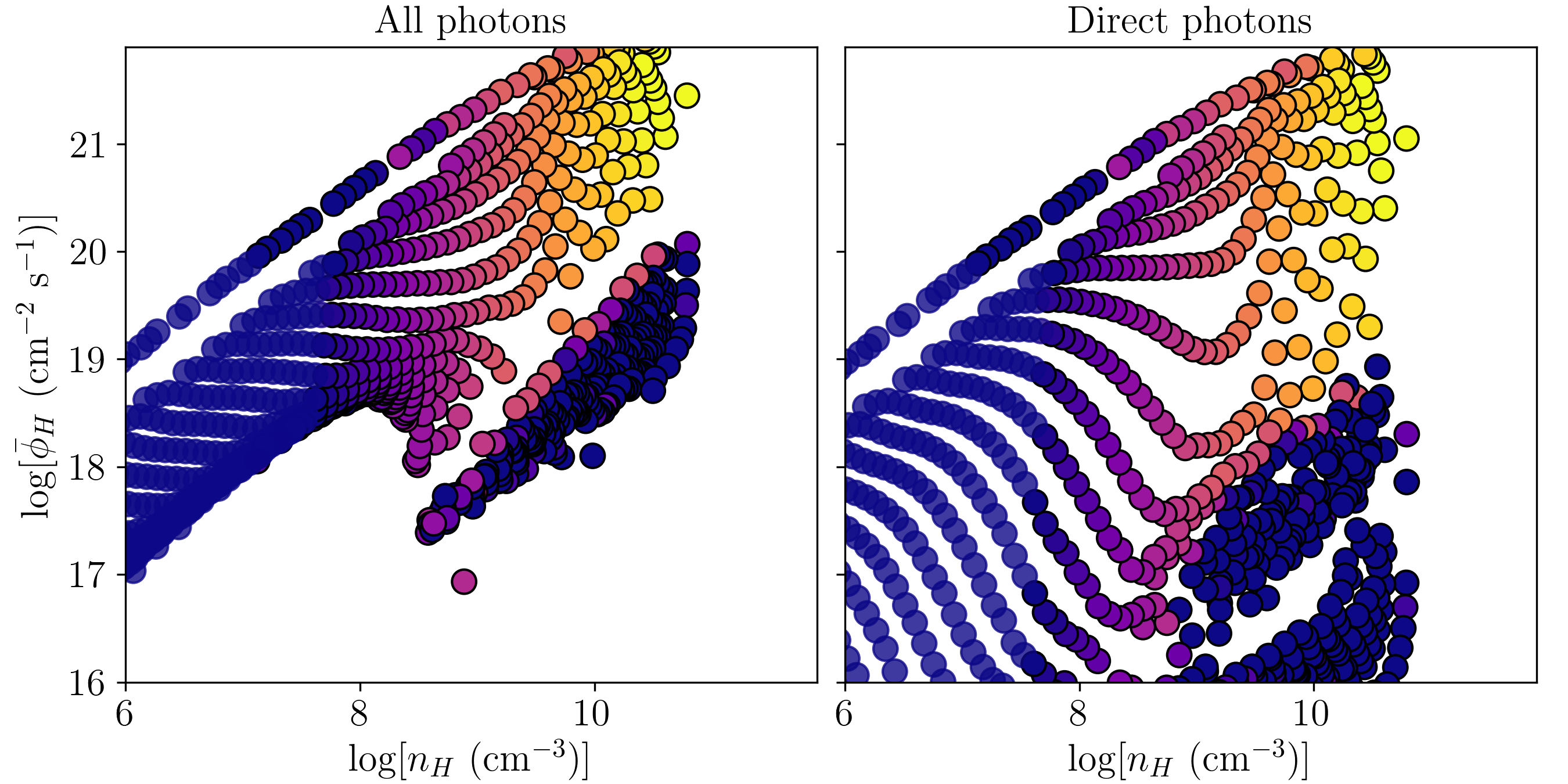}
    \caption{
    The effect of reprocessed radiation on the \phin\ plane. The points correspond to cells in the wind in model A, in the same manner as in Fig.~\ref{fig:phi-nh}, with the colour-coding indicating Ly-~$\alpha$ emissivity on the same colour scale as Fig.~\ref{fig:phi-nh}. The left-hand panel shows the actual result from the simulation, whereas the right-hand panel calculates the $\bar{\phi}_H$ estimator with only the direct photons that have not been reprocessed in anyway. The difference between the panels highlights the importance of reprocessed radiation in determining the incident flux and ionization state in the flow.
    }
    \label{fig:scattered}
\end{figure*}

\subsection{Physical Conditions and the ionizing flux-density plane}
\label{sec:conditions}
The ionization and temperature structure of disc winds is stratified, as shown by M16, but also suggested from general arguments by other authors \citep[e.g.][]{murray_accretion_1995,elvis_structure_2000,gallagher_stratified_2007,elvis_quasar_2017,yong_using_2018}. 
As we showed in section 2, we expect a disc wind to naturally populate a relatively large portion of the \phin\ plane due to absorption of the ionizing flux and variation in density. To estimate $\phi_H$, we adapt the standard MC estimator for the mean intensity \citep[e.g.][]{lucy_computing_1999}
so that the sum is over all photon packets with energy above the Lyman edge, which gives
\begin{equation}
    \bar{\phi}_H = \frac{1}{V} 
    \sum_i^{\geq 13.6~\mathrm{eV}}
    \frac{w_i \Delta s}{h\nu_i}. 
\label{eq:phibar}
\end{equation}
Here $w_i$ and $\nu_i$ are the luminosity weight and frequency of photon packet $i$, $V$ is the volume of the cell and $\Delta s$ is the path length. This estimator for $\phi_H$ is plotted against $n_H$ in Fig.~\ref{fig:phi-nh-model} for Model A.

Fig.~\ref{fig:phi-nh-model} demonstrates that a full RT simulation of a disc wind does indeed lead to a large portion of the \phin\ plane being covered. The background contours are from a \cld\ simulation, designed to match similar plots produced by, e.g., BFKD95, \cite{korista_atlas_1997} and \cite{bottorff_dynamics_1997}. The contours show the logarithm of the line flux relative to the 1215\AA\ continuum flux, covering the range $-4$ to $2$ in intervals of 0.5 dex. We used \cld\ version 17.01, last described by \cite{ferland_2017_2017}. Each grid point used to make the contours corresponds to a single cloud with column density $N_H = 10^{23}$~cm$^{-2}$ and an AGN SED generated using the \texttt{table agn} command.  The scatter points are grid cells from our simulations. In the top two panels, $\phi_H$ is calculated from the optically thin assumption (equation~\ref{eq:phi_thin}), whereas the bottom panels use $\bar{\phi}_H$ from our MCRT simulation (equation~\ref{eq:phibar}) and are colour-coded by line emissivity. We calculate emissivities for Ly~$\alpha$ (left) and \civline\ (right). The cells in the model populate a relatively large portion of the optimal parameter space due to nothing more than mass conservation and absorption effects. This plot therefore confirms our expectations from section~2 and also shows that the model has satisfactory resolution; the maximum change in $\phi_H$ from cell to cell is $\approx0.7~$dex, but normally much smaller, and the line emissivity changes relatively smoothly from cell-to-cell. In the shielded portion of the flow, at low values of $\phi_H$ and high values of $n_H$, the line emissivity decreases dramatically. Although there are still a reasonable number of ionizing photons in this region, they are all at high energies where cross-sections are small; the result is that little energy is converted from the incident radiation field into line emission. This effect illustrates the importance of taking into account frequency-dependent radiative transfer effects on the ionizing radiation field.

\begin{figure}
	\includegraphics[width=\linewidth]{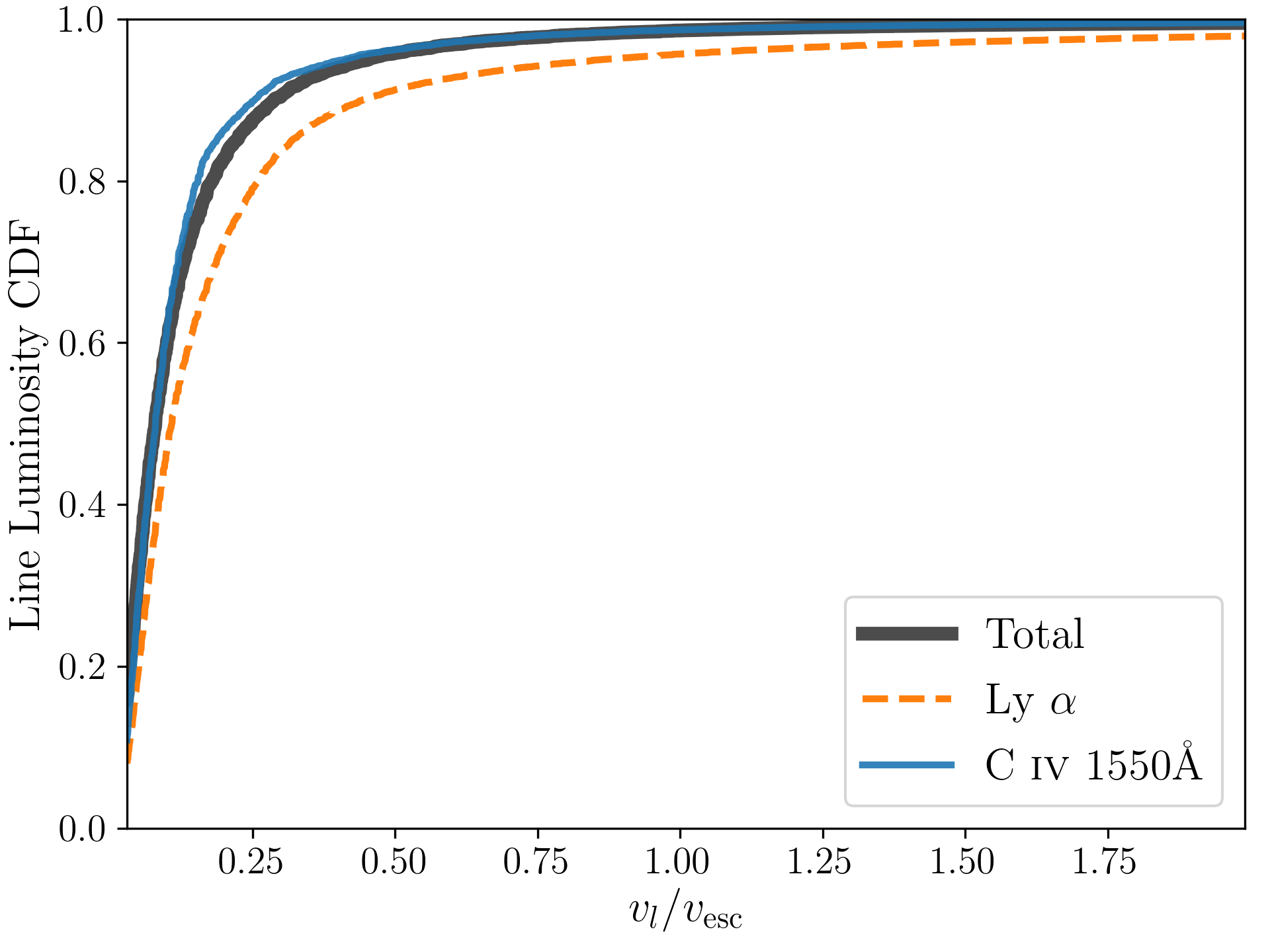}
    \caption{The velocity-ordered cumulative distribution function for the total, Ly-$\alpha$ and \civline\ line luminosity in model A. In all three cases, the majority of the line emission originates from material travelling below the local escape velocity.}
    \label{fig:cdf}
\end{figure}

The points in Fig.~\ref{fig:phi-nh-model} are also marked to indicate whether the local flow velocity is above or below the local escape velocity, $v_{\mathrm{esc}}=\sqrt{GM_{BH}/R}$, revealing an interesting result: the majority of the line emission is produced by plasma that has not yet reached escape velocity. This can be seen more clearly in Fig.~\ref{fig:cdf}, in which the normalised cumulative distribution function (CDF) of the line luminosity is plotted against the ratio of the poloidal velocity to the local escape velocity, $v_l/v_\mathrm{esc}$. To calculate the CDF, the cells are first sorted into ascending $v_l/v_\mathrm{esc}$ order, such that a CDF value of 0.9 at a velocity $v$ indicates that $90$\% of the line emission originates from plasma travelling with velocities $\leq v$. The fact that plasma that has not yet reached escape velocity dominates the line emission implies that {\em failed winds} might also be good candidates for the BLR. This conclusion may depend on the value of the acceleration length, $R_v$, for which we have adopted a relatively large value (compared to $R_\mathrm{launch}$) of $10^{19}$cm. A failed wind can occur in a line-driven scenario if it initially starts accelerating before becoming over-ionized \citep{proga_dynamics_2004}. This failed wind can be important in acting as `shield' to allow the outer wind to be driven more effectively \citep{murray_accretion_1995,proga_dynamics_2004}, and has been proposed as a possible origin of the soft X-ray excess in quasars \citep{schurch_failed_2006}. \cite{czerny_origin_2011} also proposed a failed wind as the possible origin of the BLR, albeit at larger distances \citep[see also][]{czerny_dust_2015,czerny_failed_2017}. We discuss some aspects of dusty winds and failed winds further in section~\ref{sec:failed_winds}.

It can be seen from Fig.~\ref{fig:phi-nh-model} that there are very few points at the lowest ionizing fluxes ($\bar{\phi}_H \lesssim 10^{18}$cm$^{-2}$s$^{-1}$) compared to the results from Fig.\ref{fig:phi-nh}. Fig.\ref{fig:phi-nh} instead suggests that, when $n_H \gtrsim 10^{10}$cm$^{-3}$, absorption should decrease the ionizing flux to very low values. The fact that this is not seen in the MCRT run is initially surprising, since the calculations shown in Fig.\ref{fig:phi-nh} use the Thomson cross-section, which should be a lower-limit on the opacity for an ionized plasma. The apparent discrepancy is explained by the impact of reprocessed radiation. Fig~\ref{fig:scattered} shows a $\log \bar{\phi}_H$-$\log n_H$ plane comparison between the same model, but with $\bar{\phi}_H$ computed both with all photon packets, and with just the direct photons (that is, the latter ignores scattered and reprocessed radiation). The plot illustrates that, in heavily shielded regions of the flow, the reprocessed radiation can be the dominant source of ionizing photons, as shown in other MCRT calculations that use snapshots from hydrodynamic simulations of disc winds \citep{sim_synthetic_2012,higginbottom_line-driven_2014}. The importance of reprocessed radiation also highlights the need for MCRT whenever an accurate ionization state is desired for a calculation involving a complex geometry in two or three dimensions. 

\subsection{Inclination effects and BAL sightlines}
\label{sec:bals}

\begin{figure*}
	\includegraphics[width=\linewidth]{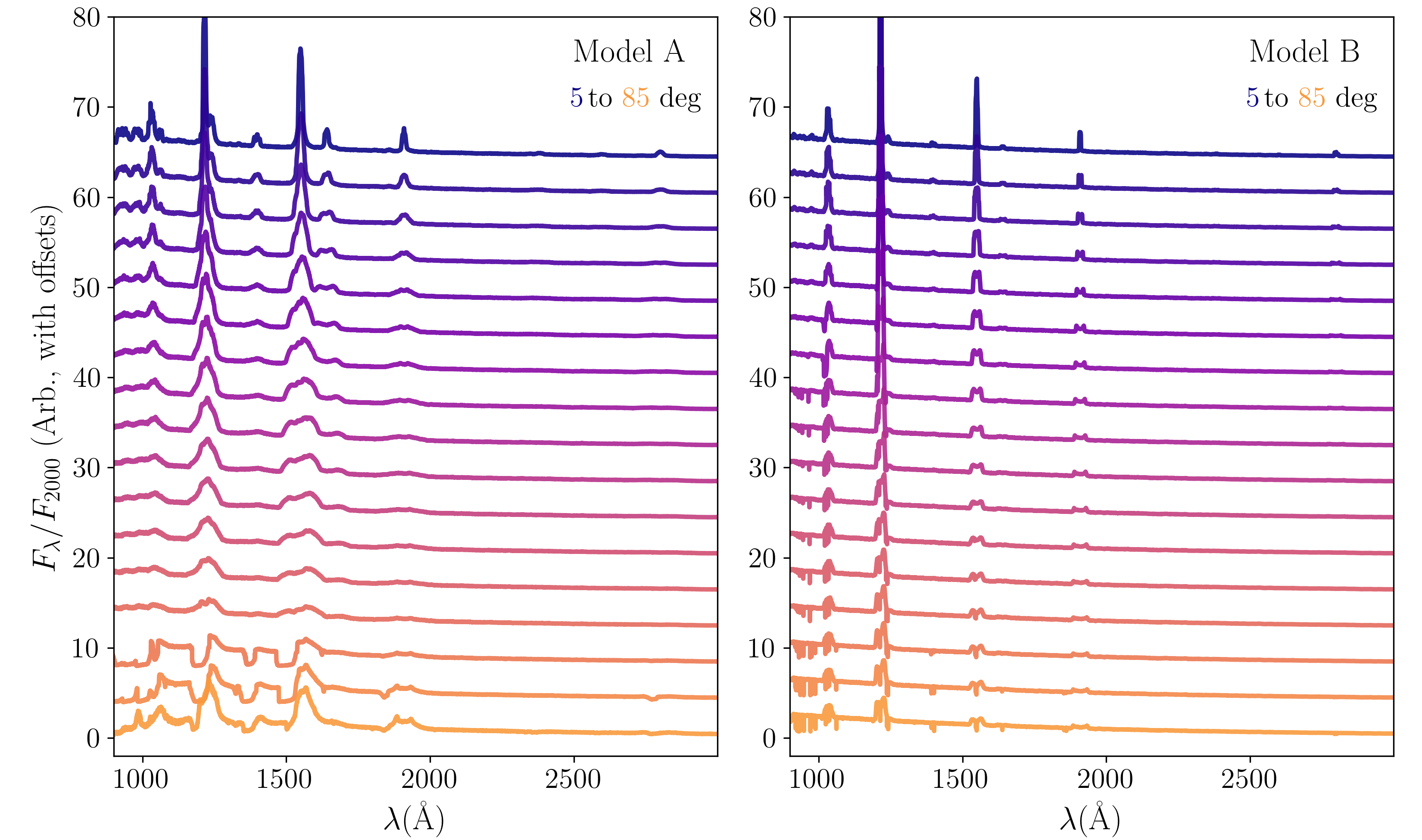}
    \caption{Ultraviolet spectra at different inclinations from Models A (Left) and B (Right). The spectra are normalised to the flux at $2000$\AA, and are shown from $i=5^\circ$ to $i=85^\circ$ at $5^\circ$ intervals. For clarity, offsets of $-4$ are applied per spectrum from low to high inclination, such that $i=85^\circ$ has $F_\lambda / F_{2000}=1$ at $2000$\AA. There are a few clear trends from low to high inclination: increasing line widths, ability to resolve double peaks and emergence of absorption lines. BALS are clearly seen in Model A, but Model B instead shows narrow absorption lines (NALs), illustrating that winds can sometimes produce strong troughs, but can also go more or less unnoticed bar the emission lines.
    }
    \label{fig:inclination}
\end{figure*}

\begin{figure*}
	\includegraphics[width=1.0\textwidth]{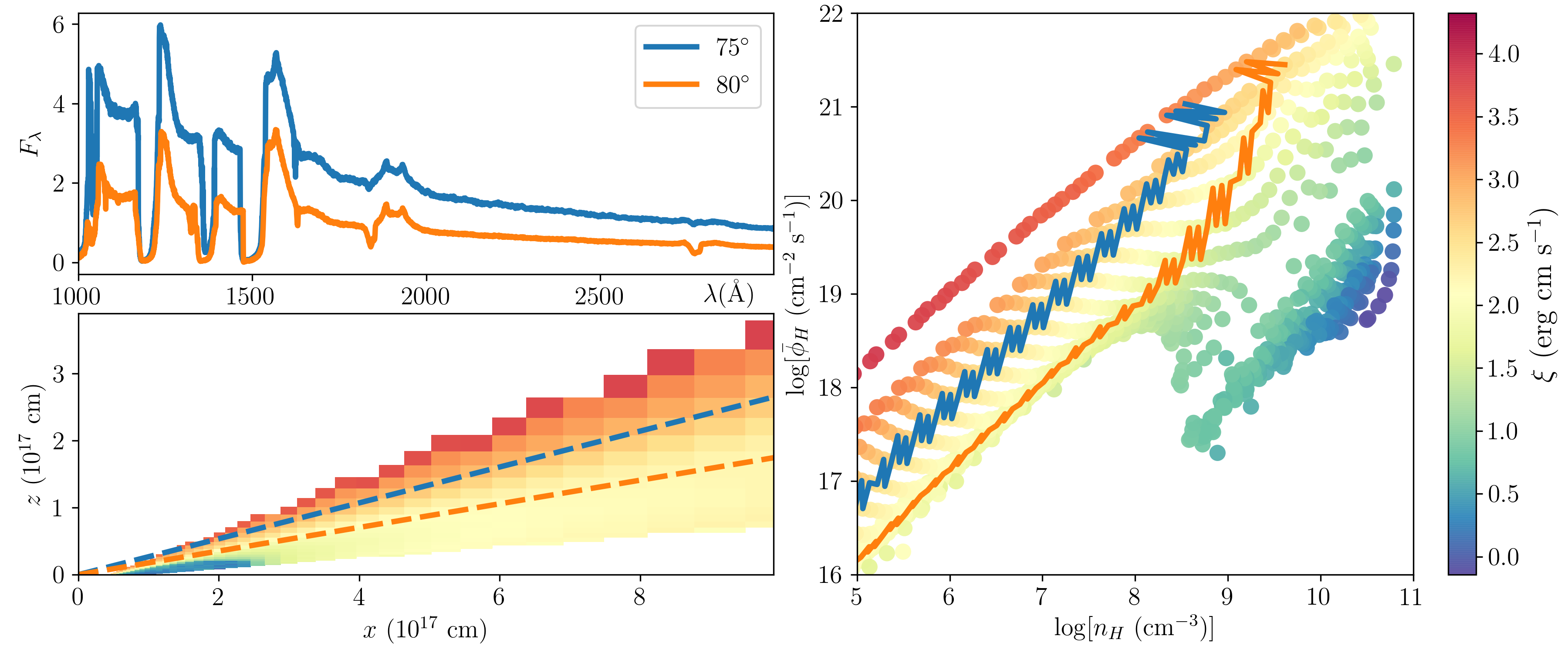}
    \caption{The formation of BALs in the wind model. {\sl Top left:} Synthetic UV spectra from model A at viewing angles of $75^\circ$ and $80^\circ$ (with wind opening angles of  $70^\circ$ and $82^\circ$). {\sl Bottom left:} A colour map of the ionization parameter, $\xi$, of the wind, in cylindrical coordinates. The two sightlines corresponding to the spectra shown in the upper panel are marked with dotted lines of the same colour. {\sl Right:} A colour map of the ionization parameter, $\xi$, of the wind, in the \phin\ plane. Each point corresponds to one cell in the model. The paths traversed by the two sightlines corresponding to the spectra shown in the upper panel are marked with lines of the same colour. The jagged structure in the lower part of the wind is due to sightlines crossing diagonally across cells in cylindrical geometry.}
    \label{fig:bals}
\end{figure*}

The inclination dependences in models A and B are depicted in Fig.~\ref{fig:inclination}, where we show spectra at angles from $5^\circ$ to $85^\circ$ at $5^\circ$ intervals. The first result is that, in contrast to the results from M16, there is {\em not} a clear increase of equivalent width with inclination. This is due to the choice of an isotropic continuum source as opposed to an anisotropic disc (see also sections~\ref{sec:params} and \ref{sec:limitations}). However, a few clear trends can still be seen in the emission lines. The lines transition from single to double-peaked as inclination increases, because the projection of $v_\phi$ increases with inclination. The overall width of the lines also increases, again due to projection effects. This effect is much more pronounced in the equatorial model A than in model B. The absolute width of the lines is also significantly higher in model A than in model B. This is because model A's wind is launched from closer in, where the Keplerian and escape velocities (and thus the poloidal and azimuthal velocities in the wind) are higher. Model B also produces narrow absorption lines (NALs) at a range of inclination, but never produces BALs. 

In model A, at angles looking into the wind $(\theta_1 < i < \theta_2)$, our models show BALs in a range of species, replicating the results of \cite{higginbottom_simple_2013} and M16, as well as earlier work by \cite{murray_accretion_1995}. The ionization parameter, $\xi$, of the wind tends to increase along streamlines (since the density drops more quickly than $1/R^2$) and decrease radially away from the ionizing source (due to absorption of ionizing photons by the wind). This stratified ionization structure means that different sightlines intersect different ranges of ionization states. We illustrate this in Fig.~\ref{fig:bals}, where we show synthetic spectra from model A at viewing angles of $75^\circ$ and $80^\circ$, together with a graphic that illustrates how the sightlines move through both physical space and \phin\ space. In both of the latter two cases, the colourmaps show the MC estimator form of $\xi$ from equation~\ref{eq:xi}. The figure demonstrates that the same stratified ionization structure that results in a BLR-like spectrum being produced also explains some of the behaviour of BALs in quasar spectra. LoBAL features are observed when the sightlines traverse denser, lower ionization state regions of the wind. LoBAL quasars, and particularly FeLoBAL quasars, are known to systematically differ from the rest of the BAL quasar population \citep{urrutia_first-2mass_2009,lazarova_nature_2012,dai_intrinsic_2012,morabito_origin_2019} and may correspond to a particular evolutionary stage of quasars \citep[e.g.][]{farrah_evidence_2007}. It is therefore unlikely that an inclination effect alone explains the different BALs observed, but our work demonstrates that it can be at least a significant factor. Generally speaking, one must consider both geometric and evolutionary effects when interpreting BAL phenomena, as discussed by, e.g. \cite{richards_unification_2011}.

\subsection{X-ray spectra}
We can also examine the X-ray spectra from the wind models. The left-hand panel of Fig.~\ref{fig:xray-spec} shows an X-ray spectrum, in units of flux density,  $F_\nu$, for models A and B at 3 different inclinations. The spectrum is characterised by broadband photoelectric absorption at angles looking into the wind cone, primarily by highly ionized (H-like and He-like) species of C, N, O and Ne. Some of the prominent edges and line transitions are marked. In Model A, in particular, the absorption is quite severe. This suggests that winds which produce BALs should also produce strong X-ray absorption, an interesting result given that BAL quasars are X-ray weak relative to their optical/UV luminosity \citep[e.g.][]{green_broad_1996,mathur_thomson_2000,brandt_nature_2000,green_chandra_2001,gallagher_x-ray_2002,gallagher_exploratory_2006,grupe_xmm-newton_2003,gibson_catalog_2009,morabito_suzaku_2011,morabito_unveiling_2014,luo_x-ray_2015,sameer_x-ray_2019}. To illustrate this further, we show a comparison with data in the right hand panel of Fig.~\ref{fig:xray-spec}. We show the relationship between the monochromatic X-ray luminosity at 2 keV, $L_{2\mathrm{keV}}$, and the monochromatic UV luminosity at 2500\AA, $L_{2500}$, from both models and observational data. The data are taken from the \cite{steffen_x-ray--optical_2006} sample of bright quasars and the \cite{saez_long-term_2012} sample of BAL quasars. We also show an alternative form of the same data in which the X-ray luminosity is replaced with a UV to X-ray spectral index, $\alpha_{\mathrm{OX}}$, defined as
\begin{equation}
    \alpha_{\mathrm{OX}} = 0.3838 \log_{10} \left(\frac{L_{2\mathrm{keV}}}{L_{2500}} \right).
\end{equation}
More negative values of $\alpha_{\mathrm{OX}}$ correspond to a lower X-ray to optical flux ratio. In model A, sightlines that produce BALs (i.e. equatorial viewing angles) show rough agreement with the level of X-ray weakness in BAL quasars at comparable UV luminosities. Model B has less severe absorption that results in the X-ray luminosity lying within the scatter of the main $L_{2\mathrm{keV}}-L_{2500}$ relation. Such a result is broadly consistent with the trend of weakening X-rays with increasing \civline\ absorption EW \cite{brandt_nature_2000,laor_luminosity_2002,richards_unification_2011} and general X-ray properties of BAL quasars. However, there is evidence that BAL quasars are {\em intrinsically} X-ray weak \citep{sabra_pg_2001,leighly2007b,luo_weak_2013,luo_weak_2014,morabito_unveiling_2014,teng_nustar_2014,liu_frequency_2018}. This result can be explained physically by considering that line-driving operates more efficiently when X-rays are weaker. We therefore expect there to be multiple factors affecting the emergent X-ray spectrum of BAL quasars \citep[see also][]{giustini_global_2019}, but the absorbing effect of the wind is clearly important.

A number of other interesting features can be seen in the X-ray spectra. The fact that the wind in model B absorbs the soft X-rays without producing BALs neatly illustrates how a disc wind can influence quasar spectra without being directly detectable in the optical or UV. Disc winds have been discussed in the context of X-ray absorption in a number of cases. For example, \cite{connolly_long-term_2016} discuss a disc wind as an explanation for the variable X-ray absorption in NGC 1365. There are other observational examples, some of which show simultaneous UV and X-ray absorption \citep[e.g.][]{reynolds_x-ray_1997,miller_absorption_2008,crenshaw_simultaneous_2003,kaspi_high-resolution_2001,kaspi_far_2004,steenbrugge_simultaneous_2005,ebrero_x-ray/uv_2013,kaastra_multiwavelength_2014}. The absorbing effects of disc winds may also prove important for explaining puzzling phenomena such as changing-look quasars and the BLR anomaly in NGC 5548 \citep{dehghanian_wind-based_2019}. The spectra from our wind models also show a number of strong emission lines in the soft X-rays, labeled on the plots, that can only be clearly seen when the continuum is absorbed. These emission lines originate in the hot ionized parts of the wind and are associated with species like Ne~\textsc{x}, O~\text{vii}, O~\text{viii} and N~\text{vii}. The presence of these lines once again indicates the relevance of ionization stratification in the flow. Emission and absorption lines in H-like and He-like species have been observed in high-resolution X-ray spectra of sources such as NGC~4051, MCG~6-30-15 and MR 2251-178 \citep{collinge_high-resolution_2001,gibson_line_2007,gibson_high-resolution_2005}. The series of absorption lines in the soft X-rays in the $50^\circ$ spectrum for model B are reminiscent of a `warm absorber' spectrum \citep[e.g.][]{kaastra_x-ray_2000,matsumoto_chandra_2004,ebrero_anatomy_2016}. Overall, the presence of simultaneous UV and X-ray absorption lines from a BLR model underlines the potential of unified schemes for AGN phenomena, especially those that invoke stratified disc winds \citep[e.g.][]{elvis_structure_2000,tombesi_unification_2013,nomura_line-driven_2017,elvis_quasar_2017,giustini_global_2019}.

\begin{figure*}
\centering
\begin{subfigure}{.5\textwidth}
  \centering
  \includegraphics[width=\linewidth]{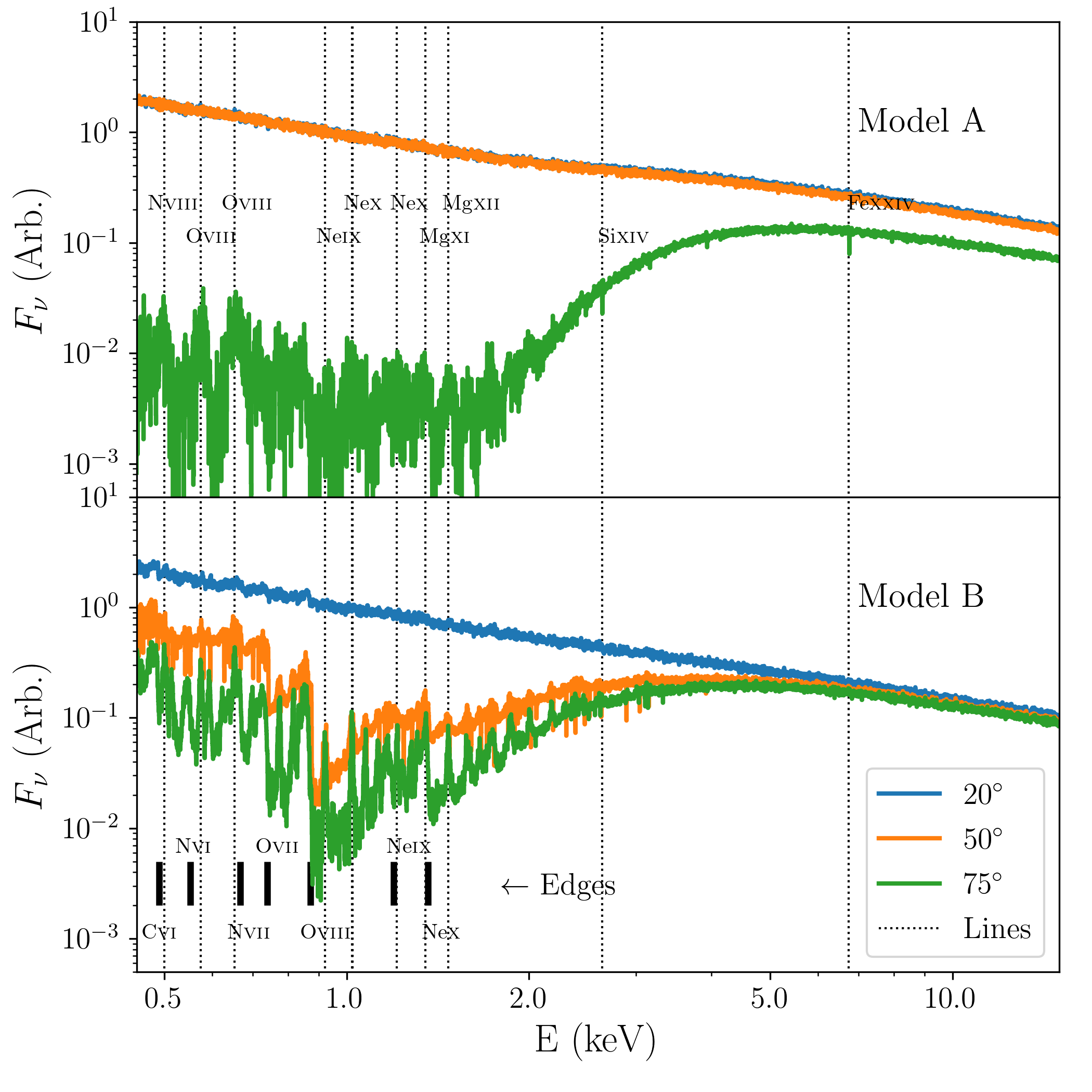}
\end{subfigure}%
\begin{subfigure}{.5\textwidth}
  \centering
  \includegraphics[width=\linewidth]{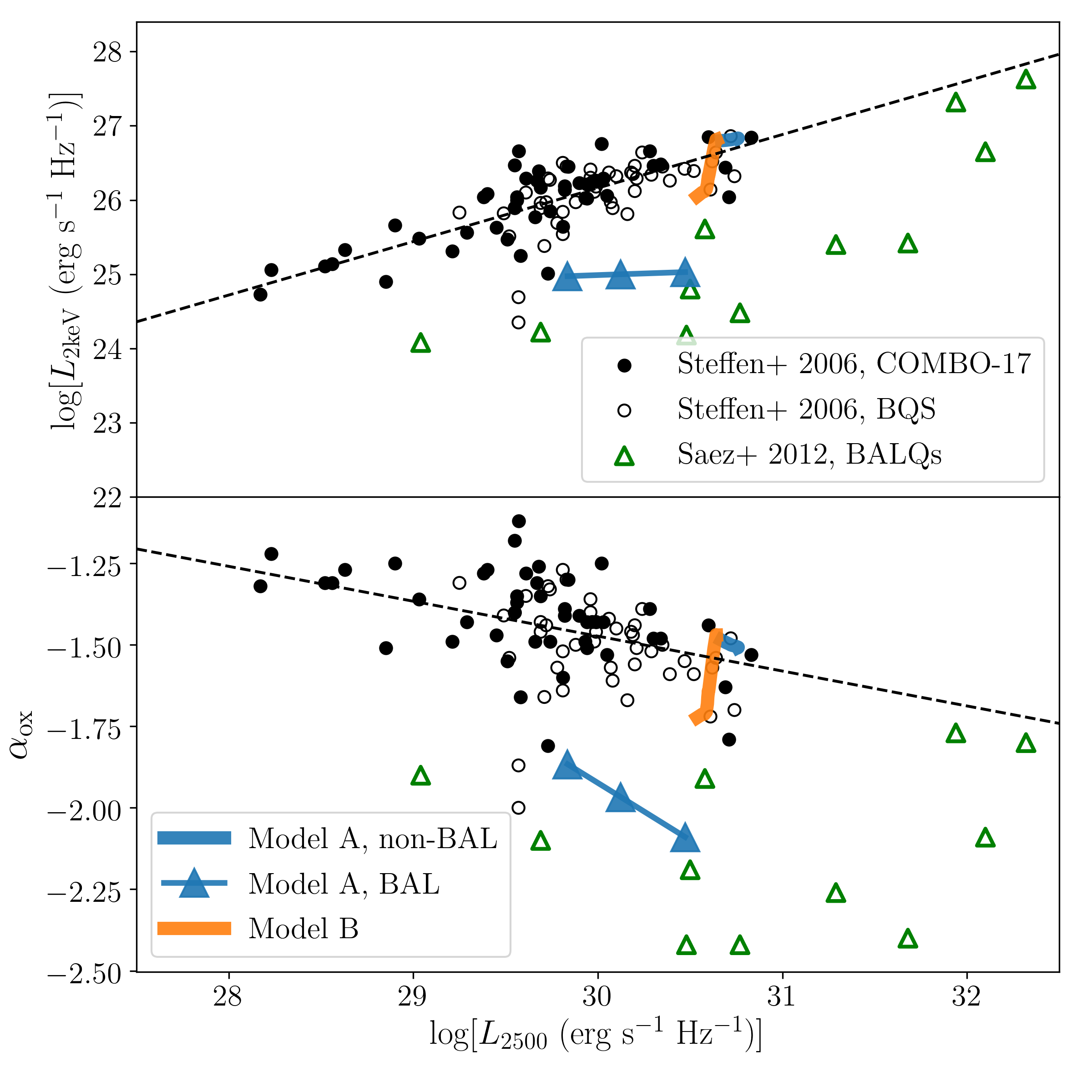}
\end{subfigure}
\caption{{\sl Left:} X-ray spectra for models A and B, in $F_\nu$ flux density units (erg~s${-1}$~cm$^{-2}$~Hz$^{-1}$) with an arbitrary normalisation applied. We show spectra for inclinations of $20^\circ,50^\circ$ and $75^\circ$. Some of the prominent absorption edges and line transitions are marked with thick solid and thin dotted lines, respectively. {\sl Right:} X-ray and UV properties of spectra from models A and B, for all viewing angles, compared to data from \protect\cite{steffen_x-ray--optical_2006} and \protect\cite{saez_long-term_2012}. The top panel shows the monochromatic 2 keV luminosity as a function of the monochromatic luminosity at 2500\AA, $L_{2500}$ and the bottom panel shows $\alpha_{\mathrm{OX}}$. Spectra featuring BALs in model A are labelled, and the \protect\cite{steffen_x-ray--optical_2006} data are labelled according to whether they come from the Bright Quasar Sample (BQS) or the COMBO-17 sample.
}
\label{fig:xray-spec}
\end{figure*}

\section{Discussion}
\label{sec:discussion} 

\subsection{Varying parameters: scalability and sensivity}
\label{sec:params}

The wind prescription we have used has various parameters that can be adjusted, relating to: (i) the scale and energetics of the system; (ii) the spectral shape, anisotropy and physical distribution of the illuminating photons; (iii) the acceleration, clumping and mass-loading of the wind; (iv) the geometry and launch radius of the wind. We have already investigated the sensitivity to some of these parameters (Table~\ref{grid_table}). We do not conduct an exhaustive test of parameter sensitivity here, but it is important to discuss how the parameters not part of the grid choice may affect the emergent spectrum. Some exploration of parameter space in this class of models has been carried out in our other work. For example, \cite{higginbottom_simple_2013} show the impact of $\dot{M}_{\mathrm{wind}}$ and $L_X$ on BAL formation in a smooth wind, M16 show the effect of clumping and high $L_X$ on line EWs and BAL strength, and \cite{mangham_reverberation_2017} show that a `Seyfert' type model, with $M_{\mathrm{BH}}=10^7 M_\odot$, produces clear broad emission lines, also with $f_V=0.01$. The results of M16 also show how disc anisotropy has a significant effect on the emergent line emission, particularly at low inclinations where the line EWs are much smaller than in our models, even for very similar wind parameters. We also ran a model with the same parameters as model A but using the mean quasar SED shown in Fig.~\ref{fig:seds}, adjusted slightly to mask out the small blue bump and mid-IR bump. We found that this model produced extremely similar spectra to that obtained when using the M16 SED shape. The spectrum from this model is shown in Appendix A, together with some preliminary investigations into the sensitivity of the spectra to $\dot{M}_{\mathrm{wind}}$ and SED anisotropy.

\subsection{Failed winds, BLR mass and emission measure}
\label{sec:failed_winds}
The emission measure of a steady-state, axisymmetric biconical wind can be written as an integral over the entire volume of the wind (equation~\ref{eq:EM}). For ionized winds, $n_H \approx n_e$, and the integrand is proportional to $n_H^2 f_V$. The total mass in a steady-state wind can either be written in a similar manner as an integral over volume, or alternatively as
\begin{equation}
    M_{\mathrm{BLR}} \approx \dot{M}_{\mathrm{wind}} t_{\mathrm{flow}},
\end{equation}
where we define the flow timescale as $t_{\mathrm{flow}} = \int_0^{R_v} (1/v_l) dl$ for a single streamline. This equation is approximate because (i) in reality there is variation in flow time for different streamlines and launch radii, and (ii) material beyond $R_v$ also contributes to the mass. For small $v_0$, this equation can be solved analytically giving $t_{\mathrm{flow}} = 3 R_v / v_\infty$ for $\alpha=0.5$. We obtain flow times of $475,1063$ and $1503$ years for our three launch radii (since $v_\infty \propto R_{\mathrm{launch}}^{-1/2}$). The corresponding BLR masses are approximately $2400 M_\odot,5300 M_\odot$ and $7500 M_\odot$. These values agree with the sum of the mass in the actual simulation wind cells to within $10\%$.

The mass-loss rate sets the normalisation of both EM and $M_{\mathrm{BLR}}$, so $\mathrm{EM}\propto \dot{M}_{\mathrm{wind}}^2$ as a result. According to \cite{king_agn_2015}, we might expect the mass of a disc, $M_d$, to be limited by the self-gravitation radius, in which case $M_d \sim (H/R) M_{\mathrm{BH}}$, giving $M_d \sim 10^7 M_\odot$ for a thin ($H/R=0.01$) disc and our canonical quasar BH mass. Estimates of the BLR mass from cloud models vary, but \cite{baldwin_mass_2003} argue that $10^{3-4}~M_\odot$ is most likely for luminous quasars. Both this estimate and our modelling imply that the BLR mass is a small fraction of the disc mass, so a single accretion episode can easily supply the mass for the entire BLR. In our wind modelling, we made the simple assumption that $\dot{M}_{\mathrm{wind}} = \dot{M}_{\mathrm{acc}}$. This is reasonable if the disc is to avoid disruption, although even at these mass-loss rates we would expect the temperature profile of a continuum emitting accretion disc to be modified \citep[e.g.][]{knigge_effective_1999,laor_line-driven_2014}. This is particularly true if the wind is magnetically driven and extracts angular momentum from the flow. If the BLR is instead composed of a failed wind, then the mass and emission measure of the line-emitting plasma is no longer constrained by $\dot{M}_{\mathrm{wind}}$, because gas that falls back on to the disc can presumably still be accreted and contribute to the radiative luminosity. One possible test of outflow or failed outflow models for the BLR would involve detecting virialised BLR gas that exists when the AGN is turned off. This is difficult, but tidal disruption events and changing-look systems might provide fruitful avenues of investigation.

\subsection{The Fe pseudocontinuum}
\label{sec:pseudo}
The $2200-2800$~\AA\ region of the observed spectra of type 1 AGN typically shows an Fe~\textsc{ii}~UV bump, consisting of a series of emission lines that merge together to form a `pseudocontinuum'. An example of an empirical template for this bump is provided by \cite{vestergaard_empirical_2001}. Our modelling does not include a sufficiently complete Fe model atom to effectively reproduce the lines that create the bump, partly because it is computationally expensive to search through -- and construct estimators for -- large numbers of lines in the MC procedure. Such a model would require a macro-atom treatment with hundreds of Fe~\textsc{ii} levels and thousands of lines. For this reason, the fact there are discrepancies between our model spectra and the composite quasar continuum in the near UV region is not surprising, but we cannot predict whether a disc wind model similar to those discussed here will create a strong Fe~\textsc{ii}~UV bump. We can, however, examine the abundance of Fe~\textsc{ii} in the wind model. The ion fraction of Fe~\textsc{ii} is around $0.3$ near the wind base ($z \lesssim 10^{16}$cm), where the densities are $n_H \approx 10^{10}$cm. 

Disc winds do possess a number of interesting characteristics relating to the Fe~\textsc{ii} pseudocontinuum. \cite{baldwin_origin_2004} studied the Fe~\textsc{ii} pseudocontinuum with LOC models, including microturbulence. They were particularly concerned with reconciling the observed EWs with the observed spike-to-gap ratio (a parameterisation of the strength of the 3 main multiplets relative to the rest of the Fe~\textsc{ii} emission). They found that, without microturbulence, the observed spike-to-gap ratio of around 0.7 was only consistent with a narrow region in the \phin\ plane. Although this region is almost always crossed by a portion of our wind models, it is unlikely that such a model would also produce the required Fe~\textsc{ii} emission EW. \cite{baldwin_origin_2004} showed that the discrepancy between spike-to-gap ratios and EWs can be explained by microturbulence with $v_{\mathrm{turb}}\gtrsim 100~$km~s$^{-1}$, or alternatively, as they note, the velocity changing by $>100$km~s$^{-1}$ over one mean free path of a continuum photon. If the continuum opacity is comparable to the Thomson opacity, an approximate necessary condition is $(dv/dl)/(\sigma_T n_H) \gtrsim 100$km~s$^{-1}$. The Thomson opacity is a lower limit on the continuum opacity in an ionized plasma and so this is a minimal requirement. This condition is plotted as a function of $n_H$ in Fig.~\ref{fig:turb} for the same five streamlines as presented in Fig.~\ref{fig:phi-nh}. The result suggests that a disc wind accelerating as slowly as model A is unlikely to be able to reproduce the right spike-to-gap ratio for the Fe~\textsc{ii} pseudocontinuum if velocity gradients are responsible, since the condition is only met once $n_H\lesssim10^{8}$cm$^{-3}$, where the line emissivity is quite low. This conclusion is strengthened if the opacity exceeds the Thomson value. A faster accelerating wind might meet the condition closer to the disc plane, but will have lower densities and a lower overall emission measure. It therefore seems likely that a disc wind must also be (micro)turbulent if it is to reproduce all aspects of the broad-line region. 

\begin{figure}
	\includegraphics[width=\linewidth]{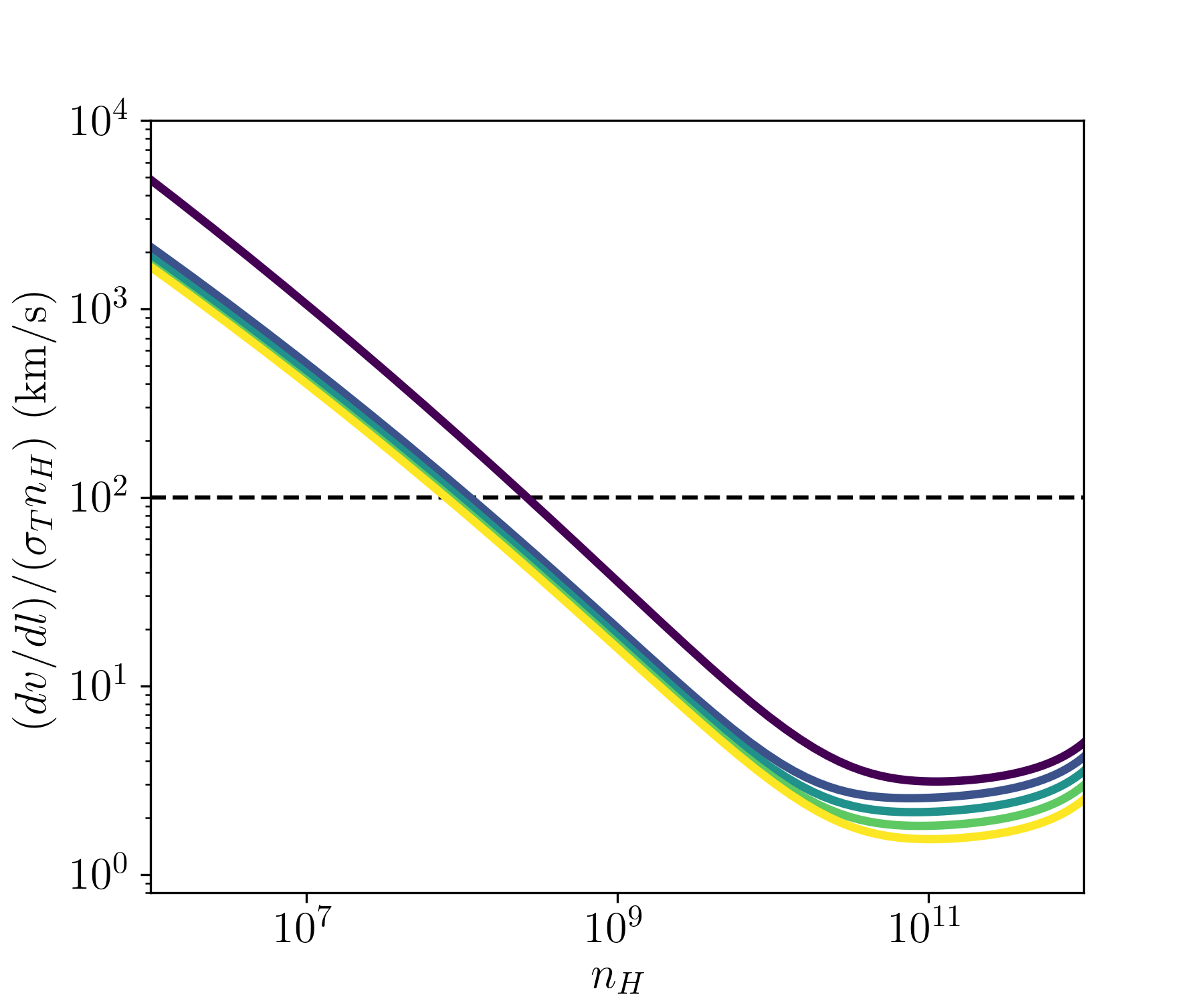}
    \caption{The ratio of the velocity gradient to the Thomson mean free path, $(dv/dl)/(\sigma_T n_H)$, as a function of $n_H$ for the same five streamlines as presented in Fig.~\ref{fig:phi-nh}. The dotted line shows the necessary condition proposed by \protect\cite{baldwin_origin_2004} for explaining the observed spike-to-gap ratio in the Fe~\textsc{ii} pseudocontinuum.}
    
    \label{fig:turb}
\end{figure}

\subsection{Limitations and caveats}
\label{sec:limitations}
There are a number of limitations and caveats of this work, some of which have already been discussed. Here we provide a focused discussion of some of the more pressing concerns.

\bigskip

{\bf (i) Tuning and wind parameters:} One of the most appealing aspects of the original BFKD95 proposal is that it dispenses with the need to tune BLR parameters to match observations. However, inspection of Figs.~\ref{fig:mean-spec} and \ref{fig:lums_hexbin}, combined with the fact that models with $f_V=1$ do not produce strong emission lines, shows that some tuning of parameters is necessary to match observations (although the required parameters are quite reasonable).  To obtain the required overall line emissivity, we seem to need volume filling factors of $0.1-0.01$ and quite large acceleration lengths for winds with $\dot{M}_{\mathrm{wind}}\approx \dot{M}_{\mathrm{acc}}$. As described in section~\ref{sec:failed_winds}, this issue is alleviated if the BLR is associated with failed winds, because the overall emission measure is then not constrained by $\dot{M}_{\mathrm{wind}}$. To obtain the required ionization state, we also need a clumpy wind or an effective `shield'. However, from BAL quasar spectra, we know that BAL outflows (somehow) manage to avoid over-ionization, so any BAL outflow will, very roughly, be in the correct region of ionization parameter space for the BLR. The behaviour described in this paper can then lead to a large portion of \phin\ parameter space being covered. While our work acts as a proof-of-concept in this regard, these statements need severe qualification. The situation would be aided by a robust physical model for BAL outflows or further detailed photoionization calculations involving physically motivated dust-driven wind models.

\smallskip

{\bf (ii) BLR size and reverberation lags:} As demonstrated by \cite{mangham_reverberation_2017}, the reverberation lags calculated using the M16 wind model are too short, by a factor of about 5, when compared to observations of similar luminosity objects. The wind models, like model A, with $R_{\mathrm{launch}}=450 r_g \approx 25$ld presented here have the same launch radius as M16, so have similar issues. Although we do not repeat a full reverberation analysis here, we expect the models with larger launch radii (e.g. model B) to produce better agreement with observed lag-luminosity relationships. However, it is not easy to explain how a disc wind might be launched from a radial distance of 1000s of $r_g$. For comparison, the self-gravitation radius is  $R_{\mathrm{sg}}\sim0.01$pc and a weak function of mass \citep{collin-souffrin_line_1990,shlosman_fuelling_1990,hure_structure_1994,king_fuelling_2007}. A launch radius of $1000$s of $r_g$ exceeds $R_{\mathrm{sg}}$ for $M_{\mathrm{BH}}=10^9M_\odot$. It is possible that the BLR could somehow be moved upwards in the flow, rather than originating further outwards along the disc plane. Nonetheless, large reverberation lags compared to theoretical wind launch radii pose a problem to line-driven wind models for the BLR -- we therefore emphasise that the LOC-type behaviour found here should apply to other models for the BLR in which the material is magnetically or dust-driven \citep[e.g.][]{blandford_hydromagnetic_1982,emmering_magnetic_1992,konigl_disk-driven_1994,czerny_origin_2011}. It is also possible that the BLR is associated with the inner edge of a dusty torus \citep[e.g.][]{clavel_steps_1991,hu_h_2008,zhu_evidence_2009,kawaguchi_orientation_2010,gandhi_dust_2015}, which could itself be dynamic and outflowing \citep{chan_radiation-driven_2016,williamson_3d_2019,leftley_parsec-scale_2019}. This class of models can also have some beneficial reverberation characteristics if they have `bowl-shaped' geometries \citep{goad_broad_2012,ramolla_simultaneous_2018}. Many of the general principles we have discussed would apply to such models and further detailed simulations are worthwhile.

It is worth noting that, in a general sense, the question of {\em scale} is a difficult one for physical models of AGN. BAL outflow distances in excess of 100 pc have been inferred from density-sensitive lines \citep{borguet_major_2012,chamberlain_strong_2015,arav_evidence_2018}. This is significantly larger than expected from disc wind theory or simulations and may require alternative scenarios involving shocks or cloud disruption \citep{faucher-giguere_physical_2012,zeilig_hess_balqso_2019}. The  accretion-disc size problem concerns the finding that effective radii from microlensing and continuum lags are a factor of a few larger than expected from standard disc theory \citep{morgan_quasar_2010,antonucci_astrophysics:_2013,edelson_space_2015,de_rosa_space_2015}. In addition, the BLR lag-luminosity relationship extends beyond the self-gravity radius above a critical luminosity. This is because the BLR size scales as roughly the square root of luminosity, whereas $R_{sg}$ only has a very weak scaling with mass \citep{collin-souffrin_line_1990,shlosman_fuelling_1990,hure_structure_1994,king_fuelling_2007}. \cite{king_fuelling_2007} and \cite{king_agn_2015} argue that material beyond this radius will form stars rather than forming part of the accretion disc. These issues hint at fundamental problems with the simplest physical models for AGN and quasars, going beyond an issue particular to wind models for the BLR.

\smallskip

{\bf (iii) Clumping:} 
In our modelling, we assume clumps are optically thin and that the clumping factor is constant throughout the flow. These approximations are unlikely to be particularly accurate. Let us first examine the optically thin assumption. Taking the Thomson opacity, the optically thin requirement gives maximum clump sizes on the order $10^{14}$cm ($10^{16}$cm) for electron densities of $10^{10}$cm$^{-3}$ ($10^{8}$cm$^{-3}$). For comparison, \cite{proga_cloud_2015} find typical cloud sizes of $6\times10^{10}$cm, \cite{waters_2019} find $\sim10^{11-12}$ cm, \cite{de_kool_radiation_1995} suggest $10^{11}$cm and \cite{laor_is_2006} obtain an upper limit of $10^{12}$ cm. \cite{mccourt_characteristic_2018} propose a universal length scale $N_H\sim10^{17}$cm for cold clumps that fragment isobarically. Thus, in regions of the flow where the opacity is relatively low, microclumping may actually be a good approximation. However, there are clearly situations where the microclumping assumption breaks down, such as where line or bound-free opacity is high, and this aspect of our work should be improved (see also section~\ref{sec:conclusion}). The value of the clumping factor $1/f_V$ is fairly high, although not unreasonable, when compared to estimates for stellar winds in O and OB stars, which range from a few to around 100 \citep[e.g.][]{hillier_constraints_1999,bouret_lower_2005,fullerton_discordance_2006,puls_bright_2006,muijres_predictions_2011}. A bigger issue may be the use of a constant clumping factor throughout the flow. It is possible that clumping is due to an instability that grows along streamlines, or that a characteristic column density or size leads to a variation of clumping factor with location. 

\section{Summary, Conclusions \& Future Work}
\label{sec:conclusion} 

We have investigated whether clumpy biconical disc winds illuminated by an isotropic AGN-like continuum produce spectra resembling those of luminous, type 1 AGN and quasars. We first used the ionizing flux-density plane as a diagnostic tool to explore some general properties of simple BLR models, before conducting MCRT and photoionization calculations using our state-of-the-art 2.5D MCRT Sobolev code, \py. Our main conclusions are as follows:
\begin{itemize}
    \item Disc winds naturally populate a large portion of the ionizing flux-density plane in a manner that intersects multiple ionization states. This is due to a combination of factors: the density drops off along streamlines faster than $1/R^2$, and absorption in the flow leads to further stratification in ionization structure, particular near the base of the wind. This means that disc winds, and more generally, `flow' models for the BLR can be expected to behave in a similar manner to LOC models in that they display a degree of `optimally emitting' behaviour.
    \item Our detailed MCRT simulations generally confirm the above finding and we have shown that the synthetic spectra we compute from our clumpy biconical disc wind models provide a reasonable match to real quasar composite spectra, with EWs and line luminosities comparable to those observed in real quasars. Volume filling factors of $f_V \lesssim 0.1$ are needed to moderate the ionization state and increase the density sufficiently to produce strong emission lines in the spectra.
    \item Scattered and reprocessed radiation is important in determining the ionization state in the wind, highlighting the need for accurate radiative transfer treatments that can treat multiple scattering and non-spherical geometries. 
    \item The majority of line emission originates near the base of the wind where the wind velocity is below the escape velocity. This result suggests that failed winds might be important BLR candidates. This is important in terms of the emission measure of the wind, because the emission measure of a successful wind is bounded by the mass-loss rate to infinity, whereas a failed wind can have a higher total emission measure that is not necessarily bounded by this value. 
    \item For some of the models, BALs are produced for sightlines passing through the wind cone. The sightlines intersect a range of ionization states and are not well characterised by a single ionization state absorber. Higher inclination sightlines generally pass through denser and cooler material such that they produce prominent LoBAL features. Geometry and inclination are therefore important variables to consider when interpreting BALs in quasar spectra, in tandem with evolutionary and environmental effects.
    \item The disc wind models presented here are effective X-ray absorbers. For models and sightlines that produce UV BALs, the broadband photoelectric absorption is sufficient to explain the X-ray weakness of BAL quasars. Even when BALs are not produced, a series of X-ray features are seen at inclination angles exceeding $\theta_1$, the inner opening angle. This behaviour demonstrates the ability of a disc wind to influence observables even when obvious wind features are not detected. These features include photoionization edges in H-like and He-like species, absorption lines and emission lines. The spectra are reasonably similar to X-ray spectra from AGN absorbed in the soft X-rays and qualitatively resemble warm absorbers. Our results suggest that a unified disc wind model can explain aspects of warm absorbers, BAL quasars and the BLR simultaneously.
    \item We have discussed our results in the context of line-driven winds, magnetically driven flows and the dusty failed wind model of \cite{czerny_origin_2011}. Although our model is based on the type used to successfully simulate BAL quasar UV spectra, the general principles discussed should apply to any biconical outflow model for the BLR.
    \item The results from our simulation grid are publicly available in the supplementary material, with the data made available in an online repository with address \url{https://github.com/jhmatthews/windy-blr-2020}.  
\end{itemize}
Our results are encouraging, in the sense that they provide support for a physically motivated wind model for the BLR,
but there are a number of limitations to the model in general and to our specific approach. These were discussed in section~\ref{sec:limitations}. Perhaps the most pressing concerns are the simple treatment of clumping, the use of an isotropic SED and the difficulties in reproducing the observed lag-luminosity relationship. There are a number of possible avenues for future work. The treatment of clumping could be modifed, to treat clumping in a statistical sense in a manner similar to \citep{stalevski_3d_2012}, or to allow $f_V$  to vary throughout the flow as implemented by \cite{puls_bright_2006} for stellar winds. Another improvement could be to include dust opacity in the flow, so that dusty wind models for the BLR can be tested. Although we have explored the sensitivity to some model parameters, it would also be interesting to investigate possible scalings with mass, Eddington fraction or SED shape, in the light of the results of \cite{richards_unification_2011} and \cite{hamann_structure_2019}. However, the use of a kinematic, prescribed model makes it hard to investigate dynamic effects -- for example, changes in wind strength due to an increased Eddington fraction or a softer SED shape. The results are also dependent on the specific choices of variables or wind parameterisation.  To improve this, we are engaged in an effort to conduct coupled MCRT-hydrodynamic simulations of line-driven winds. Our method has been successfully applied to thermal winds in X-ray binaries \cite[e.g.][]{higginbottom_radiation-hydrodynamic_2018}, and a line-driving treatment has recently been included (Higginbottom et al, submitted). Another multi-frequency line-driving scheme has also been recently outlined by \cite{dyda_line_2019}. These approaches will allow disc wind models for BAL outflows and the BLR to be tested further and from a more physics-based perspective. 

\section*{Acknowledgements}
We thank the anonymous reviewer for their time and for a helpful report. We would also like to thank David Williamson, Martin Elvis, Dom Walton, Peter Kosec, Chris Reynolds, Sergei Dyda, Gordon Richards and Paul Hewett for helpful conversations, and the attendees and organisers of the `AGN Winds on the Georgia Coast' conference in summer 2017 for useful discussions. JHM acknowledges a Herchel Smith Research Fellowship at Cambridge. NSH and CK acknowledge support from the Science and Technology Facilities Council grant ST/M001326/1. KSL acknowledges the support of NASA for this work through grant NNG15PP48P to serve as a science adviser to the Astro-H project.  EJP would like to acknowledge financial support from the EPSRC Centre for Doctoral Training in Next Generation Computational Modelling grant EP/L015382/1. Figures were generated using Matplotlib \citep{Hunter:2007}. The authors would like to acknowledge the use of the University of Oxford Advanced Research Computing (ARC) facility in carrying out this work. http://dx.doi.org/10.5281/zenodo.22558

%%%%%%%%%%%%%%%%%%%%%%%%%%%%%%%%%%%%%%%%%%%%%%%%%%


\bibliographystyle{mnras}

\begin{thebibliography}{}
\makeatletter
\relax
\def\mn@urlcharsother{\let\do\@makeother \do\$\do\&\do\#\do\^\do\_\do\%\do\~}
\def\mn@doi{\begingroup\mn@urlcharsother \@ifnextchar [ {\mn@doi@}
  {\mn@doi@[]}}
\def\mn@doi@[#1]#2{\def\@tempa{#1}\ifx\@tempa\@empty \href
  {http://dx.doi.org/#2} {doi:#2}\else \href {http://dx.doi.org/#2} {#1}\fi
  \endgroup}
\def\mn@eprint#1#2{\mn@eprint@#1:#2::\@nil}
\def\mn@eprint@arXiv#1{\href {http://arxiv.org/abs/#1} {{\tt arXiv:#1}}}
\def\mn@eprint@dblp#1{\href {http://dblp.uni-trier.de/rec/bibtex/#1.xml}
  {dblp:#1}}
\def\mn@eprint@#1:#2:#3:#4\@nil{\def\@tempa {#1}\def\@tempb {#2}\def\@tempc
  {#3}\ifx \@tempc \@empty \let \@tempc \@tempb \let \@tempb \@tempa \fi \ifx
  \@tempb \@empty \def\@tempb {arXiv}\fi \@ifundefined
  {mn@eprint@\@tempb}{\@tempb:\@tempc}{\expandafter \expandafter \csname
  mn@eprint@\@tempb\endcsname \expandafter{\@tempc}}}

\bibitem[\protect\citeauthoryear{Abbott \& Lucy}{Abbott \&
  Lucy}{1985}]{abbott_multiline_1985}
Abbott D.~C.,  Lucy L.~B.,  1985, \mn@doi [\apj] {10.1086/162834}, 288, 679

\bibitem[\protect\citeauthoryear{Allen, Hewett, Maddox, Richards  \&
  Belokurov}{Allen et~al.}{2011}]{allen_strong_2011}
Allen J.~T.,  Hewett P.~C.,  Maddox N.,  Richards G.~T.,   Belokurov V.,  2011,
  \mn@doi [\mnras] {10.1111/j.1365-2966.2010.17489.x}, 410, 860

\bibitem[\protect\citeauthoryear{Antonucci}{Antonucci}{1993}]{antonucci_unified_1993}
Antonucci R.,  1993, \mn@doi [ARA\&A]
  {10.1146/annurev.aa.31.090193.002353}, 31, 473

\bibitem[\protect\citeauthoryear{Antonucci}{Antonucci}{2013}]{antonucci_astrophysics:_2013}
Antonucci R.,  2013, \mn@doi [\nat] {10.1038/495165a}, 495, 165

\bibitem[\protect\citeauthoryear{Arav, Liu, Xu, Stidham, Benn  \&
  Chamberlain}{Arav et~al.}{2018}]{arav_evidence_2018}
Arav N.,  Liu G.,  Xu X.,  Stidham J.,  Benn C.,   Chamberlain C.,  2018,
  \mn@doi [\apj] {10.3847/1538-4357/aab494}, 857, 60

\bibitem[\protect\citeauthoryear{Baldwin, Ferland, Korista  \& Verner}{Baldwin
  et~al.}{1995}]{baldwin_locally_1995}
Baldwin J.,  Ferland G.,  Korista K.,   Verner D.,  1995, \mn@doi [\apjl]
  {10.1086/309827}, 455, L119

\bibitem[\protect\citeauthoryear{Baldwin et al.}{2003}]{baldwin_mass_2003} 
Baldwin J.~A., Ferland G.~J., Korista K.~T., Hamann F., Dietrich M., 2003, \apj, 582, 590

\bibitem[\protect\citeauthoryear{Baldwin, Ferland, Korista, Hamann  \&
  LaCluyzé}{Baldwin et~al.}{2004}]{baldwin_origin_2004}
Baldwin J.~A.,  Ferland G.~J.,  Korista K.~T.,  Hamann F.,   LaCluyzé A.,
  2004, \mn@doi [\apj] {10.1086/424683}, 615, 610

\bibitem[\protect\citeauthoryear{Barvainis}{1987}]{barvainis_1987} Barvainis R., 1987, ApJ, 320, 537

\bibitem[\protect\citeauthoryear{Baskin \& Laor}{Baskin \&
  Laor}{2005}]{baskin_what_2005}
Baskin A.,  Laor A.,  2005, \mn@doi [\mnras]
  {10.1111/j.1365-2966.2004.08525.x}, 356, 1029

\bibitem[\protect\citeauthoryear{Baskin \& Laor}{Baskin \&
  Laor}{2018}]{baskin_dust_2018}
Baskin A.,  Laor A.,  2018, \mn@doi [\mnras] {10.1093/mnras/stx2850}, 474, 1970

\bibitem[\protect\citeauthoryear{Bentz, Peterson, Pogge, Vestergaard  \&
  Onken}{Bentz et~al.}{2006}]{bentz_radius-luminosity_2006}
Bentz M.~C.,  Peterson B.~M.,  Pogge R.~W.,  Vestergaard M.,   Onken C.~A.,
  2006, \mn@doi [\apj] {10.1086/503537}, 644, 133

\bibitem[\protect\citeauthoryear{Bentz, Peterson, Netzer, Pogge  \&
  Vestergaard}{Bentz et~al.}{2009}]{bentz_radius-luminosity_2009}
Bentz M.~C.,  Peterson B.~M.,  Netzer H.,  Pogge R.~W.,   Vestergaard M.,
  2009, \mn@doi [\apj] {10.1088/0004-637X/697/1/160}, 697, 160

\bibitem[\protect\citeauthoryear{Bentz et~al.,}{Bentz
  et~al.}{2013}]{bentz_low-luminosity_2013}
Bentz M.~C.,  et~al., 2013, \mn@doi [\apj] {10.1088/0004-637X/767/2/149}, 767,
  149

\bibitem[\protect\citeauthoryear{Blandford \& McKee}{Blandford \&
  McKee}{1982}]{blandford_reverberation_1982}
Blandford R.~D.,  McKee C.~F.,  1982, \mn@doi [\apj] {10.1086/159843}, 255, 419

\bibitem[\protect\citeauthoryear{Blandford \& Payne}{Blandford \&
  Payne}{1982}]{blandford_hydromagnetic_1982}
Blandford R.~D.,  Payne D.~G.,  1982, \mn@doi [\mnras] {10.1093/mnras/199.4.883}, 199, 883

\bibitem[\protect\citeauthoryear{Blustin, Page, Fuerst, Branduardi-Raymont  \&
  Ashton}{Blustin et~al.}{2005}]{blustin_nature_2005}
Blustin A.~J.,  Page M.~J.,  Fuerst S.~V.,  Branduardi-Raymont G.,   Ashton
  C.~E.,  2005, \mn@doi [A\&A] {10.1051/0004-6361:20041775}, 431, 111

\bibitem[\protect\citeauthoryear{Borguet, Arav, Edmonds, Chamberlain  \&
  Benn}{Borguet et~al.}{2012}]{borguet_major_2012}
Borguet B. C.~J.,  Arav N.,  Edmonds D.,  Chamberlain C.,   Benn C.,  2012,
  \mn@doi [\apj] {10.1088/0004-637X/762/1/49}, 762, 49

\bibitem[\protect\citeauthoryear{Bottorff \& Ferland}{Bottorff \&
  Ferland}{2001}]{bottorff_fractal_2001}
Bottorff M.,  Ferland G.,  2001, \mn@doi [\apj] {10.1086/319083}, 549, 118

\bibitem[\protect\citeauthoryear{Bottorff, Korista, Shlosman  \&
  Blandford}{Bottorff et~al.}{1997}]{bottorff_dynamics_1997}
Bottorff M.,  Korista K.~T.,  Shlosman I.,   Blandford R.~D.,  1997, \mn@doi
  [\apj] {10.1086/303867}, 479, 200

\bibitem[\protect\citeauthoryear{Bottorff, Korista  \& Shlosman}{Bottorff
  et~al.}{2000}]{bottorff_dynamics_2000}
Bottorff M.~C.,  Korista K.~T.,   Shlosman I.,  2000, \mn@doi [\apj]
  {10.1086/309006}, 537, 134

\bibitem[\protect\citeauthoryear{Bottorff, Baldwin, Ferland, Ferguson  \&
  Korista}{Bottorff et~al.}{2002}]{bottorff_he_2002}
Bottorff M.~C.,  Baldwin J.~A.,  Ferland G.~J.,  Ferguson J.~W.,   Korista
  K.~T.,  2002, \mn@doi [\apj] {10.1086/344408}, 581, 932

\bibitem[\protect\citeauthoryear{Bouret, Lanz  \& Hillier}{Bouret
  et~al.}{2005}]{bouret_lower_2005}
Bouret J.-C.,  Lanz T.,   Hillier D.~J.,  2005, \mn@doi [A\&A]
  {10.1051/0004-6361:20042531}, 438, 301

\bibitem[\protect\citeauthoryear{Brandt, Laor  \& Wills}{Brandt
  et~al.}{2000}]{brandt_nature_2000}
Brandt W.~N.,  Laor A.,   Wills B.~J.,  2000, \mn@doi [\apj] {10.1086/308207}, 528, 637

\bibitem[\protect\citeauthoryear{Cappi, Mihara, Matsuoka, Hayashida, Weaver  \&
  Otani}{Cappi et~al.}{1996}]{cappi_warm_1996}
Cappi M.,  Mihara T.,  Matsuoka M.,  Hayashida K.,  Weaver K.~A.,   Otani C.,
  1996, \mn@doi [\apj] {10.1086/176799}, 458, 149

\bibitem[\protect\citeauthoryear{Chajet \& Hall}{Chajet \&
  Hall}{2013}]{chajet_magnetohydrodynamic_2013}
Chajet L.~S.,  Hall P.~B.,  2013, \mn@doi [\mnras] {10.1093/mnras/sts580}, 429, 3214

\bibitem[\protect\citeauthoryear{Chajet \& Hall}{Chajet \&
  Hall}{2017}]{chajet_magnetohydrodynamic_2017}
Chajet L.~S.,  Hall P.~B.,  2017, \mn@doi [\mnras] {10.1093/mnras/stw2626}, 465, 1741

\bibitem[\protect\citeauthoryear{Chamberlain, Arav  \& Benn}{Chamberlain
  et~al.}{2015}]{chamberlain_strong_2015}
Chamberlain C.,  Arav N.,   Benn C.,  2015, \mn@doi [\mnras]
  {10.1093/mnras/stv572}, 450, 1085

\bibitem[\protect\citeauthoryear{Chan \& Krolik}{Chan \&
  Krolik}{2016}]{chan_radiation-driven_2016}
Chan C.-H.,  Krolik J.~H.,  2016, \mn@doi [\apj] {10.3847/0004-637X/825/1/67},
  825, 67

\bibitem[\protect\citeauthoryear{Chartas, Brandt  \& Gallagher}{Chartas
  et~al.}{2003}]{chartas_xmm-newton_2003}
Chartas G.,  Brandt W.~N.,   Gallagher S.~C.,  2003, \mn@doi [\apj]
  {10.1086/377299}, 595, 85

\bibitem[\protect\citeauthoryear{Clavel et~al.,}{Clavel
  et~al.}{1991}]{clavel_steps_1991}
Clavel J.,  et~al., 1991, \mn@doi [\apj] {10.1086/169540}, 366, 64

\bibitem[\protect\citeauthoryear{Coatman, Hewett, Banerji  \& Richards}{Coatman
  et~al.}{2016}]{coatman_c_2016}
Coatman L.,  Hewett P.~C.,  Banerji M.,   Richards G.~T.,  2016, \mn@doi
  [\mnras] {10.1093/mnras/stw1360}, 461, 647

\bibitem[\protect\citeauthoryear{Collin-Souffrin \& Dumont}{Collin-Souffrin \&
  Dumont}{1990}]{collin-souffrin_line_1990}
Collin-Souffrin S.,  Dumont A.~M.,  1990, \aap,
  229, 292

\bibitem[\protect\citeauthoryear{Collinge et~al.,}{Collinge
  et~al.}{2001}]{collinge_high-resolution_2001}
Collinge M.~J.,  et~al., 2001, \mn@doi [ApJ] {10.1086/321635}, 557, 2

\bibitem[\protect\citeauthoryear{Connolly, McHardy, Skipper  \&
  Emmanoulopoulos}{Connolly et~al.}{2016}]{connolly_long-term_2016}
Connolly S.~D.,  McHardy I.~M.,  Skipper C.~J.,   Emmanoulopoulos D.,  2016,
  \mn@doi [\mnras] {10.1093/mnras/stw878}, 459, 3963

\bibitem[\protect\citeauthoryear{Crenshaw et~al.,}{Crenshaw
  et~al.}{2003}]{crenshaw_simultaneous_2003}
Crenshaw D.~M.,  et~al., 2003, \mn@doi [\apj] {10.1086/376792}, 594, 116

\bibitem[\protect\citeauthoryear{Czerny \& Hryniewicz}{Czerny \&
  Hryniewicz}{2011}]{czerny_origin_2011}
Czerny B.,  Hryniewicz K.,  2011, \mn@doi [\aap] {10.1051/0004-6361/201016025},
  525, L8

\bibitem[\protect\citeauthoryear{Czerny et~al.,}{Czerny
  et~al.}{2014}]{czerny_dust_2014}
Czerny B.,  et~al., 2014, \mn@doi [arXiv:1409.7312 [astro-ph]]
  {10.1016/j.asr.2015.01.004}

\bibitem[\protect\citeauthoryear{Czerny et~al.,}{Czerny
  et~al.}{2015}]{czerny_dust_2015}
Czerny B.,  et~al., 2015, \mn@doi [Advances in Space Research]
  {10.1016/j.asr.2015.01.004}, 55, 1806

\bibitem[\protect\citeauthoryear{Czerny et~al.,}{Czerny
  et~al.}{2017}]{czerny_failed_2017}
Czerny B.,  et~al., 2017, \mn@doi [\apj] {10.3847/1538-4357/aa8810}, 846, 154

\bibitem[\protect\citeauthoryear{Dai, Shankar  \& Sivakoff}{Dai
  et~al.}{2008}]{dai_2mass_2008}
Dai X.,  Shankar F.,   Sivakoff G.~R.,  2008, \mn@doi [\apj] {10.1086/523688},
  672, 108

\bibitem[\protect\citeauthoryear{Dai, Shankar  \& Sivakoff}{Dai
  et~al.}{2012}]{dai_intrinsic_2012}
Dai X.,  Shankar F.,   Sivakoff G.~R.,  2012, \mn@doi [\apj]
  {10.1088/0004-637X/757/2/180}, 757, 180

\bibitem[\protect\citeauthoryear{Dannen et al.}{2020}]{dannen2020} 
Dannen R., Proga D., Waters T., Dyda S., 2020, arXiv, arXiv:2001.00133


\bibitem[\protect\citeauthoryear{De~Rosa et~al.,}{De~Rosa
  et~al.}{2015}]{de_rosa_space_2015}
De~Rosa G.,  et~al., 2015, \mn@doi [\apj] {10.1088/0004-637X/806/1/128}, 806, 128

\bibitem[\protect\citeauthoryear{de Kool \& Begelman}{de~Kool \&
  Begelman}{1995}]{de_kool_radiation_1995}
de Kool M.,  Begelman M.~C.,  1995, \mn@doi [\apj]
  {10.1086/176594}, 455, 448

\bibitem[\protect\citeauthoryear{Dehghanian et~al.,}{Dehghanian
  et~al.}{2019}]{dehghanian_wind-based_2019}
Dehghanian M.,  et~al., 2019, \mn@doi [\apj] {10.3847/2041-8213/ab3d41}, 882,
  L30

\bibitem[\protect\citeauthoryear{Denney et~al.,}{Denney
  et~al.}{2009}]{denney_diverse_2009}
Denney K.~D.,  et~al., 2009, \mn@doi [\apj] {10.1088/0004-637X/704/2/L80}, 704,
  L80

\bibitem[\protect\citeauthoryear{Dere, Landi, Mason, Monsignori~Fossi  \&
  Young}{Dere et~al.}{1997}]{dere_chianti_1997}
Dere K.~P.,  Landi E.,  Mason H.~E.,  Monsignori~Fossi B.~C.,   Young P.~R.,
  1997, \mn@doi [A\&AS]
  {10.1051/aas:1997368}, 125, 149

\bibitem[\protect\citeauthoryear{DiPompeo, Brotherton, Cales  \&
  Runnoe}{DiPompeo et~al.}{2012}]{dipompeo_rest-frame_2012}
DiPompeo M.~A.,  Brotherton M.~S.,  Cales S.~L.,   Runnoe J.~C.,  2012, \mn@doi
  [\mnras]
  {10.1111/j.1365-2966.2012.21971.x}, 427, 1135

\bibitem[\protect\citeauthoryear{Du \& Wang}{Du \&
  Wang}{2019}]{du_radius-luminosity_2019}
Du P.,  Wang J.-M.,  2019, arXiv e-prints, p. arXiv:1909.06735

\bibitem[\protect\citeauthoryear{Du et~al.,}{Du
  et~al.}{2016}]{du_supermassive_2016}
Du P.,  et~al., 2016, \mn@doi [\apj] {10.3847/0004-637X/820/1/27}, 820, 27

\bibitem[\protect\citeauthoryear{Dyda \& Proga}{2018}]{dyda_geometry_2018} Dyda S., Proga D., 2018, \mnras, 481, 2745

\bibitem[\protect\citeauthoryear{Dyda \& Proga}{2018}]{dyda_axisymmetry_2018} Dyda S., Proga D., 2018, \mnras, 475, 3786

\bibitem[\protect\citeauthoryear{Dyda, Reynolds \& Jiang}{2019}]{dyda_line_2019} Dyda S., Reynolds C.~S., Jiang Y.-F., 2019, arXiv, arXiv:1911.09183

\bibitem[\protect\citeauthoryear{Ebrero, Kaastra, Kriss, de Vries  \&
  Costantini}{Ebrero et~al.}{2013}]{ebrero_x-ray/uv_2013}
Ebrero J.,  Kaastra J.~S.,  Kriss G.~A.,  de Vries C.~P.,   Costantini E.,
  2013, \mn@doi [MNRAS] {10.1093/mnras/stt1497}, 435, 3028

\bibitem[\protect\citeauthoryear{Ebrero et~al.,}{Ebrero
  et~al.}{2016}]{ebrero_anatomy_2016}
Ebrero J.,  et~al., 2016, \mn@doi [A\&A] {10.1051/0004-6361/201527808}, 587,
  A129

\bibitem[\protect\citeauthoryear{Edelson et~al.,}{Edelson
  et~al.}{2015}]{edelson_space_2015}
Edelson R.,  et~al., 2015, \mn@doi [\apj] {10.1088/0004-637X/806/1/129}, 806, 129

\bibitem[\protect\citeauthoryear{Elvis}{Elvis}{2000}]{elvis_structure_2000}
Elvis M.,  2000, \mn@doi [\apj]
  {10.1086/317778}, 545, 63

\bibitem[\protect\citeauthoryear{Elvis}{Elvis}{2017}]{elvis_quasar_2017}
Elvis M.,  2017, \mn@doi [\apj] {10.3847/1538-4357/aa82b6}, 847, 56

\bibitem[\protect\citeauthoryear{Emmering, Blandford  \& Shlosman}{Emmering
  et~al.}{1992}]{emmering_magnetic_1992}
Emmering R.~T.,  Blandford R.~D.,   Shlosman I.,  1992, \mn@doi [\apj]
  {10.1086/170955}, 385, 460

\bibitem[\protect\citeauthoryear{Eracleous \& Halpern}{Eracleous \&
  Halpern}{1994}]{eracleous_double-peaked_1994}
Eracleous M.,  Halpern J.~P.,  1994, \mn@doi [\apjs]
  {10.1086/191856}, 90, 1

\bibitem[\protect\citeauthoryear{Eracleous \& Halpern}{Eracleous \&
  Halpern}{2003}]{eracleous_completion_2003}
Eracleous M.,  Halpern J.~P.,  2003, \mn@doi [\apj] {10.1086/379540}, 599, 886

\bibitem[\protect\citeauthoryear{Everett}{Everett}{2005}]{everett_radiative_2005}
Everett J.~E.,  2005, \mn@doi [\apj] {10.1086/432678}, 631, 689

\bibitem[\protect\citeauthoryear{Fabian et~al.,}{Fabian
  et~al.}{1994}]{fabian_asca_1994}
Fabian A.~C.,  et~al., 1994, \pasj, 46, L59

\bibitem[\protect\citeauthoryear{Farrah, Lacy, Priddey, Borys  \&
  Afonso}{Farrah et~al.}{2007}]{farrah_evidence_2007}
Farrah D.,  Lacy M.,  Priddey R.,  Borys C.,   Afonso J.,  2007, \mn@doi [\apj]
  {10.1086/519492}, 662, L59

\bibitem[\protect\citeauthoryear{Faucher-Gigu\'{e}re, Quataert  \&
  Murray}{Faucher-Gigu\'{e}re et~al.}{2012}]{faucher-giguere_physical_2012}
Faucher-Gigu\'{e}re C.-A.,  Quataert E.,   Murray N.,  2012, \mn@doi [\mnras]
  {10.1111/j.1365-2966.2011.20120.x}, 420, 1347

\bibitem[\protect\citeauthoryear{Feldmeier, Oskinova  \& Hamann}{Feldmeier
  et~al.}{2003}]{feldmeier_x-ray_2003}
Feldmeier A.,  Oskinova L.,   Hamann W.-R.,  2003, \mn@doi [\aap]
  {10.1051/0004-6361:20030231}, 403, 217

\bibitem[\protect\citeauthoryear{Ferland et~al.,}{Ferland
  et~al.}{2017}]{ferland_2017_2017}
Ferland G.~J.,  et~al., 2017, Revista Mexicana de Astronomia y Astrofisica, 53,
  385

\bibitem[\protect\citeauthoryear{Flohic, Eracleous  \& Bogdanović}{Flohic
  et~al.}{2012}]{flohic_effects_2012}
Flohic H. M. L.~G.,  Eracleous M.,   Bogdanović T.,  2012, \mn@doi [\apj]
  {10.1088/0004-637X/753/2/133}, 753, 133

\bibitem[\protect\citeauthoryear{Fullerton, Massa  \& Prinja}{Fullerton
  et~al.}{2006}]{fullerton_discordance_2006}
Fullerton A.~W.,  Massa D.~L.,   Prinja R.~K.,  2006, \mn@doi [ApJ]
  {10.1086/498560}, 637, 1025

\bibitem[\protect\citeauthoryear{Galianni \& Horne}{Galianni \&
  Horne}{2013}]{galianni_test_2013}
Galianni P.,  Horne K.,  2013, \mn@doi [\mnras] {10.1093/mnras/stt1507}, 435,
  3122

\bibitem[\protect\citeauthoryear{Gallagher \& Everett}{Gallagher \&
  Everett}{2007}]{gallagher_stratified_2007}
Gallagher S.~C.,  Everett J.~E.,  2007, in Ho L.~C.,  Wang J.-W.,  eds,
  Astronomical {Society} of the {Pacific} {Conference} {Series} Vol. 373, The
  {Central} {Engine} of {Active} {Galactic} {Nuclei}. p.~305

\bibitem[\protect\citeauthoryear{Gallagher, Brandt, Chartas  \&
  Garmire}{Gallagher et~al.}{2002}]{gallagher_x-ray_2002}
Gallagher S.~C.,  Brandt W.~N.,  Chartas G.,   Garmire G.~P.,  2002, \mn@doi
  [ApJ] {10.1086/338485}, 567, 37

\bibitem[\protect\citeauthoryear{Gallagher, Brandt, Chartas, Priddey, Garmire
  \& Sambruna}{Gallagher et~al.}{2006}]{gallagher_exploratory_2006}
Gallagher S.~C.,  Brandt W.~N.,  Chartas G.,  Priddey R.,  Garmire G.~P.,
  Sambruna R.~M.,  2006, \mn@doi [\apj]
  {10.1086/503762}, 644, 709

\bibitem[\protect\citeauthoryear{Gandhi, H\"{o}nig  \& Kishimoto}{Gandhi
  et~al.}{2015}]{gandhi_dust_2015}
Gandhi P.,  H\"{o}nig S.~F.,   Kishimoto M.,  2015, \mn@doi [\apj]
  {10.1088/0004-637X/812/2/113}, 812, 113

\bibitem[\protect\citeauthoryear{Gaskell}{1982}]{gaskell1982} Gaskell C.~M., 1982, ApJ, 263, 79

\bibitem[\protect\citeauthoryear{Gaskell}{Gaskell}{1988}]{gaskell_direct_1988}
Gaskell C.~M.,  1988, \mn@doi [\apj] {10.1086/165986}, 325, 114

\bibitem[\protect\citeauthoryear{Gaskell}{Gaskell}{2009}]{gaskell_what_2009}
Gaskell C.~M.,  2009, \mn@doi [New Astronomy Reviews]
  {10.1016/j.newar.2009.09.006}, 53, 140

\bibitem[\protect\citeauthoryear{Gaskell \& Goosmann}{Gaskell \&
  Goosmann}{2013}]{gaskell_line_2013}
Gaskell C.~M.,  Goosmann R.~W.,  2013, \mn@doi [\apj]
  {10.1088/0004-637X/769/1/30}, 769, 30

\bibitem[\protect\citeauthoryear{Gaskell \& Sparke}{Gaskell \&
  Sparke}{1986}]{gaskell_line_1986}
Gaskell C.~M.,  Sparke L.~S.,  1986, \mn@doi [\apj] {10.1086/164238}, 305, 175

\bibitem[\protect\citeauthoryear{Gibson, Marshall, Canizares  \& Lee}{Gibson
  et~al.}{2005}]{gibson_high-resolution_2005}
Gibson R.~R.,  Marshall H.~L.,  Canizares C.~R.,   Lee J.~C.,  2005, \mn@doi
  [ApJ] {10.1086/430199}, 627, 83

\bibitem[\protect\citeauthoryear{Gibson, Canizares, Marshall, Young  \&
  Lee}{Gibson et~al.}{2007}]{gibson_line_2007}
Gibson R.~R.,  Canizares C.~R.,  Marshall H.~L.,  Young A.~J.,   Lee J.~C.,
  2007, \mn@doi [ApJ] {10.1086/510441}, 655, 749

\bibitem[\protect\citeauthoryear{Gibson et~al.,}{Gibson
  et~al.}{2009}]{gibson_catalog_2009}
Gibson R.~R.,  et~al., 2009, \mn@doi [ApJ] {10.1088/0004-637X/692/1/758}, 692,
  758

\bibitem[\protect\citeauthoryear{Giustini \& Proga}{Giustini \&
  Proga}{2019}]{giustini_global_2019}
Giustini M.,  Proga D.,  2019, arXiv e-prints, p. arXiv:1904.07341

\bibitem[\protect\citeauthoryear{Goad, O`Brien  \& Gondhalekar}{Goad
  et~al.}{1993}]{goad_response_1993}
Goad M.~R.,  O`Brien P.~T.,   Gondhalekar P.~M.,  1993, \mn@doi [\mnras] {10.1093/mnras/263.1.149}, 263, 149

\bibitem[\protect\citeauthoryear{Goad, Korista  \& Ruff}{Goad
  et~al.}{2012}]{goad_broad_2012}
Goad M.~R.,  Korista K.~T.,   Ruff A.~J.,  2012, \mn@doi [MNRAS]
  {10.1111/j.1365-2966.2012.21808.x}, 426, 3086

\bibitem[\protect\citeauthoryear{Goad et~al.,}{Goad
  et~al.}{2016}]{goad_space_2016}
Goad M.~R.,  et~al., 2016, \mn@doi [\apj] {10.3847/0004-637X/824/1/11}, 824, 11

\bibitem[\protect\citeauthoryear{Gofford, Reeves, Tombesi, Braito, Turner,
  Miller  \& Cappi}{Gofford et~al.}{2013}]{gofford_suzaku_2013}
Gofford J.,  Reeves J.~N.,  Tombesi F.,  Braito V.,  Turner T.~J.,  Miller L.,
   Cappi M.,  2013, \mn@doi [\mnras] {10.1093/mnras/sts481}, 430, 60

\bibitem[\protect\citeauthoryear{{Gravity Collaboration} et~al.,}{{Gravity
  Collaboration} et~al.}{2018}]{gravity_collaboration_detection_2018}
{Gravity Collaboration} et~al., 2018, \mn@doi [\aap] {10.1051/0004-6361/201834294}, 618, L10

\bibitem[\protect\citeauthoryear{Green \& Mathur}{Green \&
  Mathur}{1996}]{green_broad_1996}
Green P.~J.,  Mathur S.,  1996, \mn@doi [\apj] {10.1086/177178},
  462, 637

\bibitem[\protect\citeauthoryear{Green, Aldcroft, Mathur, Wilkes  \&
  Elvis}{Green et~al.}{2001}]{green_chandra_2001}
Green P.~J.,  Aldcroft T.~L.,  Mathur S.,  Wilkes B.~J.,   Elvis M.,  2001,
  \mn@doi [\apj] {10.1086/322311}, 558, 109

\bibitem[\protect\citeauthoryear{Greene \& Ho}{Greene \&
  Ho}{2005}]{greene_estimating_2005}
Greene J.~E.,  Ho L.~C.,  2005, \mn@doi [\apj] {10.1086/431897}, 630, 122

\bibitem[\protect\citeauthoryear{Grupe, Mathur  \& Elvis}{Grupe
  et~al.}{2003}]{grupe_xmm-newton_2003}
Grupe D.,  Mathur S.,   Elvis M.,  2003, \mn@doi [\aj] {10.1086/377141}, 126, 1159

\bibitem[\protect\citeauthoryear{Hagino, Odaka, Done, Tomaru, Watanabe  \&
  Takahashi}{Hagino et~al.}{2016}]{hagino_disc_2016}
Hagino K.,  Odaka H.,  Done C.,  Tomaru R.,  Watanabe S.,   Takahashi T.,
  2016, \mn@doi [\mnras] {10.1093/mnras/stw1579}, 461, 3954

\bibitem[\protect\citeauthoryear{Hamann \& Koesterke}{Hamann \&
  Koesterke}{1998}]{hamann_spectrum_1998}
Hamann W.~R.,  Koesterke L.,  1998, \aap, 335, 1003

\bibitem[\protect\citeauthoryear{Hamann, Oskinova  \& Feldmeier}{Hamann
  et~al.}{2008}]{hamann_spectrum_2008}
Hamann W.-R.,  Oskinova L.~M.,   Feldmeier A.,  2008. p.~75, \url
  {https://ui.adsabs.harvard.edu/abs/2008cihw.conf...75H}

\bibitem[\protect\citeauthoryear{Hamann, Chartas, McGraw, Rodriguez~Hidalgo,
  Shields, Capellupo, Charlton  \& Eracleous}{Hamann
  et~al.}{2013}]{hamann_extreme-velocity_2013}
Hamann F.,  Chartas G.,  McGraw S.,  Rodriguez~Hidalgo P.,  Shields J.,
  Capellupo D.,  Charlton J.,   Eracleous M.,  2013, \mn@doi [\mnras] {10.1093/mnras/stt1231}, 435, 133

\bibitem[\protect\citeauthoryear{Hamann, Chartas, Reeves  \& Nardini}{Hamann
  et~al.}{2018}]{hamann_does_2018}
Hamann F.,  Chartas G.,  Reeves J.,   Nardini E.,  2018, \mn@doi [\mnras]
  {10.1093/mnras/sty043}, 476, 943

\bibitem[\protect\citeauthoryear{Hamann, Herbst, Paris  \& Capellupo}{Hamann
  et~al.}{2019a}]{hamann_structure_2019}
Hamann F.,  Herbst H.,  Paris I.,   Capellupo D.,  2019a, \mn@doi [\mnras]
  {10.1093/mnras/sty2900}, 483, 1808

\bibitem[\protect\citeauthoryear{Hamann, Tripp, Rupke  \& Veilleux}{Hamann
  et~al.}{2019b}]{hamann_emergence_2019}
Hamann F.,  Tripp T.~M.,  Rupke D.,   Veilleux S.,  2019b, \mn@doi [\mnras]
  {10.1093/mnras/stz1408}, 487, 5041

\bibitem[\protect\citeauthoryear{Hamann et al.}{2002}]{hamann2002} 
Hamann F., Korista K.~T., Ferland G.~J., Warner C., Baldwin J., 2002, ApJ, 564, 592


\bibitem[\protect\citeauthoryear{Higginbottom}{Higginbottom}{2014}]{nick_thesis}
Higginbottom N.,  2014, PhD thesis, University of Southampton, \url
  {https://eprints.soton.ac.uk/368584/}

\bibitem[\protect\citeauthoryear{Higginbottom, Knigge, Long, Sim  \&
  Matthews}{Higginbottom et~al.}{2013}]{higginbottom_simple_2013}
Higginbottom N.,  Knigge C.,  Long K.~S.,  Sim S.~A.,   Matthews J.~H.,  2013,
  \mn@doi [\mnras] {10.1093/mnras/stt1658},
  436, 1390

\bibitem[\protect\citeauthoryear{Higginbottom, Proga, Knigge, Long, Matthews
  \& Sim}{Higginbottom et~al.}{2014}]{higginbottom_line-driven_2014}
Higginbottom N.,  Proga D.,  Knigge C.,  Long K.~S.,  Matthews J.~H.,   Sim
  S.~A.,  2014, \mn@doi [\apj] {10.1088/0004-637X/789/1/19}, 789, 19

\bibitem[\protect\citeauthoryear{Higginbottom, Knigge, Long, Matthews, Sim  \&
  Hewitt}{Higginbottom et~al.}{2018}]{higginbottom_radiation-hydrodynamic_2018}
Higginbottom N.,  Knigge C.,  Long K.~S.,  Matthews J.~H.,  Sim S.~A.,   Hewitt
  H.~A.,  2018, \mn@doi [\mnras] {10.1093/mnras/sty1599}, 479, 3651

\bibitem[\protect\citeauthoryear{Hillier}{Hillier}{1991}]{hillier_effects_1991}
Hillier D.~J.,  1991, \aap, 247, 455

\bibitem[\protect\citeauthoryear{Hillier \& Miller}{Hillier \&
  Miller}{1999}]{hillier_constraints_1999}
Hillier D.~J.,  Miller D.~L.,  1999, \mn@doi [\apj] {10.1086/307339}, 519, 354

\bibitem[\protect\citeauthoryear{H{\"o}nig \& Beckert}{2007}]{honig_2007} H{\"o}nig S.~F., Beckert T., 2007, \mnras, 380, 1172

\bibitem[\protect\citeauthoryear{Hu, Wang, Ho, Chen, Bian  \& Xue}{Hu
  et~al.}{2008}]{hu_h_2008}
Hu C.,  Wang J.-M.,  Ho L.~C.,  Chen Y.-M.,  Bian W.-H.,   Xue S.-J.,  2008,
  \mn@doi [ApJL] {10.1086/591848}, 683, L115

\bibitem[\protect\citeauthoryear{Hunter}{Hunter}{2007}]{Hunter:2007}
Hunter J.~D.,  2007, \mn@doi [Computing in Science \& Engineering]
  {10.1109/MCSE.2007.55}, 9, 90

\bibitem[\protect\citeauthoryear{Hure, Collin-Souffrin, Le~Bourlot  \&
  Pineau~des Forets}{Hure et~al.}{1994}]{hure_structure_1994}
Hure J.~M.,  Collin-Souffrin S.,  Le~Bourlot J.,   Pineau~des Forets G.,  1994,
  \aap, 290, 19

\bibitem[\protect\citeauthoryear{Hutsem{\'e}kers, Sluse \& Kumar}{2019}]{hutsemekers_2019} Hutsem{\'e}kers D., Sluse D., Kumar P., 2019, arXiv, arXiv:1912.04336

\bibitem[\protect\citeauthoryear{Kaastra, Mewe, Liedahl, Komossa  \&
  Brinkman}{Kaastra et~al.}{2000}]{kaastra_x-ray_2000}
Kaastra J.~S.,  Mewe R.,  Liedahl D.~A.,  Komossa S.,   Brinkman A.~C.,  2000,
  A\&A, 354, L83

\bibitem[\protect\citeauthoryear{Kaastra et~al.,}{Kaastra
  et~al.}{2014}]{kaastra_multiwavelength_2014}
Kaastra J.~S.,  et~al., 2014, \mn@doi [A\&A] {10.1051/0004-6361/201424662},
  570, A73

\bibitem[\protect\citeauthoryear{Kaspi et~al.,}{Kaspi
  et~al.}{2001}]{kaspi_high-resolution_2001}
Kaspi S.,  et~al., 2001, \mn@doi [ApJ] {10.1086/321333}, 554, 216

\bibitem[\protect\citeauthoryear{Kaspi, Brandt, Collinge, Elvis  \&
  Reynolds}{Kaspi et~al.}{2004}]{kaspi_far_2004}
Kaspi S.,  Brandt W.~N.,  Collinge M.~J.,  Elvis M.,   Reynolds C.~S.,  2004,
  \mn@doi [AJ] {10.1086/383555}, 127, 2631

\bibitem[\protect\citeauthoryear{Kawaguchi \& Mori}{Kawaguchi \&
  Mori}{2010}]{kawaguchi_orientation_2010}
Kawaguchi T.,  Mori M.,  2010, \mn@doi [ApJL] {10.1088/2041-8205/724/2/L183},
  724, L183

\bibitem[\protect\citeauthoryear{King \& Nixon}{King \&
  Nixon}{2015}]{king_agn_2015}
King A.,  Nixon C.,  2015, \mn@doi [\mnras: Letters] {10.1093/mnrasl/slv098},
  453, L46

\bibitem[\protect\citeauthoryear{King \& Pringle}{King \&
  Pringle}{2007}]{king_fuelling_2007}
King A.~R.,  Pringle J.~E.,  2007, \mn@doi [\mnras Letters] {10.1111/j.1745-3933.2007.00296.x}, 377, L25

\bibitem[\protect\citeauthoryear{Knigge}{Knigge}{1999}]{knigge_effective_1999}
Knigge C.,  1999, \mn@doi [\mnras] {10.1046/j.1365-8711.1999.02839.x}, 309, 409

\bibitem[\protect\citeauthoryear{Knigge, Woods  \& Drew}{Knigge
  et~al.}{1995}]{knigge_application_1995}
Knigge C.,  Woods J.~A.,   Drew J.~E.,  1995, \mnras, 273, 225

\bibitem[\protect\citeauthoryear{Knigge, Scaringi, Goad  \& Cottis}{Knigge
  et~al.}{2008}]{knigge_intrinsic_2008}
Knigge C.,  Scaringi S.,  Goad M.~R.,   Cottis C.~E.,  2008, \mn@doi [\mnras] {10.1111/j.1365-2966.2008.13081.x}, 386,
  1426

\bibitem[\protect\citeauthoryear{Kollatschny \& Zetzl}{Kollatschny \&
  Zetzl}{2013}]{kollatschny_accretion_2013}
Kollatschny W.,  Zetzl M.,  2013, \mn@doi [\aap] {10.1051/0004-6361/201220923},
  551, L6

\bibitem[\protect\citeauthoryear{Konigl \& Kartje}{Konigl \&
  Kartje}{1994}]{konigl_disk-driven_1994}
Konigl A.,  Kartje J.~F.,  1994, \mn@doi [\apj] {10.1086/174746}, 434, 446

\bibitem[\protect\citeauthoryear{Koratkar \& Gaskell}{Koratkar \&
  Gaskell}{1989}]{koratkar_emission-line_1989}
Koratkar A.~P.,  Gaskell C.~M.,  1989, \mn@doi [\apj] {10.1086/167937}, 345,
  637

\bibitem[\protect\citeauthoryear{Koratkar \& Gaskell}{Koratkar \&
  Gaskell}{1991}]{koratkar_radius-luminosity_1991}
Koratkar A.~P.,  Gaskell C.~M.,  1991, \mn@doi [\apj] {10.1086/185977}, 370,
  L61

\bibitem[\protect\citeauthoryear{Korista \& Goad}{Korista \&
  Goad}{2000}]{korista_locally_2000}
Korista K.~T.,  Goad M.~R.,  2000, \mn@doi [\apj] {10.1086/308930}, 536, 284

\bibitem[\protect\citeauthoryear{Korista \& Goad}{Korista \&
  Goad}{2004}]{korista_what_2004}
Korista K.~T.,  Goad M.~R.,  2004, \mn@doi [\apj] {10.1086/383193}, 606, 749

\bibitem[\protect\citeauthoryear{Korista, Baldwin, Ferland  \& Verner}{Korista
  et~al.}{1997}]{korista_atlas_1997}
Korista K.,  Baldwin J.,  Ferland G.,   Verner D.,  1997, \mn@doi [\apjs] {10.1086/312966}, 108, 401

\bibitem[\protect\citeauthoryear{Krolik \& Kriss}{Krolik \&
  Kriss}{1995}]{krolik_observable_1995}
Krolik J.~H.,  Kriss G.~A.,  1995, \mn@doi [\apj] {10.1086/175896}, 447, 512

\bibitem[\protect\citeauthoryear{Krolik \& Kriss}{Krolik \&
  Kriss}{2001}]{krolik_warm_2001}
Krolik J.~H.,  Kriss G.~A.,  2001, \mn@doi [\apj] {10.1086/323442}, 561, 684

\bibitem[\protect\citeauthoryear{Krolik, Horne, Kallman, Malkan, Edelson  \&
  Kriss}{Krolik et~al.}{1991}]{krolik_ultraviolet_1991}
Krolik J.~H.,  Horne K.,  Kallman T.~R.,  Malkan M.~A.,  Edelson R.~A.,   Kriss
  G.~A.,  1991, \mn@doi [\apj] {10.1086/169918}, 371, 541

\bibitem[\protect\citeauthoryear{LaMassa et~al.,}{LaMassa
  et~al.}{2015}]{lamassa_discovery_2015}
LaMassa S.~M.,  et~al., 2015, \mn@doi [\apj] {10.1088/0004-637X/800/2/144},
  800, 144

\bibitem[\protect\citeauthoryear{Landi, Young, Dere, Zanna  \& Mason}{Landi
  et~al.}{2013}]{landi_chiantiatomic_2013}
Landi E.,  Young P.~R.,  Dere K.~P.,  Zanna G.~D.,   Mason H.~E.,  2013,
  \mn@doi [\apj] {10.1088/0004-637X/763/2/86}, 763, 86

\bibitem[\protect\citeauthoryear{Laor \& Brandt}{Laor \&
  Brandt}{2002}]{laor_luminosity_2002}
Laor A.,  Brandt W.~N.,  2002, \mn@doi [\apj] {10.1086/339476}, 569, 641

\bibitem[\protect\citeauthoryear{Laor \& Davis}{Laor \&
  Davis}{2014}]{laor_line-driven_2014}
Laor A.,  Davis S.~W.,  2014, \mn@doi [\mnras] {10.1093/mnras/stt2408}, 438,
  3024

\bibitem[\protect\citeauthoryear{Laor, Barth, Ho  \& Filippenko}{Laor
  et~al.}{2006}]{laor_is_2006}
Laor A.,  Barth A.~J.,  Ho L.~C.,   Filippenko A.~V.,  2006, \mn@doi [\apj]
  {10.1086/497908}, 636, 83

\bibitem[\protect\citeauthoryear{Lawther, Goad, Korista, Ulrich  \&
  Vestergaard}{Lawther et~al.}{2018}]{lawther_quantifying_2018}
Lawther D.,  Goad M.~R.,  Korista K.~T.,  Ulrich O.,   Vestergaard M.,  2018,
  \mn@doi [\mnras] {10.1093/mnras/sty2242}, 481, 533

\bibitem[\protect\citeauthoryear{Lazarova, Canalizo, Lacy  \& Sajina}{Lazarova
  et~al.}{2012}]{lazarova_nature_2012}
Lazarova M.~S.,  Canalizo G.,  Lacy M.,   Sajina A.,  2012, \mn@doi [\apj]
  {10.1088/0004-637X/755/1/29}, 755, 29

\bibitem[\protect\citeauthoryear{Leftley, H\"{o}nig, Asmus, Tristram, Gandhi,
  Kishimoto, Venanzi  \& Williamson}{Leftley
  et~al.}{2019}]{leftley_parsec-scale_2019}
Leftley J.~H.,  H\"{o}nig S.~F.,  Asmus D.,  Tristram K. R.~W.,  Gandhi P.,
  Kishimoto M.,  Venanzi M.,   Williamson D.~J.,  2019, arXiv e-prints, p.
  arXiv:1910.00600

\bibitem[\protect\citeauthoryear{Leighly}{2004}]{leighly2004} 
Leighly K.~M., 2004, ApJ, 611, 125

\bibitem[\protect\citeauthoryear{Leighly \& Casebeer}{2007}]{leighly2007a} 
Leighly K.~M., Casebeer D., 2007, ASPC,  365, ASPC..373

\bibitem[\protect\citeauthoryear{Leighly et al.}{2007}]{leighly2007b} 
Leighly K.~M., Halpern J.~P., Jenkins E.~B., Grupe D., Choi J., Prescott K.~B., 2007, ApJ, 663, 103

\bibitem[\protect\citeauthoryear{Liu, Luo, Brandt, Gallagher  \& Garmire}{Liu
  et~al.}{2018}]{liu_frequency_2018}
Liu H.,  Luo B.,  Brandt W.~N.,  Gallagher S.~C.,   Garmire G.~P.,  2018,
  \mn@doi [ApJ] {10.3847/1538-4357/aabe8d}, 859, 113

\bibitem[\protect\citeauthoryear{Long \& Knigge}{Long \&
  Knigge}{2002}]{long_modeling_2002}
Long K.~S.,  Knigge C.,  2002, \mn@doi [\apj] {10.1086/342879}, 579, 725

\bibitem[\protect\citeauthoryear{Lu et~al.,}{Lu et~al.}{2019}]{lu_active_2019}
Lu K.-X.,  et~al., 2019, arXiv e-prints, p. arXiv:1911.01852

\bibitem[\protect\citeauthoryear{Lucy}{Lucy}{1999a}]{lucy_computing_1999}
Lucy L.~B.,  1999a, \aap, 344, 282

\bibitem[\protect\citeauthoryear{Lucy}{Lucy}{1999b}]{lucy_improved_1999}
Lucy L.~B.,  1999b, \aap, 345, 211

\bibitem[\protect\citeauthoryear{Lucy}{Lucy}{2002}]{lucy_monte_2002}
Lucy L.~B.,  2002, \mn@doi [\aap]
  {10.1051/0004-6361:20011756}, 384, 725

\bibitem[\protect\citeauthoryear{Lucy}{Lucy}{2003}]{lucy_monte_2003}
Lucy L.~B.,  2003, \mn@doi [\aap]
  {10.1051/0004-6361:20030357}, 403, 261

\bibitem[\protect\citeauthoryear{Lucy \& Abbott}{Lucy \&
  Abbott}{1993}]{lucy_multiline_1993}
Lucy L.~B.,  Abbott D.~C.,  1993, \mn@doi [\apj] {10.1086/172402}, 405, 738

\bibitem[\protect\citeauthoryear{Luo et~al.,}{Luo et~al.}{2013}]{luo_weak_2013}
Luo B.,  et~al., 2013, \mn@doi [ApJ] {10.1088/0004-637X/772/2/153}, 772, 153

\bibitem[\protect\citeauthoryear{Luo et~al.,}{Luo et~al.}{2014}]{luo_weak_2014}
Luo B.,  et~al., 2014, \mn@doi [ApJ] {10.1088/0004-637X/794/1/70}, 794, 70

\bibitem[\protect\citeauthoryear{Luo et~al.,}{Luo
  et~al.}{2015}]{luo_x-ray_2015}
Luo B.,  et~al., 2015, \mn@doi [\apj] {10.1088/0004-637X/805/2/122}, 805, 122

\bibitem[\protect\citeauthoryear{MacGregor, Hartmann  \& Raymond}{MacGregor
  et~al.}{1979}]{macgregor_radiative_1979}
MacGregor K.~B.,  Hartmann L.,   Raymond J.~C.,  1979, \mn@doi [\apj] {10.1086/157213}, 231, 514

\bibitem[\protect\citeauthoryear{MacLeod et~al.,}{MacLeod
  et~al.}{2016}]{macleod_systematic_2016}
MacLeod C.~L.,  et~al., 2016, \mn@doi [\mnras] {10.1093/mnras/stv2997}, 457,
  389

\bibitem[\protect\citeauthoryear{Mangham, Knigge, Matthews, Long, Sim  \&
  Higginbottom}{Mangham et~al.}{2017}]{mangham_reverberation_2017}
Mangham S.~W.,  Knigge C.,  Matthews J.~H.,  Long K.~S.,  Sim S.~A.,
  Higginbottom N.,  2017, \mn@doi [\mnras] {10.1093/mnras/stx1863}, 471, 4788

\bibitem[\protect\citeauthoryear{Mas-Ribas}{Mas-Ribas}{2019}]{mas-ribas_radiation-pressure_2019}
Mas-Ribas L.,  2019, \mn@doi [\apj] {10.3847/1538-4357/ab4181}, 885, 95

\bibitem[\protect\citeauthoryear{Mathur et~al.,}{Mathur
  et~al.}{2000}]{mathur_thomson_2000}
Mathur S.,  et~al., 2000, \mn@doi [\apj]
  {10.1086/312617}, 533, L79

\bibitem[\protect\citeauthoryear{Matsumoto, Leighly  \& Marshall}{Matsumoto
  et~al.}{2004}]{matsumoto_chandra_2004}
Matsumoto C.,  Leighly K.~M.,   Marshall H.~L.,  2004, \mn@doi [ApJ]
  {10.1086/381666}, 603, 456

\bibitem[\protect\citeauthoryear{Matthews}{Matthews}{2016}]{matthews_disc_2016}
Matthews J.~H.,  2016, ] {10.5281/zenodo.1256805}

\bibitem[\protect\citeauthoryear{Matthews, Knigge, Long, Sim  \&
  Higginbottom}{Matthews et~al.}{2015}]{matthews_impact_2015}
Matthews J.~H.,  Knigge C.,  Long K.~S.,  Sim S.~A.,   Higginbottom N.,  2015,
  \mn@doi [\mnras] {10.1093/mnras/stv867},
  450, 3331

\bibitem[\protect\citeauthoryear{Matthews, Knigge, Long, Sim, Higginbottom  \&
  Mangham}{Matthews et~al.}{2016}]{matthews_testing_2016}
Matthews J.~H.,  Knigge C.,  Long K.~S.,  Sim S.~A.,  Higginbottom N.,
  Mangham S.~W.,  2016, \mn@doi [\mnras]
  {10.1093/mnras/stw323}, 458, 293

\bibitem[\protect\citeauthoryear{Matthews, Knigge  \& Long}{Matthews
  et~al.}{2017}]{matthews_quasar_2017}
Matthews J.~H.,  Knigge C.,   Long K.~S.,  2017, \mn@doi [\mnras]
  {10.1093/mnras/stx231}, 467, 2571

\bibitem[\protect\citeauthoryear{McCourt, Oh, O'Leary  \& Madigan}{McCourt
  et~al.}{2018}]{mccourt_characteristic_2018}
McCourt M.,  Oh S.~P.,  O'Leary R.,   Madigan A.-M.,  2018, \mn@doi [\mnras]
  {10.1093/mnras/stx2687}, 473, 5407

\bibitem[\protect\citeauthoryear{McLure \& Jarvis}{McLure \&
  Jarvis}{2002}]{mclure_measuring_2002}
McLure R.~J.,  Jarvis M.~J.,  2002, \mn@doi [\mnras]
  {10.1046/j.1365-8711.2002.05871.x}, 337, 109

\bibitem[\protect\citeauthoryear{Miller, Turner  \& Reeves}{Miller
  et~al.}{2008}]{miller_absorption_2008}
Miller L.,  Turner T.~J.,   Reeves J.~N.,  2008, \mn@doi [Astronomy \&
  Astrophysics] {10.1051/0004-6361:200809590}, 483, 437

\bibitem[\protect\citeauthoryear{Mizumoto, Ebisawa, Tsujimoto, Done, Hagino  \&
  Odaka}{Mizumoto et~al.}{2018}]{mizumoto_line-driven_2018}
Mizumoto M.,  Ebisawa K.,  Tsujimoto M.,  Done C.,  Hagino K.,   Odaka H.,
  2018, \mn@doi [arXiv e-prints] {10.1093/mnras/sty3056}, p. arXiv:1811.01966

\bibitem[\protect\citeauthoryear{Mizumoto, Done, Tomaru  \& Edwards}{Mizumoto
  et~al.}{2019}]{mizumoto_thermally_2019}
Mizumoto M.,  Done C.,  Tomaru R.,   Edwards I.,  2019, \mn@doi [\mnras]
  {10.1093/mnras/stz2225}, p.~2149

\bibitem[\protect\citeauthoryear{Morabito, Dai, Leighly, Sivakoff  \&
  Shankar}{Morabito et~al.}{2011}]{morabito_suzaku_2011}
Morabito L.~K.,  Dai X.,  Leighly K.~M.,  Sivakoff G.~R.,   Shankar F.,  2011,
  \mn@doi [\apj]
  {10.1088/0004-637X/737/1/46}, 737, 46

\bibitem[\protect\citeauthoryear{Morabito, Dai, Leighly, Sivakoff  \&
  Shankar}{Morabito et~al.}{2014}]{morabito_unveiling_2014}
Morabito L.~K.,  Dai X.,  Leighly K.~M.,  Sivakoff G.~R.,   Shankar F.,  2014,
  \mn@doi [\apj]
  {10.1088/0004-637X/786/1/58}, 786, 58

\bibitem[\protect\citeauthoryear{Morabito et~al.,}{Morabito
  et~al.}{2019}]{morabito_origin_2019}
Morabito L.~K.,  et~al., 2019, \mn@doi [\aap] {10.1051/0004-6361/201833821},
  622, A15

\bibitem[\protect\citeauthoryear{Morgan, Kochanek, Morgan  \& Falco}{Morgan
  et~al.}{2010}]{morgan_quasar_2010}
Morgan C.~W.,  Kochanek C.~S.,  Morgan N.~D.,   Falco E.~E.,  2010, \mn@doi
  [\apj]
  {10.1088/0004-637X/712/2/1129}, 712, 1129

\bibitem[\protect\citeauthoryear{Muijres, de Koter, Vink, Krtička, Kubát  \&
  Langer}{Muijres et~al.}{2011}]{muijres_predictions_2011}
Muijres L.~E.,  de Koter A.,  Vink J.~S.,  Krtička J.,  Kubát J.,   Langer
  N.,  2011, \mn@doi [\aap] {10.1051/0004-6361/201014290}, 526, A32

\bibitem[\protect\citeauthoryear{Murray \& Chiang}{Murray \&
  Chiang}{1996}]{murray_wind-dominated_1996}
Murray N.,  Chiang J.,  1996, \mn@doi [\nat] {10.1038/382789a0}, 382, 789

\bibitem[\protect\citeauthoryear{Murray \& Chiang}{Murray \&
  Chiang}{1997}]{murray_disk_1997}
Murray N.,  Chiang J.,  1997, \mn@doi [\apj]
  {10.1086/303443}, 474, 91

\bibitem[\protect\citeauthoryear{Murray, Chiang, Grossman  \& Voit}{Murray
  et~al.}{1995}]{murray_accretion_1995}
Murray N.,  Chiang J.,  Grossman S.~A.,   Voit G.~M.,  1995, \mn@doi [\apj]
  {10.1086/176238}, 451, 498

\bibitem[\protect\citeauthoryear{Nagao, Marconi \& Maiolino}{2006}]{nagao2006} 
Nagao T., Marconi A., Maiolino R., 2006, \aap, 447, 157

\bibitem[\protect\citeauthoryear{Netzer}{Netzer}{1990}]{netzer_agn_1990}
Netzer H.,  1990. pp 57--160, \url
  {https://ui.adsabs.harvard.edu/abs/1990agn..conf...57N}

\bibitem[\protect\citeauthoryear{Netzer, Laor  \& Gondhalekar}{Netzer
  et~al.}{1992}]{netzer_quasar_1992}
Netzer H.,  Laor A.,   Gondhalekar P.~M.,  1992, \mn@doi [\mnras]
  {10.1093/mnras/254.1.15}, 254, 15

\bibitem[\protect\citeauthoryear{Netzer}{1987}]{netzer_discs_1987} Netzer H., 1987, MNRAS, 225, 55


\bibitem[\protect\citeauthoryear{Noebauer \& Sim}{Noebauer \&
  Sim}{2019}]{noebauer_monte_2019}
Noebauer U.~M.,  Sim S.~A.,  2019, \mn@doi [Living Reviews in Computational
  Astrophysics] {10.1007/s41115-019-0004-9}, 5, 1

\bibitem[\protect\citeauthoryear{Noebauer, Long, Sim  \& Knigge}{Noebauer
  et~al.}{2010}]{noebauer_geometry_2010}
Noebauer U.~M.,  Long K.~S.,  Sim S.~A.,   Knigge C.,  2010, \mn@doi [\apj] {10.1088/0004-637X/719/2/1932},
  719, 1932

\bibitem[\protect\citeauthoryear{Nomura \& Ohsuga}{Nomura \&
  Ohsuga}{2017}]{nomura_line-driven_2017}
Nomura M.,  Ohsuga K.,  2017, \mn@doi [\mnras] {10.1093/mnras/stw2877}, 465,
  2873

\bibitem[\protect\citeauthoryear{Orr, Molendi, Fiore, Grandi, Parmar  \&
  Owens}{Orr et~al.}{1997}]{orr_soft_1997}
Orr A.,  Molendi S.,  Fiore F.,  Grandi P.,  Parmar A.~N.,   Owens A.,  1997,
  \aap, 324, L77

\bibitem[\protect\citeauthoryear{Otani et~al.,}{Otani
  et~al.}{1996}]{otani_variable_1996}
Otani C.,  et~al., 1996, \mn@doi [\pasj] {10.1093/pasj/48.2.211}, 48, 211

\bibitem[\protect\citeauthoryear{Owocki \& Cohen}{Owocki \&
  Cohen}{2006}]{owocki_effect_2006}
Owocki S.~P.,  Cohen D.~H.,  2006, \mn@doi [\apj] {10.1086/505698}, 648, 565

\bibitem[\protect\citeauthoryear{Owocki \& Rybicki}{Owocki \&
  Rybicki}{1984}]{owocki_instabilities_1984}
Owocki S.~P.,  Rybicki G.~B.,  1984, \mn@doi [\apj] {10.1086/162412}, 284, 337

\bibitem[\protect\citeauthoryear{Owocki \& Rybicki}{Owocki \&
  Rybicki}{1985}]{owocki_instabilities_1985}
Owocki S.~P.,  Rybicki G.~B.,  1985, \mn@doi [\apj] {10.1086/163697}, 299, 265

\bibitem[\protect\citeauthoryear{Peterson}{Peterson}{1993}]{peterson_reverberation_1993}
Peterson B.~M.,  1993, \mn@doi [\pasp] {10.1086/133140}, 105, 247

\bibitem[\protect\citeauthoryear{Peterson}{Peterson}{2014}]{peterson_measuring_2014}
Peterson B.~M.,  2014, \mn@doi [Space Science Reviews]
  {10.1007/s11214-013-9987-4}, 183, 253

\bibitem[\protect\citeauthoryear{Peterson \& Wandel}{Peterson \&
  Wandel}{1999}]{peterson_keplerian_1999}
Peterson B.~M.,  Wandel A.,  1999, \mn@doi [\apjl] {10.1086/312190}, 521, L95

\bibitem[\protect\citeauthoryear{Pounds, Reeves, King, Page, O`Brien  \&
  Turner}{Pounds et~al.}{2003}]{pounds_high-velocity_2003}
Pounds K.~A.,  Reeves J.~N.,  King A.~R.,  Page K.~L.,  O`Brien P.~T.,   Turner
  M. J.~L.,  2003, \mn@doi [\mnras] {10.1046/j.1365-8711.2003.07006.x}, 345,
  705

\bibitem[\protect\citeauthoryear{Proga \& Kallman}{Proga \&
  Kallman}{2004}]{proga_dynamics_2004}
Proga D.,  Kallman T.~R.,  2004, \mn@doi [\apj] {10.1086/425117}, 616, 688

\bibitem[\protect\citeauthoryear{Proga \& Waters}{Proga \&
  Waters}{2015}]{proga_cloud_2015}
Proga D.,  Waters T.,  2015, \mn@doi [\apj] {10.1088/0004-637X/804/2/137}, 804,
  137

\bibitem[\protect\citeauthoryear{Proga, Stone  \& Kallman}{Proga
  et~al.}{2000}]{proga_dynamics_2000}
Proga D.,  Stone J.~M.,   Kallman T.~R.,  2000, \mn@doi [\apj]
  {10.1086/317154}, 543, 686

\bibitem[\protect\citeauthoryear{Puls, Markova, Scuderi, Stanghellini,
  Taranova, Burnley  \& Howarth}{Puls et~al.}{2006}]{puls_bright_2006}
Puls J.,  Markova N.,  Scuderi S.,  Stanghellini C.,  Taranova O.~G.,  Burnley
  A.~W.,   Howarth I.~D.,  2006, \mn@doi [\aap] {10.1051/0004-6361:20065073},
  454, 625

\bibitem[\protect\citeauthoryear{Ramolla et~al.,}{Ramolla
  et~al.}{2018}]{ramolla_simultaneous_2018}
Ramolla M.,  et~al., 2018, \mn@doi [\aap] {10.1051/0004-6361/201732081}, 620,
  A137

\bibitem[\protect\citeauthoryear{Rankine et al.}{2019}]{rankine2019} 
Rankine A.~L., Hewett P.~C., Banerji M., Richards G.~T., 2019, arXiv, arXiv:1912.08700

\bibitem[\protect\citeauthoryear{Rees, Netzer  \& Ferland}{Rees
  et~al.}{1989}]{rees_small_1989}
Rees M.~J.,  Netzer H.,   Ferland G.~J.,  1989, \mn@doi [\apj]
  {10.1086/168155}, 347, 640

\bibitem[\protect\citeauthoryear{Reeves, O`Brien  \& Ward}{Reeves
  et~al.}{2003}]{reeves_massive_2003}
Reeves J.~N.,  O`Brien P.~T.,   Ward M.~J.,  2003, \mn@doi [\apjl]
  {10.1086/378218}, 593, L65

\bibitem[\protect\citeauthoryear{Reynolds}{Reynolds}{1997}]{reynolds_x-ray_1997}
Reynolds C.~S.,  1997, \mn@doi [MNRAS] {10.1093/mnras/286.3.513}, 286, 513

\bibitem[\protect\citeauthoryear{Reynolds \& Fabian}{Reynolds \&
  Fabian}{1995}]{reynolds_warm_1995}
Reynolds C.~S.,  Fabian A.~C.,  1995, \mn@doi [\mnras]
  {10.1093/mnras/273.4.1167}, 273, 1167

\bibitem[\protect\citeauthoryear{Richards, Vanden~Berk, Reichard, Hall,
  Schneider, SubbaRao, Thakar  \& York}{Richards
  et~al.}{2002}]{richards_broad_2002}
Richards G.~T.,  Vanden~Berk D.~E.,  Reichard T.~A.,  Hall P.~B.,  Schneider
  D.~P.,  SubbaRao M.,  Thakar A.~R.,   York D.~G.,  2002, \mn@doi [\aj]
  {10.1086/341167}, 124, 1

\bibitem[\protect\citeauthoryear{Richards et~al.,}{Richards
  et~al.}{2006}]{richards_spectral_2006}
Richards G.~T.,  et~al., 2006b, \mn@doi [\apjs] {10.1086/506525}, 166, 470

\bibitem[\protect\citeauthoryear{Richards et~al.,}{Richards
  et~al.}{2011}]{richards_unification_2011}
Richards G.~T.,  et~al., 2011, \mn@doi [\aj] {10.1088/0004-6256/141/5/167},
  141, 167

\bibitem[\protect\citeauthoryear{Ruff, Floyd, Webster, Korista  \& Landt}{Ruff
  et~al.}{2012}]{ruff_new_2012}
Ruff A.~J.,  Floyd D. J.~E.,  Webster R.~L.,  Korista K.~T.,   Landt H.,  2012,
  \mn@doi [\apj] {10.1088/0004-637X/754/1/18}, 754, 18

\bibitem[\protect\citeauthoryear{Runnoe, Shang \& Brotherton}{2013}]{runnoe_orientation_2013} Runnoe J.~C., Shang Z., Brotherton M.~S., 2013, MNRAS, 435, 3251

\bibitem[\protect\citeauthoryear{Runnoe et al.}{2013}]{runnoe_bals_2013} Runnoe J.~C., Ganguly R., Brotherton M.~S., DiPompeo M.~A., 2013, MNRAS, 433, 1778


\bibitem[\protect\citeauthoryear{Sabra \& Hamann}{Sabra \&
  Hamann}{2001}]{sabra_pg_2001}
Sabra B.~M.,  Hamann F.,  2001, \mn@doi [\apj] {10.1086/324043}, 563, 555

\bibitem[\protect\citeauthoryear{Saez, Brandt, Gallagher, Bauer  \&
  Garmire}{Saez et~al.}{2012}]{saez_long-term_2012}
Saez C.,  Brandt W.~N.,  Gallagher S.~C.,  Bauer F.~E.,   Garmire G.~P.,  2012,
  \mn@doi [\apj]
  {10.1088/0004-637X/759/1/42}, 759, 42

\bibitem[\protect\citeauthoryear{Sameer et~al.,}{Sameer
  et~al.}{2019}]{sameer_x-ray_2019}
Sameer et~al., 2019, \mn@doi [\mnras] {10.1093/mnras/sty2718}, 482, 1121

\bibitem[\protect\citeauthoryear{Schurch \& Done}{Schurch \&
  Done}{2006}]{schurch_failed_2006}
Schurch N.~J.,  Done C.,  2006, \mn@doi [\mnras]
  {10.1111/j.1365-2966.2006.10645.x}, 371, 81

\bibitem[\protect\citeauthoryear{Selsing, Fynbo, Christensen  \&
  Krogager}{Selsing et~al.}{2016}]{selsing_x-shooter_2016}
Selsing J.,  Fynbo J. P.~U.,  Christensen L.,   Krogager J.-K.,  2016, \mn@doi
  [\aap] {10.1051/0004-6361/201527096}, 585, A87

\bibitem[\protect\citeauthoryear{Shen et~al.,}{Shen
  et~al.}{2011}]{shen_catalog_2011}
Shen Y.,  et~al., 2011, \mn@doi [\apjs] {10.1088/0067-0049/194/2/45}, 194, 45

\bibitem[\protect\citeauthoryear{Shlosman \& Vitello}{Shlosman \&
  Vitello}{1993}]{shlosman_winds_1993}
Shlosman I.,  Vitello P.,  1993, \mn@doi [\apj] {10.1086/172670},
  409, 372

\bibitem[\protect\citeauthoryear{Shlosman, Vitello  \& Shaviv}{Shlosman
  et~al.}{1985}]{shlosman_active_1985}
Shlosman I.,  Vitello P.~A.,   Shaviv G.,  1985, \mn@doi [\apj]
  {10.1086/163278}, 294, 96

\bibitem[\protect\citeauthoryear{Shlosman, Begelman  \& Frank}{Shlosman
  et~al.}{1990}]{shlosman_fuelling_1990}
Shlosman I.,  Begelman M.~C.,   Frank J.,  1990, \mn@doi [\nat] {10.1038/345679a0}, 345, 679

\bibitem[\protect\citeauthoryear{Sim, Drew  \& Long}{Sim
  et~al.}{2005}]{sim_two-dimensional_2005}
Sim S.~A.,  Drew J.~E.,   Long K.~S.,  2005, \mn@doi [\mnras]
  {10.1111/j.1365-2966.2005.09472.x}, 363, 615

\bibitem[\protect\citeauthoryear{Sim, Long, Miller  \& Turner}{Sim
  et~al.}{2008}]{sim_multidimensional_2008}
Sim S.~A.,  Long K.~S.,  Miller L.,   Turner T.~J.,  2008, \mn@doi [\mnras] {10.1111/j.1365-2966.2008.13466.x}, 388,
  611

\bibitem[\protect\citeauthoryear{Sim, Proga, Miller, Long  \& Turner}{Sim
  et~al.}{2010}]{sim_multidimensional_2010}
Sim S.~A.,  Proga D.,  Miller L.,  Long K.~S.,   Turner T.~J.,  2010, \mn@doi
  [\mnras] {10.1111/j.1365-2966.2010.17215.x}, 408, 1396

\bibitem[\protect\citeauthoryear{Sim, Proga, Kurosawa, Long, Miller  \&
  Turner}{Sim et~al.}{2012}]{sim_synthetic_2012}
Sim S.~A.,  Proga D.,  Kurosawa R.,  Long K.~S.,  Miller L.,   Turner T.~J.,
  2012, \mn@doi [\mnras] {10.1111/j.1365-2966.2012.21816.x}, 426, 2859

\bibitem[\protect\citeauthoryear{Stalevski, Fritz, Baes, Nakos  \&
  Popović}{Stalevski et~al.}{2012}]{stalevski_3d_2012}
Stalevski M.,  Fritz J.,  Baes M.,  Nakos T.,   Popović L.~?,  2012, \mn@doi
  [\mnras] {10.1111/j.1365-2966.2011.19775.x}, 420, 2756

\bibitem[\protect\citeauthoryear{Steenbrugge et~al.,}{Steenbrugge
  et~al.}{2005}]{steenbrugge_simultaneous_2005}
Steenbrugge K.~C.,  et~al., 2005, \mn@doi [A\&A] {10.1051/0004-6361:20047138},
  434, 569

\bibitem[\protect\citeauthoryear{Steffen, Strateva, Brandt, Alexander,
  Koekemoer, Lehmer, Schneider  \& Vignali}{Steffen
  et~al.}{2006}]{steffen_x-ray--optical_2006}
Steffen A.~T.,  Strateva I.,  Brandt W.~N.,  Alexander D.~M.,  Koekemoer A.~M.,
   Lehmer B.~D.,  Schneider D.~P.,   Vignali C.,  2006, \mn@doi [\aj] {10.1086/503627}, 131, 2826

\bibitem[\protect\citeauthoryear{Storchi-Bergmann, Baldwin  \&
  Wilson}{Storchi-Bergmann et~al.}{1993}]{storchi-bergmann_double-peaked_1993}
Storchi-Bergmann T.,  Baldwin J.~A.,   Wilson A.~S.,  1993, \mn@doi [\apjl]
  {10.1086/186867}, 410, L11

\bibitem[\protect\citeauthoryear{Storchi-Bergmann, Schimoia, Peterson, Elvis,
  Denney, Eracleous  \& Nemmen}{Storchi-Bergmann
  et~al.}{2017}]{storchi-bergmann_double-peaked_2017}
Storchi-Bergmann T.,  Schimoia J.~S.,  Peterson B.~M.,  Elvis M.,  Denney
  K.~D.,  Eracleous M.,   Nemmen R.~S.,  2017, \mn@doi [\apj] {10.3847/1538-4357/835/2/236}835, 236

\bibitem[\protect\citeauthoryear{Strateva et~al.,}{Strateva
  et~al.}{2003}]{strateva_double-peaked_2003}
Strateva I.~V.,  et~al., 2003, \mn@doi [\aj] {10.1086/378367}, 126, 1720

\bibitem[\protect\citeauthoryear{Suganuma et al.}{2006}]{suganama_2006} 
Suganuma M., et al., 2006, \apj, 639, 46

\bibitem[\protect\citeauthoryear{Sulentic, Marziani  \&
  Dultzin-Hacyan}{Sulentic et~al.}{2000}]{sulentic_phenomenology_2000}
Sulentic J.~W.,  Marziani P.,   Dultzin-Hacyan D.,  2000, \mn@doi [ARA\&A] {10.1146/annurev.astro.38.1.521}, 38, 521

\bibitem[\protect\citeauthoryear{Sulentic, Bachev, Marziani, Negrete  \&
  Dultzin}{Sulentic et~al.}{2007}]{sulentic_c_2007}
Sulentic J.~W.,  Bachev R.,  Marziani P.,  Negrete C.~A.,   Dultzin D.,  2007,
  \mn@doi [\apj] {10.1086/519916}, 666, 757

\bibitem[\protect\citeauthoryear{Sundqvist, Owocki  \& Puls}{Sundqvist
  et~al.}{2018}]{sundqvist_2d_2018}
Sundqvist J.~O.,  Owocki S.~P.,   Puls J.,  2018, \mn@doi [\aap]
  {10.1051/0004-6361/201731718}, 611, A17

\bibitem[\protect\citeauthoryear{\v{S}urlan, Hamann, Kubát, Oskinova  \&
  Feldmeier}{\v{S}urlan et~al.}{2012}]{surlan_three-dimensional_2012}
\v{S}urlan B.,  Hamann W.-R.,  Kubát J.,  Oskinova L.~M.,   Feldmeier A.,
  2012, \mn@doi [\aap] {10.1051/0004-6361/201118590}, 541,
  A37

\bibitem[\protect\citeauthoryear{Teng et~al.,}{Teng
  et~al.}{2014}]{teng_nustar_2014}
Teng S.~H.,  et~al., 2014, \mn@doi [ApJ] {10.1088/0004-637X/785/1/19}, 785, 19

\bibitem[\protect\citeauthoryear{Tombesi, Cappi, Reeves, Nemmen, Braito,
  Gaspari  \& Reynolds}{Tombesi et~al.}{2013}]{tombesi_unification_2013}
Tombesi F.,  Cappi M.,  Reeves J.~N.,  Nemmen R.~S.,  Braito V.,  Gaspari M.,
  Reynolds C.~S.,  2013, \mn@doi [\mnras]
  {10.1093/mnras/sts692}, 430, 1102

\bibitem[\protect\citeauthoryear{Tuccillo, Bruni, DiPompeo, Brotherton,
  Pasetto, Kraus, Gonz\'{a}lez-Serrano  \& Mack}{Tuccillo
  et~al.}{2017}]{tuccillo_multiwavelength_2017}
Tuccillo D.,  Bruni G.,  DiPompeo M.~A.,  Brotherton M.~S.,  Pasetto A.,  Kraus
  A.,  Gonz\'{a}lez-Serrano J.~I.,   Mack K.~H.,  2017, \mn@doi [\mnras] {10.1093/mnras/stx333}, 467, 4763

\bibitem[\protect\citeauthoryear{Ulrich \& Horne}{Ulrich \&
  Horne}{1996}]{ulrich_month_1996}
Ulrich M.-H.,  Horne K.,  1996, \mn@doi [\mnras] {10.1093/mnras/283.3.748},
  283, 748

\bibitem[\protect\citeauthoryear{Urrutia, Becker, White, Glikman, Lacy, Hodge
  \& Gregg}{Urrutia et~al.}{2009}]{urrutia_first-2mass_2009}
Urrutia T.,  Becker R.~H.,  White R.~L.,  Glikman E.,  Lacy M.,  Hodge J.,
  Gregg M.~D.,  2009, \mn@doi [\apj] {10.1088/0004-637X/698/2/1095}, 698, 1095

\bibitem[\protect\citeauthoryear{Urry \& Padovani}{Urry \&
  Padovani}{1995}]{urry_unified_1995}
Urry C.~M.,  Padovani P.,  1995, \mn@doi [\pasp] {10.1086/133630}, 107, 803

\bibitem[\protect\citeauthoryear{van Regemorter}{van
  Regemorter}{1962}]{van_regemorter_rate_1962}
van Regemorter H.,  1962, \mn@doi [\apj] {10.1086/147445}, 136, 906

\bibitem[\protect\citeauthoryear{Vanden~Berk et~al.,}{Vanden~Berk
  et~al.}{2001}]{vanden_berk_composite_2001}
Vanden~Berk D.~E.,  et~al., 2001, \mn@doi [\aj] {10.1086/321167}, 122, 549

\bibitem[\protect\citeauthoryear{Verner \& Yakovlev}{Verner \&
  Yakovlev}{1995}]{verner_analytic_1995}
Verner D.~A.,  Yakovlev D.~G.,  1995, A\&AS, 109, 125

\bibitem[\protect\citeauthoryear{Verner, Barthel \& Tytler}{1994}]{verner1994} 
Verner D.~A., Barthel P.~D., Tytler D., 1994, \aaps, 108, 287

\bibitem[\protect\citeauthoryear{Vestergaard \& Wilkes}{Vestergaard \&
  Wilkes}{2001}]{vestergaard_empirical_2001}
Vestergaard M.,  Wilkes B.~J.,  2001, \mn@doi [\apjs] {10.1086/320357}, 134, 1

\bibitem[\protect\citeauthoryear{Waters \& Proga}{Waters \&
  Proga}{2016}]{waters_efficient_2016}
Waters T.,  Proga D.,  2016, \mn@doi [\mnras] {10.1093/mnrasl/slw056}, 460, L79

\bibitem[\protect\citeauthoryear{Waters \& Proga}{Waters \&
  Proga}{2019}]{waters_non-isobaric_2019}
Waters T.,  Proga D.,  2019, \mn@doi [\apj] {10.3847/1538-4357/ab10e1}, 875,
  158

\bibitem[\protect\citeauthoryear{Waters, Kashi, Proga, Eracleous, Barth  \&
  Greene}{Waters et~al.}{2016}]{waters_reverberation_2016}
Waters T.,  Kashi A.,  Proga D.,  Eracleous M.,  Barth A.~J.,   Greene J.,
  2016, \mn@doi [\apj] {10.3847/0004-637X/827/1/53}, 827, 53

\bibitem[\protect\citeauthoryear{Waters, Proga, Dannen  \& Kallman}{Waters
  et~al.}{2017}]{waters_synthetic_2017}
Waters T.,  Proga D.,  Dannen R.,   Kallman T.~R.,  2017, \mn@doi [\mnras]
  {10.1093/mnras/stx238}, 467, 3160

\bibitem[\protect\citeauthoryear{Waters \& Li}{2019}]{waters_2019} Waters T., Li H., 2019, arXiv, arXiv:1912.03382


\bibitem[\protect\citeauthoryear{Weymann, Morris, Foltz  \& Hewett}{Weymann
  et~al.}{1991}]{weymann_comparisons_1991}
Weymann R.~J.,  Morris S.~L.,  Foltz C.~B.,   Hewett P.~C.,  1991, \mn@doi
  [\apj] {10.1086/170020}, 373, 23

\bibitem[\protect\citeauthoryear{Williamson, H\"{o}nig  \& Venanzi}{Williamson
  et~al.}{2019}]{williamson_3d_2019}
Williamson D.,  H\"{o}nig S.,   Venanzi M.,  2019, \mn@doi [\apj]
  {10.3847/1538-4357/ab17d5}, 876, 137

\bibitem[\protect\citeauthoryear{Yong, Webster  \& King}{Yong
  et~al.}{2016}]{yong_black_2016}
Yong S.~Y.,  Webster R.~L.,   King A.~L.,  2016, \mn@doi [\pasa]
  {10.1017/pasa.2016.8}, 33, e009

\bibitem[\protect\citeauthoryear{Yong, Webster, King, Bate, O`Dowd  \&
  Labrie}{Yong et~al.}{2017}]{yong_kinematics_2017}
Yong S.~Y.,  Webster R.~L.,  King A.~L.,  Bate N.~F.,  O`Dowd M.~J.,   Labrie
  K.,  2017, \mn@doi [\pasa] {10.1017/pasa.2017.37}, 34

\bibitem[\protect\citeauthoryear{Yong, King, Webster, Bate, O'Dowd  \&
  Labrie}{Yong et~al.}{2018}]{yong_using_2018}
Yong S.~Y.,  King A.~L.,  Webster R.~L.,  Bate N.~F.,  O'Dowd M.~J.,   Labrie
  K.,  2018, \mn@doi [\mnras] {10.1093/mnras/sty1540}, 479, 4153

\bibitem[\protect\citeauthoryear{Young, Axon, Robinson, Hough  \& Smith}{Young
  et~al.}{2007}]{young_rotating_2007}
Young S.,  Axon D.~J.,  Robinson A.,  Hough J.~H.,   Smith J.~E.,  2007,
  \mn@doi [Nature] {10.1038/nature06319}, 450, 74

\bibitem[\protect\citeauthoryear{Zeilig~Hess, Levinson, Xu  \&
  Arav}{Zeilig~Hess et~al.}{2019}]{zeilig_hess_balqso_2019}
Zeilig~Hess M.,  Levinson A.,  Xu X.,   Arav N.,  2019, arXiv e-prints, p.
  arXiv:1910.09376

\bibitem[\protect\citeauthoryear{Zhu, Zhang  \& Tang}{Zhu
  et~al.}{2009}]{zhu_evidence_2009}
Zhu L.,  Zhang S.~N.,   Tang S.,  2009, \mn@doi [ApJ]
  {10.1088/0004-637X/700/2/1173}, 700, 1173


\makeatother
\end{thebibliography}

\appendix

\section{Parameter Sensitivity}
As discussed in section~\ref{sec:params}, here we show a limited exploration of parameter space. Fig.~\ref{fig:sed-comp} shows an investigation of the effect of both disc anisotropy and SED shape on the emergent spectrum at $20^\circ$. Fig.~\ref{fig:mdot-comp} shows the effect of changing $\dot{M}_{\mathrm{wind}}$, in tandem with the clumping factor. In both cases, Model A is used as the starting point.

\begin{figure*}
	\includegraphics[width=1.0\textwidth]{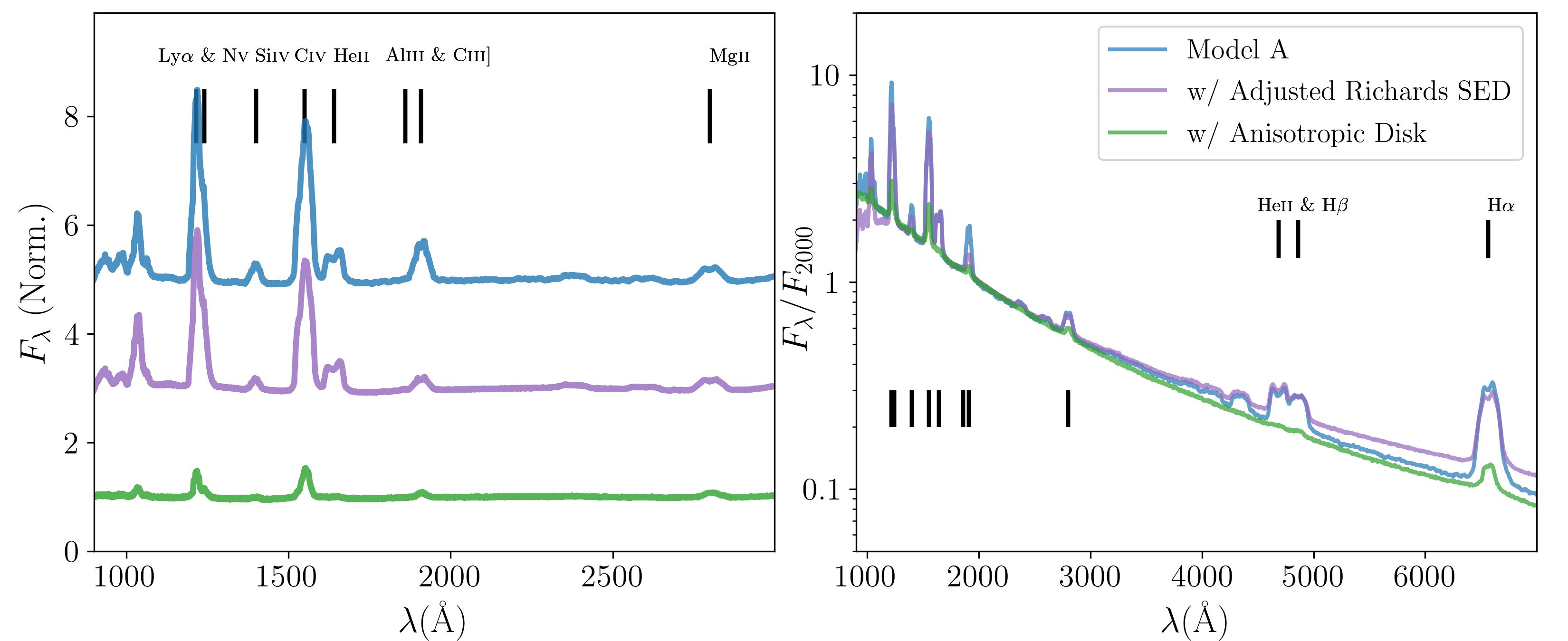}
    \caption{The effect of SED shape and anisotropy on the emergent UV and optical spectrum at $20^\circ$. The layout is the same as in Figs.~\ref{fig:modela} and \ref{fig:modelb}, except that the composite spectrum is not shown and three models are compared. One is Model A as presented earlier in the paper, one is the same model but with the adjusted mean quasar SED shown in Fig.~\ref{fig:seds}, based on the \protect\cite{richards_spectral_2006} mean quasar SED, and one is the same model but with the disc component originating from an anisotropic, thin disc that is both limb-darkened and foreshortened, as in M16. The disc anisotropy causes significantly more continuum flux to emerge at lower inclinations and also causes less incident flux to hit the outflow. As a result, the line EW at low inclinations is significantly decreased.
    }
    \label{fig:sed-comp}
\end{figure*}

\begin{figure*}
	\includegraphics[width=1.0\textwidth]{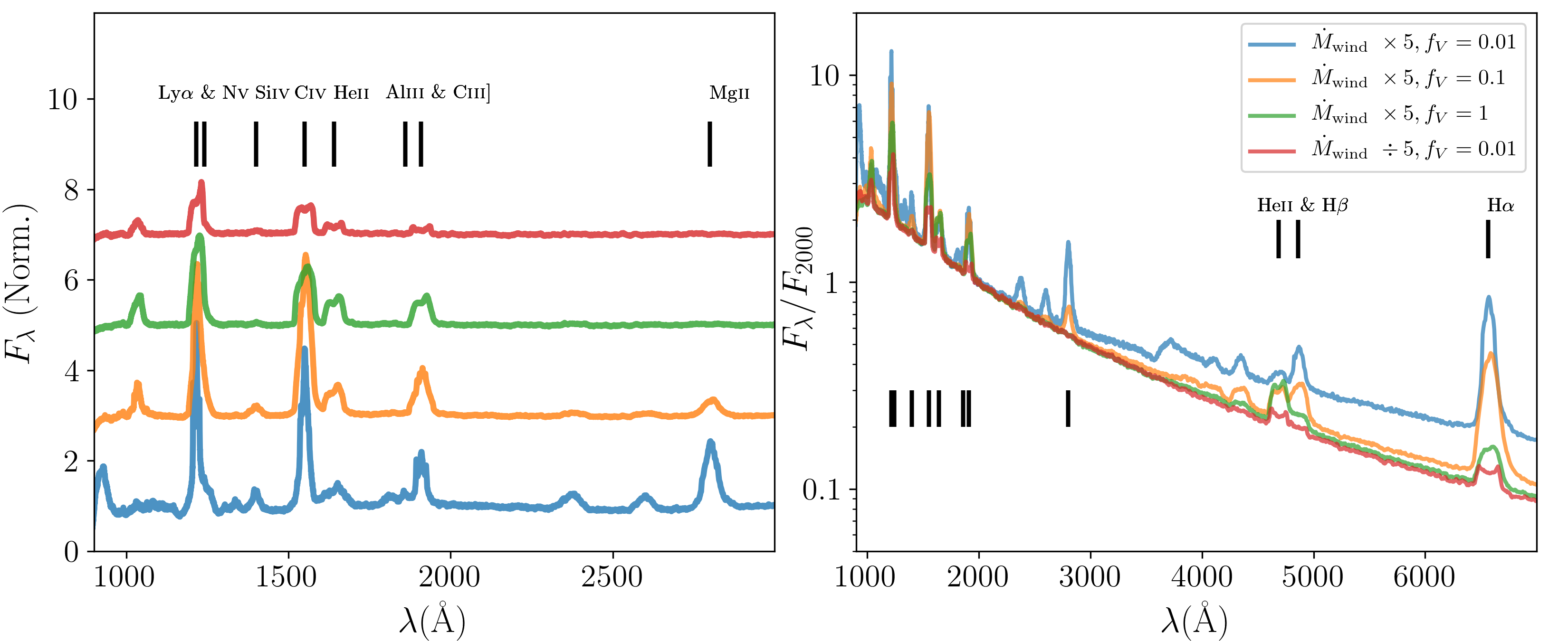}
    \caption{The effect of varying $\dot{M}_{\mathrm{wind}}$ on the emergent UV and optical spectrum at $20^\circ$. Model A is taken as the starting point. The layout is the same as in Figs.~\ref{fig:modela} and \ref{fig:modelb}, except that the composite spectrum is not shown and four models are compared. The four models use different combinations of values for $\dot{M}_{\mathrm{wind}}$ and $f_V$. 
    }
    \label{fig:mdot-comp}
\end{figure*}

% Don't change these lines
\bsp	% typesetting comment
\label{lastpage}
\end{document}